\documentclass[aoas]{imsart}
\usepackage{amsmath}
\usepackage{lstbayes}
\usepackage{amsthm}
\usepackage{amssymb}
\usepackage{amsfonts}
\RequirePackage[colorlinks,citecolor=blue,urlcolor=blue]{hyperref}
\usepackage{framed}
\usepackage[normalem]{ulem}
\usepackage{enumerate}
\usepackage{enumitem}
\usepackage{tikz}
\usetikzlibrary{bayesnet}
\usetikzlibrary{arrows}
\usepackage{xcolor}
\usepackage{graphicx}
\usepackage[utf8]{inputenc}
\usepackage{bbm}
\usepackage{mathtools}
\usepackage[font=small,labelfont=bf]{caption}
\usepackage{latexsym}
\usepackage{epsfig}
\usepackage{color}
\usepackage{xparse}
\usepackage{float}
\usepackage{lscape}
\usepackage{subfigure}
\usepackage{MnSymbol}
\usepackage{fancyhdr}
\usepackage{lastpage}
\usepackage{url}
\usepackage{mathrsfs}
\usepackage{cleveref}
\usepackage{multirow}

\DeclareSymbolFont{AMSb}{U}{msb}{m}{n}  
\DeclareMathSymbol{\Sph}{\mathbin}{AMSb}{"53} \DeclareMathSymbol{\R}{\mathbin}{AMSb}{"52}
\DeclareMathSymbol{\T}{\mathbin}{AMSb}{"54} \DeclareMathSymbol{\Z}{\mathbin}{AMSb}{"5A}
\DeclareMathSymbol{\K}{\mathbin}{AMSb}{"4B} \DeclareMathSymbol{\Pb}{\mathbin}{AMSb}{"50}
\DeclareMathSymbol{\Q}{\mathbin}{AMSb}{"51}
\DeclareMathSymbol{\N}{\mathbin}{AMSb}{"49}

\newcommand{\abs}[1]{\ensuremath{\left| #1 \right|}}

\newcommand{\norm}[1]{\ensuremath{\mathcal \| #1 \|}}

\newcommand{\lp}{\ensuremath{\left(}}
\newcommand{\rp}{\ensuremath{\right)}}
\newcommand{\lb}{\ensuremath{\left[}}
\newcommand{\rb}{\ensuremath{\right]}}

\newcommand{\ind}[1]{\ensuremath{\mathbbm{1}_{#1}}}

\NewDocumentCommand\Prob{mg}{\ensuremath{\Pb \IfNoValueTF{#2}{}{_{#2}} \lp #1 \rp}}
\NewDocumentCommand\Exp{mg}{\ensuremath{\mathbb{E} \IfNoValueTF{#2}{}{_{#2}} \lb #1 \rb}}

\newcommand{\inv}{\ensuremath{^{-1}}}

\newcommand{\iprod}[2]{\ensuremath{\llangle #1, #2 \rrangle}}
\newcommand{\Var}[1]{\ensuremath{\text{Var}\lp #1 \rp}}
\newcommand{\Cov}[2]{\ensuremath{\text{Cov}\lp #1, #2 \rp}}

\NewDocumentCommand\func{mg}{\ensuremath{f \IfNoValueTF{#2}{}{^{#2}} \lp #1 \rp}}
\NewDocumentCommand\charfunc{mg}{\ensuremath{\varphi \IfNoValueTF{#2}{}{_{#2}} \lp #1 \rp}}

\newcommand{\tikzcircle}[2][red,fill=red]{\tikz[baseline=-0.5ex]\draw[#1,radius=#2] (0,0) circle ;}%

\newtheorem{theorem}{Theorem}[section]
\newtheorem{lemma}[theorem]{Lemma}

\theoremstyle{plain}

\usepackage{amsbsy}

\crefname{enumi}{condition}{conditions}
\Crefname{enumi}{Condition}{Conditions}

\def\independenT#1#2{\mathrel{\rlap{$#1#2$}\mkern2mu{#1#2}}}

\newcommand{\indy}{\ensuremath{\perp\!\!\!\perp}}

\RequirePackage[authoryear]{natbib}
\begin{document}

\begin{frontmatter}
\title{Modeling racial/ethnic differences in COVID-19 incidence with covariates subject to non-random missingness}
\runtitle{Disease data with missing categorical data}

\begin{aug}
\author[A]{\fnms{Rob} \snm{Trangucci}\ead[label=e1]{trangucc@umich.edu}},
\author[B]{\fnms{Yang} \snm{Chen}\ead[label=e2]{ychenang@umich.edu}},
\and
\author[C]{\fnms{Jon} \snm{Zelner}\ead[label=e3]{jzelner@umich.edu}\ead[label=u1,url]{epibayes.io}}

\address[A]{Rob Trangucci is PhD candidate, Dept. of Statistics,
University of Michigan, Ann Arbor, MI, USA \printead{e1}.}

\address[B]{Yang Chen is Assistant Professor, Dept. of Statistics,
University of Michigan, Ann Arbor, MI, USA \printead{e2}.}

\address[C]{Jon Zelner is Associate Professor, Dept. of  Epidemiology \& Center for Social Epidemiology and Population Health,
University of Michigan School of Public Health, Ann Arbor, MI, USA  \printead{e3,u1}.}

\end{aug}

\begin{abstract}
Characterizing the cumulative burden of COVID-19 by race/ethnicity is of the utmost importance for public health researchers and policy makers in order to design effective mitigation measures. 
This analysis is hampered, however, by surveillance case data with substantial missingness in race and ethnicity covariates. 
Worse yet, this missingness likely depends on the values of these missing covariates, i.e. they are not missing at random (NMAR).
We propose a Bayesian parametric model that leverages joint information on spatial variation in the disease  
and covariate missingness processes and can accommodate both MAR and NMAR missingness. 
We show that the model is locally identifiable when the spatial distribution of the population covariates is known and observed cases can be associated with a spatial unit of observation.
We also use a simulation study to investigate the model's finite-sample performance. We compare our model's performance on NMAR data against complete-case analysis and multiple imputation (MI), both of which are commonly used by public health researchers when confronted with missing categorical covariates.
Finally, we model spatial variation in cumulative COVID-19 incidence in Wayne County, Michigan using data from the Michigan Department and Health and Human Services. 
The analysis suggests that population relative risk estimates by race during the early part of the COVID-19 pandemic in Michigan were understated for non-white residents compared to white residents when cases missing race were dropped or had these values imputed using MI.

\end{abstract}


\end{frontmatter}

\section{Introduction}

Complete and detailed surveillance data are critical sources of information for decision-making and communication in public health emergencies like the COVID-19 pandemic.
Under ideal conditions, these data can provide an indication of emerging trends, e.g. growth in socioeconomic inequity in infection and disease risk, which can be used to craft policies and target resources.\footnote{
Surveillance data are aggregated sets of disease cases, often subject to timely reporting requirements, meeting a common set of diagnostic characteristics so as to aid in the monitoring of disease outbreaks \citep{held2019handbook}.}  
For example, after surveillance data pointed to wide racial/ethnic disparities in incidence and mortality in the COVID-19 pandemic in the United States \citep{millettAssessingDifferentialImpacts2020,zelnerRacialDisparitiesCoronavirus2021}, policies intended to narrow the gap were put in place \cite{MI_taskforce,MI_policy}. However, without adequate information on the distribution of infection within and between different socioeconomic and race/ethnic groups, the impact of such policy measures is difficult to evaluate.

Missing covariates have long been a challenge associated with administrative datasets, such as public health surveillance data, and the scale and importance of this problem has only grown during the COVID-19 pandemic \citep{labgoldEstimatingUnknownGreater2021,millettAssessingDifferentialImpacts2020}. 
Covariate missingness in surveillance data may result from a variety of mechanisms,  ranging from non-response on an intake form, refusal to participate in tracing interviews, or data-entry errors after these data are collected. Often this missingness is implicitly or explicitly assumed to occur at random, i.e. not as a function of the disease process or the attributes of individual cases. But if the process causing case data to be missing important categorical variables, e.g. age, race, sex, or neighborhood, is dependent on the disease process, excluding cases that are missing  these covariates may result in estimates that are overconfident and biased. Furthermore, the direction of this bias is not easy to characterize, and may result in over- or under-estimates of group-level differences in risk as epidemic conditions shift. 

In emergency situations, such as a surging pandemic, it is easy to see how the disease process itself may induce non-random missingness of covariates. For example, during a period of rapidly increasing caseloads, such as the Delta and Omicron surges of the COVID-19 pandemic, the overwhelming number of cases is likely to limit the ability of case investigators to collect data that are as detailed as those collected during lower-incidence periods. These differences may also be more pronounced when comparing wealthier and poorer jurisdictions with differential resources for case-finding and intervention.  When these differential risks and resources are concentrated in communities with large proportions of non-White residents,  the likelihood that the missingness of key demographic information, including race/ethnicity, will depend on the race of respondents is high. The intensity of this missingness is also likely to vary in space, reflecting numerous factors including differences in epidemic conditions as well as varying data quality across public health jurisdictions. 

Both of these characteristics point to a nonignorable missing data problem, as presented in \cite{rubinInferenceMissingData1976}.
Both issues make quantifying the relative risk of infection among population strata during the COVID-19 pandemic potentially error-prone: high proportions of cases are missing demographic data, with missingness that is likely differential across population strata. It is in this scenario that omitting cases with missing demographic data may yield biased estimates of relative risk.
Tools that assume ignorability, like multiple imputation methods typically do \citep{audigier2018multiple}, cannot correct for missingness that depends on the value of the covariate, and will thus incur bias as well. 
This problem is not exclusive to the challenge of characterizing sociodemographic disparities in infection risk: incomplete reporting of vaccination status may lead to difficulty in estimating risks of breakthrough COVID-19 infections among vaccinated people, and missing information on comorbid conditions increasing the risk of death may complicate efforts at estimating risks of death associated with infection. 

In order to employ statistical methods that appropriately account for missingness, such as those presented in \cite{littleStatisticalAnalysisMissing2002}, one must make the modeling assumptions explicit in a joint probability model for the outcome variable, the covariate subject to missingness, and the missingness process for that covariate.
When the missingness process is nonignorable, two broad classes of models can be used to encode the assumptions about the joint distribution: selection models and pattern mixture models \citep{littleStatisticalAnalysisMissing2002}.
There is much literature on the theoretical and practical applications of both classes of models: \cite{diggle1994informative,clark2014validation,roy2008general}.
For a review of selection and pattern mixture models see \cite{littleSelectionPatternmixtureModels2008} and \citet{little1995modeling}.

We develop a novel model that accounts for nonignorable missingness of demographic covariates for which there is known population data, as in \cite{zangenehModelbasedMethodsRobusta}, but we take a selection model approach instead of a pattern mixture model approach.
Our probabilistic model is similar to that of \cite{stasnyHierarchicalModelsProbabilities1991}, wherein Stasny develops a selection model for nonignorable missingness in binary survey data, though we incorporate ideas from \cite{zangenehModelbasedMethodsRobusta} in using known census demographic data.  
Our approach is to develop a model that allows for simultaneous modeling of the disease and missingness processes, and that  incorporates information on spatial clustering of risk in addition to sociodemographic risk factors. 
Given the ubiquity of missing categorical covariates in public health surveillance data and the generality of our model, there are many potential applications of this class of models.

\subsection{Alternative approaches}

Because missing data can lead to ineffective and potentially life-threatening decision-making in public health and medicine, analysis of epidemiological data subject to missingness is an area of active research.
This work, however, is often focused on accounting for data missing-at-random  or on imputing values of continuous covariates.
Recent work focusing on accounting for missing covariates when modeling disease data in space and time like \cite{hollandSpatiotemporalModellingDisease2015a,gomez-rubioMissingDataAnalysis2019,bakerMissingSpaceEvaluation2014} suffer from several limitations. 
\cite{gomez-rubioMissingDataAnalysis2019} presents a framework for joint modeling of the disease process and missingness process, which can incorporate NMAR missingness, but only for continuous covariates.
When missing covariates are discrete, \citeauthor{gomez-rubioMissingDataAnalysis2019} resort to multiple imputation, which compromises statistical efficiency gained from joint modeling, increases the computational burden, and assumes MAR missingness.
\cite{hollandSpatiotemporalModellingDisease2015a} does include a model for discrete missing covariates with an outcome model, but the missing data mechanism is assumed to be MAR.
In other work, \cite{bakerMissingSpaceEvaluation2014} developed a cross-validation approach to missing data imputation, but assume MAR missingness.
Recent work in applications for missing data continue to assume MAR \citep{aguayoIdentifyingHotspotsCardiometabolic2020,labgoldEstimatingUnknownGreater2021}. 
As we argue above, assuming missingness is MAR can bias inferences;  \cite{perkinsPrincipledApproachesMissing2018,sidiTreatmentIncompleteData2018,stavsethHowHandlingMissing2019} explore how mistaken assumptions in the missing data model impact inferences. 
When missing data are inherently social in nature, MAR assumptions become even more tenuous than they might be in other settings. In the context of the COVID-19 pandemic, missingness of race/ethnicity data reflects a host of factors, including socioeconomic biases in the quality and thoroughness of public health data systems, which effectively guarantee correlation between the race/ethnicity of the respondent and their likelihood of missing these data. 
This paucity of recent work in spatial epidemiology employing an NMAR missingness model for discrete missing covariates, and the urgency of improving the quality of inferences from public health surveillance data provided the primary motivation for the development of our model.

The most germane work is \cite{labgoldEstimatingUnknownGreater2021}, which applies Bayesian Improved Surname Geocoding (BISG) to the estimation of race/ethnicity disparities in COVID-19 incidence using data from Fulton County, Georgia.
BISG was originally developed in \cite{elliottUsingCensusBureau2009} for understanding disparities in health outcomes when race data are not available.
The approach is an extension of a geocoding model for race, which generates a categorical distribution over race using the location of the unit of analysis (in \citeauthor{labgoldEstimatingUnknownGreater2021} the unit of analysis is a case-patient notified of a positive SARS-CoV-2 test).
BISG adds surname information to this categorical distribution, with the intention of more accurately imputing race.
The weakness of this approach is that the imputation is not informed by the outcome model or vice versa.
In the infectious disease context, the information that many cases for which one observes race or ethnicity should inform the categorical distribution for cases missing race information.
\citeauthor{labgoldEstimatingUnknownGreater2021} addresses this limitation by further modeling the misclassification rate for BISG by comparing BISG's imputed race to that of race for case-patients not missing race.
The procedure, however, does not correspond to a probabilistic model, which makes it challenging to validate its implicit assumptions. 
Furthermore, BISG assumes that the missingness process is missing-at-random, which may not be a good assumption in the context of missing race data.
\cite{zhangMultipleImputationMissing2022} also accounts for missing race data in COVID-19 cases via multiple imputation, again assuming MAR missingness.

Despite the wide-ranging literature on analyzing case with missing covariates, a common practice among academic and applied researchers is to omit observations with missing covariates, performing what is known as ``complete case analysis'', when confronted with missing data \citep{eekhoutMissingDataSystematic2012}.
For example, complete case analysis has been used in studies of racial/ethnic disparities of COVID-19 burden when race/ethnicity data are missing \citep{millettAssessingDifferentialImpacts2020,zelnerRacialDisparitiesCoronavirus2021} despite the authors' acknowledgement of the risks inherent in dropping incomplete cases.
This is an indication of the pervasiveness of the practice, in part due to the ease of performing complete case analysis in most statistical packages (e.g. the ubiquitous \texttt{na.rm=TRUE} argument in the R language) and in part due to the lack of methods available to researchers for nonignorable missingness.
That complete case analysis is widely employed should galvanize methodologists to develop techniques that are more finely attuned to infectious disease epidemiology.

\subsection{Considerations when imputing missing demographic information}

Imputing missing demographic data presents multiple challenges at the intersection of ethics, sociology, and statistics \citep{kennedy2020using}.
\citeauthor{kennedy2020using} note that, beyond the formidable statistical challenges, dealing with missing data of this nature requires understanding how demographic categories in administrative data have changed over time, the relationship between official categories and categories that individuals use to identify themselves, and the fact that attributes like sex, gender, and race/ethnicity must be understood through an intersectional lens rather than as independent dimensions of identity. 
Furthermore, the authors note that to the extent that there is an imbalance in how groups are misrepresented in surveys, even imputation with uncertainty confers statistical bias and, ultimately, discrimination.
The authors argue that despite these realities and being bound by data availability and antiquated study designs, researchers must take responsibility for the choices they make in handling missing data.

Omitting cases that have missing data may be mistaken as a safe practice when missingness of demographic information is assumed - implicitly or explicitly - to occur at random.
However, in real-world public health surveillance data, it is unlikely that such information will be missing at random. Instead, there is a high likelihood that the rate of missingness will be correlated with the burden of disease in a community and the resources local authorities have to address it, which are in turn often reflected in race/ethnic disparities in disease outcomes. As this burden increases and the financial and material tools to find new cases dwindle, it becomes increasingly likely that race/ethnic minority groups will be subject to higher rates of missingness which are positively associated with disease risk.
This intuition is reflected in our results, which show that when missingness does not occur completely random, dropping cases can result in biased estimates and overstatement of certainty in these estimates. 

Furthermore, even when the data are missing completely at random - the most innocuous scenario - random variation in which cases are missing can amount to statistical bias in finite samples.  
\citeauthor{kennedy2020using} note that this can serve as a form of discrimination if the conclusions from the analysis are used to draw inferences about the population and make decisions. This is particularly problematic when addressing missingness for groups that represent a small share of the observed data or overall population: In this case, dropping even a small number of cases missing data can result in diminished power to make valid inferences about group-specific risks. Because of this, making every effort to account for all sources of information on race/ethnicity, even those that are plausibly missing at random, is an ethical imperative. 
  
Exact probabilistic imputation, like the approach we present below, avoids some, but not all, of the risks of wrongly imputing demographic characteristics associated with deterministic approaches.
As \citeauthor{kennedy2020using} point out, probabilistic imputation does not guarantee bias-free conclusions, especially if the procedure's misclassification rates are not equally distributed across demographic subgroups.
For example, a model that mistakenly assumes that the missingness process is ignorable risks under-representing groups for which the baseline rate of missingness is higher or for whom missingness is positively correlated with the disease process. 
A well-designed procedure, with results interpreted with awareness of simplifying assumptions and their potential to induce bias, can facilitate increased relevance of study results for underrepresented groups, while also providing a more appropriate representation of uncertainty in these conclusions.

\subsection{Paper roadmap}
In this paper, we present a new joint model that allows researchers to account for relationships between variation in the disease  outcome measure of interest and the missingness process. This approach makes the flow of information and model assumptions clear, while at the same time affording researchers all the tools that have been developed to interrogate, summarize and present results from a coherent probabilistic model.

In the following sections, we will 1) Describe the justification for our approach and theoretical properties of the model, 2) Conduct a detailed simulation study to investigate the finite-sample performance of the model under several known data-generating processes, and finally, 3) Apply this model to detailed COVID-19 data from southeastern Michigan.



\section{Methods}\label{sec:method}
Suppose for each resident, indexed by $n$, in a large population with size $E$, the variable $U_n$ is a binary random variable that represents a diagnostic test result (e.g. COVID-19 polymerase chain reaction (PCR) test), $C_n$ is a categorical variable with $J$ levels that encodes race/ethnicity information which may be missing for some residents, $R_n$ is an indicator variable equal to $1$ if $C_n$ is observed and $0$ otherwise, and $S_n$ is a categorical variable encoding stratum information, like age or sex information for that resident.  
In other words, each resident is associated with the vector $(U_n, C_n, R_n, S_n)$, of which $U_n$ and $R_n$ are assumed to be random variables, while $C_n$ and $S_n$ are fixed characteristics of each resident. 

Let the variable $Y_{ij}$ be the total cases in the population for which $S_n = i$ and $C_n = j$, or more explicitly,
$$
Y_{ij} = \textstyle\sum_{\{n \,|\, S_n = i, C_n = j\}} U_n ,
$$
and let $E_{ij}$ be the total count of the population in stratum $i$ and race/ethnicity $j$:
$$E_{ij} = \sum_{n=1}^E \ind{S_n = i}\ind{C_n = j}.$$
Let the set of test results $U_{n}$ for the population be $\mathcal{U} \in \{0,1\}^E$.
Define $X_{ij}$ as the number of cases in stratum $i$ for which race/ethnicity $C_n$ is observed to be $j$ and $M_i$ as the number of cases in stratum $i$ as the number of cases missing race/ethnicity information:
$$
X_{ij} \,|\, \mathcal{U} = \textstyle\sum_{\{n \,|\, S_n = i, C_n = j, U_n = 1\}} R_n \quad\text{and}\quad M_{i} \,|\, \mathcal{U} = \textstyle\sum_{j} \sum_{\{n \,|\, S_n = i, C_n = j, U_n = 1\}} (1 - R_n)
$$
Let $Y_{ij}$ be conditionally independent Poisson random variables:\footnote{We discuss the Poisson distribution and the conditional independence assumption in \cref{subsec:future}}
$
Y_{ij} | \mu_{ij} \sim \text{Poisson}(\mu_{ij}).
$
where we define \emph{incidence} as $$\mu_{ij}/E_{ij},$$ or the per-capita rate of disease.
We further assume that race/ethnicity observation indicators $R_n$ are conditionally independent Bernoulli distributed random variables with probability observing race/ethnicity information denoted as $p_{ij}$, which depends solely on stratum $i$ and race/ethnicity category $j$. Then
$$
X_{ij} | \mathcal{U} \overset{d}{=} X_{ij} | Y_{ij}, \implies \quad X_{ij} | Y_{ij}, p_{ij} \sim \text{Binomial}(Y_{ij}, p_{ij})
$$
The distributional assumptions imply that marginalizing over total cases of race/ethnicity $j$ in stratum $i$, $Y_{ij}$, yields conditionally independent Poisson random variables: 
$$
X_{ij} | p_{ij}, \mu_{ij} \sim \text{Poisson}(p_{ij} \mu_{ij})
$$
for cases of race/ethnicity $j$ observed with race/ethnicity and missing cases are mutually independent of $X_{ij}$ and conditionally independent Poisson random variables:
$$
M_{i} | (p_{i1},\mu_{i1}), \dots,(p_{iJ},\mu_{iJ})  \sim \text{Poisson}(\textstyle\sum_j (1 - p_{ij}) \mu_{ij})
$$
We show the connection between our model and the missing data modeling paradigm introduced by \cite{rubinInferenceMissingData1976} and further developed in \cite{littleStatisticalAnalysisMissing2002} in \cref{app:selection-model}, and also show that the model implies that the missingness can be not missing at random (NMAR) if $p_{ij}$ vary by $j$. 

Given that $p_{i1} = p_{i2} = \dots = p_{iJ}$ for all $i$ is a strong constraint, allowing the model to learn the extent to which probability of observing race/ethnicity varies by race/ethnicity (i.e. allowing the model to learn how far missingness deviates from MAR) is the most judicious modeling choice.

\subsection{Modeling incidence when missingness is dependent on race/ethnicity}

We present a simple example of the model below, in which we assume that race/ethnicity is the only characteristic that predicts both disease incidence and the rate of  missingness for individual-level race/ethnicity information.
As above, we summarise population counts by age-sex stratum $i$ and race/ethnicity category $j$: 
$$E_{ij} = \sum_{n=1}^E \ind{S_n = i}\ind{C_n = j}.$$
Let $\mathbf{e}_i$ be the vector $(E_{i1}, \dots, E_{iJ})^T$ with the $j^{\rm th}$-element $E_{ij}$, and let 
$$\mathbf{E} \in \R^{I \times J} \, \text{such that } \, \mathbf{E}_{[i,:]} = \mathbf{e}_i^T.$$
If exposure to the disease is governed solely by race/ethnicity, and infection probability and exposure is constant across age-sex strata $i$, then we may assume that $\mu_{ij} = \lambda_j E_{ij}$ and that $p_{ij} = p_j, \forall i$. 
The observed data model, letting $\boldsymbol{\lambda} = (\lambda_1, \dots, \lambda_J)$ and $\mathbf{p} = (p_1, \dots, p_J)$, simplifies to the following:
\begin{align}\label{model:no-cov-obs}
  \begin{split}
X_{ij} | p_j, \lambda_j, E_{ij} & \sim \text{Poisson}(p_{j} \lambda_j E_{ij}), \\
M_{i} | \mathbf{p}, \boldsymbol{\lambda}, \mathbf{e}_{i} & \sim \text{Poisson}(\textstyle \sum_{j} \lambda_{j} (1 - p_j) E_{ij});
\end{split}
\end{align}
where we have made the conditioning on parameters and population counts explicit.


\subsubsection{Identifiability properties of the model}

We can show that model \eqref{model:no-cov-obs} is globally identifiable by appealing to Theorem 4 of \cite{rothenbergIdentificationParametricModels1971}. Theorem~\ref{thm:glob-id-no-cov} shows that the model parameters $(\mathbf{p},\boldsymbol{\lambda})$ are globally identifiable given the observed data under minimal conditions on the parameters, and an easily verifiable condition on the population count matrix $\mathbf{E}$.
\begin{theorem}\label{thm:glob-id-no-cov}
  The observational model \eqref{model:no-cov-obs} is globally identifiable under the following conditions:
  \begin{enumerate}[label=(E.\alph*)]
  \item $\mathbf{E}$ is rank $J$,\label{cond:E-mat}
  \item $\lambda_j \in (0,\infty) \forall j \in [1, \dots, J]$,\label{cond:lambda}
  \item $p_j \in (0,1) \forall j \in [1, \dots, J]$.\label{cond:p}
    \end{enumerate}
  \end{theorem}
  \begin{proof}
    Reparameterize the model from $(\lambda_j, p_j)$ to $(v_j, u_j)$ where $v_j =  p_j\lambda_j$ and $u_j =  (1 - p_j)\lambda_j$.
    Given \crefrange{cond:lambda}{cond:p}, the mapping is one-to-one and onto.
    Let $\mathbf{u}$ be the $J$-vector with element $j$ equal to $u_j$ and let $\mathbf{v}$ be similarly defined for $v_j$.
The reparameterized model becomes:
\begin{align}
  X_{ij} | v_j, E_{ij} & \sim \text{Poisson}(v_j E_{ij}) \\
  M_{i} | \mathbf{u}, \mathbf{e}_{i} & \sim \text{Poisson}(\textstyle \sum_{j} u_j E_{ij})
\end{align}
We know that $\hat{v}_j = \frac{\sum_i X_{ij}}{\sum_i E_{ij}}$ is unbiased for
$v_j$.
Let $\mathbf{m}$ be the $I$-vector with element $i$ equal to $M_i$.
Then 
$$
\Exp{\mathbf{m}} = \mathbf{E} \mathbf{u}
$$
By \cref{cond:E-mat} 
\begin{align}
  \lp \mathbf{E}^T\mathbf{E} \rp^{-1} \mathbf{E}^T \Exp{\mathbf{m}} = \mathbf{u}.
\end{align}
Given that we can define unbiased estimators for $\mathbf{v}$ and $\mathbf{u}$ , by Theorem 4 in \cite{rothenbergIdentificationParametricModels1971}, the model is globally identifiable in $(v_j, u_j)$.
Given that our mapping from $(\lambda_j, p_j)$ to $(v_j, u_j)$ is one-to-one and onto, global identifiability in $(v_j, u_j)$ implies global identifiability in $(\lambda_j, p_j)$ because we can define an inverse mapping from $(v_j, u_j)$ to $(\lambda_j, p_j)$. 
\end{proof}
It can also be seen that the variance-covariance matrix for the estimator for $\hat{\boldsymbol{\lambda}}$ is a sum of two components: the variance of $\hat{\mathbf{v}}$
and the variance-covariance matrix of the unbiased linear estimator for $\mathbf{u}$, $\lp\mathbf{E}^T\mathbf{E} \rp^{-1} \mathbf{E}^T \mathbf{m}$. 
This coincides with the Fisher information matrix, as derived in Appendix Section \ref{app:id-simple}, where we also show that the Fisher information matrix is positive definite under \crefrange{cond:E-mat}{cond:p}. 

\subsubsection{Model intuition} \label{subsec:model-intuit}
We examine a simple setting in which there are two race/ethnicity groups (or equivalently, $J = 2$) subject to missingness.
The unbiased estimator, $\hat{u}_1$, for $u_1 = (1 - p_1) \lambda_1$ can be expressed in terms of a projection matrix:
$$
\mathbf{P}_2 = \mathbf{e}_2 (\mathbf{e}_2^T \mathbf{e}_2)^{-1}\mathbf{e}_2^T
$$
which is the projection for a vector in $\R^I$ to the subspace spanned by $\mathbf{e}_2$.
Then
\begin{align} \label{eq:unbiased-u}
\hat{u}_1 = \frac{\mathbf{m}^T (\mathbf{I} - \mathbf{P}_2)\mathbf{e}_1}{\norm{(\mathbf{I} - \mathbf{P}_2)\mathbf{e}_1}_2^2}.
\end{align}
which can be understood as a relative measure of the strength of the covariance between the number of cases with missing race/ethnicity information and population counts for group $1$ after accounting for the variation in population counts attributable to group $2$.
The following estimator is unbiased for $\lambda_1$:
\begin{align} \label{eq:unbiased-lambda}
\frac{\sum_{i=1}^I X_{i1}}{\sum_{i=1}^I E_{i1}} + \hat{u}_1.
\end{align}
The first term in \cref{eq:unbiased-lambda} is the estimator for $\lambda_1 p_1$ for a Poisson distribution with rate $\lambda_1 p_1$, while the second term is a correction to account for missingness. 
If the covariance of $\mathbf{m}$ and a residualized $\mathbf{e}_1$ (by regressing $\mathbf{e}_1$ on $\mathbf{e}_2$) is large relative to the variance of the residualized $\mathbf{e}_1$ then the correction will be large. 
If, on the contrary, this quantity is small, the correction factor will be small.
The estimator depends on the conditional expectation of the number of cases for each race/ethnicity being proportional to the number of residents in that category, which is a common assumption in modeling count data in epidemiology \citep{fromeAnalysisRatesUsing1983,fromeUSEPOISSONREGRESSION1985,lashmodernepi}. 
\subsubsection{Bayesian inference and prior sensitivity}\label{subsec:prior-sens}
The two group setting in which one group's population is small compared to the other group's population motivates the careful choice of priors when doing Bayesian inference.
We will show that in this setting the  posterior mean for the rate of disease in the minority group is sensitive to priors. 
As above, let $j \in \{1, 2\}$ and let the rate of disease in group $j$ be $\lambda_j$ while the probability of observing race/ethnicity $j$ is $p_j$.
Under the following priors:
\begin{align}
\begin{split}\label{eq:simple-priors}
    p_j & \overset{\text{iid}}{\sim} \text{Beta}(\alpha_j, \beta_j) \\
    \lambda_j &\overset{\text{iid}}{\sim} \text{Gamma}(\alpha_j + \beta_j, r_j),
    \end{split}
\end{align}
$v_j = p_j \lambda_j \indy u_j = (1 - p_j) \lambda_j$.
As shown in \cref{thm:glob-id-no-cov}, we can write the observed-data likelihood in terms of $u_j$ and $v_j$.
If we assume, without loss of generality, that the majority group is $2$ for all $i$ (i.e. that $E_{i1} \ll E_{i2}$ for all $i$) we can make a likelihood approximation, detailed in Appendix Section \ref{app:post-mean-u}, that allows us to compute a closed-form approximate posterior mean and variance for $\lambda_1$, as well as the partial derivative of the posterior mean with respect to the prior rate parameter, $r_1$, for $\lambda_1$ when $\beta_1 = 1$, 
which is the second shape parameter for the beta prior over $p_1$ as well as part of the shape parameter for $\lambda_1$.

Let $s_1 = \textstyle\sum_i \frac{m_i E_{i1}}{E_{i2}}, s_2 = \textstyle\sum_i \frac{m_i E_{i1}^2}{E_{i2}^2}$, and $E_{+1} = \textstyle \sum_i E_{i1}$. Further, let $\phi$ and $\Phi$ be the density and distribution function of the standard normal distribution, respectively, and $z = \frac{s_1 - u_2(r_1  + E_{+1})}{\sqrt{s_2}}$.
If $\beta_1 = 1$, the posterior mean for $\lambda_1$ given $u_2$ is then
\begin{align}\label{eq:post-mean-lambda}
\Exp{\lambda_1 | u_2, r_1, \beta_1 = 1} = \frac{\alpha_1 + \sum_i x_{i1}}{r_1 + E_{+1}} + 
   \frac{s_1 - u_2(r_1  + E_{+1})}{s_2/u_2} + \frac{u_2}{\sqrt{s_2}} \phi(z)\Phi(z)^{-1}
\end{align}
with variance:
\begin{align}\label{eq:post-var-lambda}
\Var{\lambda_1 | u_2, r_1, \beta_1 = 1} = \frac{\alpha_1 + \sum_i x_{i1}}{(r_1 + \sum_i E_{i1})^2} + \frac{u_2^2}{s_2}\lp 1 - z\phi(z)\Phi(z)^{-1} - \phi(z)^2\Phi(z)^{-2})\rp
\end{align}
Like our unbiased estimator for $\lambda_1$ in \cref{eq:unbiased-lambda}, the first term in \cref{eq:post-mean-lambda} is the posterior mean for the rate of a Poisson random variable with rate $\lambda_1 p_1$, while the second term is the correction for missing data.
The first term of the correction, $\frac{s_1}{s_2/u_2}$\footnote{$\frac{s_1}{I} = \textstyle\sum_i \frac{m_i E_{i1}}{E_{i2}}/I$ is an approximate empirical covariance between the vector $\mathbf{m}$ and a vector with elements $\frac{E_{i1}}{E_{i2}}$ because $\frac{E_{i1}}{E_{i2}} \to 0$.} can be seen as an approximate least squares estimator for $u_1$ scaled by the weighted average of $m_i$ by dividing top and bottom by $\sum_i \frac{E_{i1}^2}{E_{i2}^2}$. The estimate is shrunk towards zero with magnitude dependent on $u_2$ and $r_1$. In fact, for an increase in $u_2$ the posterior conditional mean is shrunk towards zero, while an increase in $r_1$ similarly shrinks the posterior mean towards zero. 
This agrees with intuition that as the prior rate parameter for $\lambda_1$ increases, the prior mean decreases, and so too does the posterior mean.
This can be seen from the partial derivative of the posterior mean with respect to $r_1$, which we show in \cref{app:post-mean-u} to be
$\frac{\partial \Exp{\lambda_1 | u_2, r_1, \beta_1 = 1}}{\partial r_1} = -\Var{\lambda_1 | u_2, r_1, \beta_1 = 1}$.
The magnitude of the derivative is equal to that of the variance, \cref{eq:post-var-lambda}, which implies that the posterior mean is sensitive to $r_1$.
This sensitivity does not decline as $E_{i1}, E_{i2} \to \infty$ such that group 1 remains a minority to group 2.
Suppose that we take $E_{i1}, E_{i2} \to \infty$ such that $\frac{E_{i1}}{E_{i2}} = O(\frac{1}{E_{i1}})$. 
Then $\frac{E_{i1}^2}{E_{i2}} \to K < \infty$ for all $i$. Let $u_2^\star$, and $u_1^\star$ be the true data generating parameters, and let $z^\star = {(u_1^\star I K - u_2^\star r_1)}/{\sqrt{u_2^\star I K}} + \mathcal{Z}$, where $\mathcal{Z} \sim \mathrm{N}(0,1)$. The posterior mean and variance for $\lambda_1$ have the following convergence in distribution as $E_{i1}, E_{i2}$ goes to infinity in the same order:
\begin{align*}
\Exp{\lambda_1 | u_2, r_1, \beta_1 = 1} & \overset{d}{\to} v_1^\star + \frac{u_1^\star I K - r_1 u_2^\star}{I K}
+ \frac{\sqrt{u_2^\star} \phi(z^\star)\Phi(z^\star)^{-1}}{\sqrt{I K}} + \sqrt{\frac{u_2^\star}{IK}}\mathcal{Z}, \\
\Var{\lambda_1 | u_2, r_1, \beta_1 = 1} & \overset{d}{\to} \frac{u_2^\star}{IK}\lp 1 - z^\star\phi(z^\star)\Phi(z^\star)^{-1} - \phi(z^\star)^2\Phi(z^\star)^{-2})\rp. 
\end{align*}
We can see that as $I \to \infty$, the posterior mean for $\lambda_1$ converges in probability to $v^\star + u^\star$, or the true data generating parameter, as we would expect for a globally identifiable model.
However, for fixed $I$, the posterior mean remains both dependent on $u_2^\star$ and $r_1$, and, moreover, the derivative of the posterior mean with respect to $r_1$ can be seen to remain bounded away from zero and of the same magnitude as the posterior variance.

Nor does the sensitivity of the posterior mean for $\lambda_1$ to changes in $\beta_1$ diminish.
We can calculate the limit for the expression $(\Exp{\lambda_1 | u_2, r_1, \beta_1 = 2} - \allowbreak\Exp{\lambda_1 | u_2, r_1, \beta_1 = 1}) / \allowbreak\sqrt{\Var{\lambda_1 | u_2, r_1, \beta_1 = 1}}$, or the change in posterior mean scaled by the posterior standard deviation.
The form is shown in the appendix to be
$$
\frac{\sqrt{1  - z^\star\phi(z^\star)\Phi(z^\star)^{-1} - \phi(z^\star)^2\Phi(z^\star)^{-2}}}{z^\star + \phi(z^\star)\Phi(z^\star)^{-1}}.
$$
which approaches $1$ as $\Exp{z^\star}\to -\infty$.

This analysis implies that posterior inferences can be sensitive to both prior hyperparameters $\beta_j$ and $r_j$ for minority groups and can behave like a partially identified model asymptotically \citep{gustafson2015bayesian}.
In terms of classical approaches to prior sample size, like \cite{gelman2013bayesian}, a change from $p_j \sim \text{Beta}(1, 1)$ to $p_j \sim \text{Beta}(1, 2)$, or from $\lambda_j \sim \text{Gamma}(2, 100)$ to $\lambda_j \sim \text{Gamma}(3, 150)$ represents a small increase in prior information, but this can translate to large changes in the posterior mean for $\lambda_j$ for small minority populations.

Despite this prior sensitivity, we show in \Cref{fig:asympt-rmse} in \Cref{app:post-mean-u}  that for reasonable values of cumulative incidence and race/ethnicity reporting rate for the majority group, the posterior mean dominates the maximum likelihood estimator in terms of root mean squared error for a large range of $u_1^\star$.
Moreover, the asymptotic MLE for $u_1^\star$ has a non-negligible probability of being zero, whereas the posterior mean is almost surely positive.
These results demonstrate the benefits of using Bayesian inference over classical maximum likelihood estimation.

\subsection{Modeling incidence when missingness is dependent on both age-sex and  race/ethnicity}\label{sec:mode-w-covariates}    

Now assume that the rate of incident cases for age-sex stratum $i$ in race/ethnicity group  $j$, or 
observation $(i,j)$,
depends on fully-observed covariates, $\mathbf{z}_i \in \R^K$, associated with each stratum $i$. 
In the context of COVID-19, we expect that age-sex stratum will predict both exposure and probability of infection and disease given exposure, as well as the probability of race/ethnicity being recorded, so it is important to extend our model to incorporate this information.
We assume that coefficients for $\mathbf{z}_i$, $\boldsymbol{\beta}$ for incidence and  $\boldsymbol{\gamma}$ for race/ethnicity missingness, both in $\R^K$, are shared between race/ethnicity groups, which amounts to assuming there is no interaction between race and age-sex strata for predicting incidence and missingness. As above, we allow average incidence, $\lambda_j$, and log-odds of observing race/ethnicity or $\eta_j$ to vary by group $j$.
Let $\mathbf{p}_i$ be the length-$J$ vector with $j^\text{th}$ element equal to $p_{ij}$.
Then we can define the following observed-data model:
\begin{align}
  \begin{split}\label{model:full}
    X_{ij} | \lambda_j, \mathbf{z}_i, \boldsymbol{\beta}, p_{ij}, E_{ij} & \sim \text{Poisson}(p_{ij} \lambda_j \exp(\mathbf{z}_i^T \boldsymbol{\beta}) E_{ij}),\\
    M_{i} | \boldsymbol{\lambda}, \mathbf{z}_i, \mathbf{\beta}, \mathbf{p}_{i}, \mathbf{e}_i & \sim \text{Poisson}(\exp(\mathbf{z}_i^T \boldsymbol{\beta}) \textstyle \sum_j (1 - p_{ij}) \lambda_j E_{ij}), \\
    p_{ij} & = \lp 1 + \exp(-(\mathbf{z}_i^T \boldsymbol{\gamma} + \eta_j) \rp^{-1};
  \end{split}
\end{align}
where the $X_{ij}$ are independent of $M_i$ since the missingness process is conditionally independent of the disease process.
See \ref{app:model-diagram} for a graphical depiction of the model.

\begin{theorem}\label{full-model-id}

  Let the model be defined as in \eqref{model:full} and let $\mathbf{E}$ be the $I$ by $J$ matrix where the $i$-th row is $\mathbf{e}_i = (E_{i1},
  E_{i2}, \dots, E_{iJ})^T$, and let $\mathbf{Z}$ be the $I$ by $K$ matrix
  where the $i$-th row is $\mathbf{z}_i$. If all of the following conditions hold:
  \begin{enumerate}[label=(S.\alph*)]\label{conds:full-model-id}
  \item $\mathbf{E}$ is rank $J$\label{cond:full-model-rank-E}
  \item $\mathbf{Z}$ is rank $K$\label{cond:full-model-rank-Z}
  \item $I \geq J + K$\label{cond:full-model-dim-I}
    \item $p_j \in (0, 1)$ for all $j$\label{cond:full-model-p}
    \item $\lambda_j \in (0, \infty)$ for all $j$\label{cond:full-model-l}
    \item $e^{\mathbf{z}_i \boldsymbol{\beta}}\sum_j^J E_{ij} \in (0, \infty)$ for all $i$\label{cond:full-model-E}
    \item 
      ${\rm rank}\lp \begin{bmatrix}
        {\rm diag}(\mathbf{E}_{[:,1]})\mathbf{Z} & \dots & {\rm diag}(\mathbf{E}_{[:,J]})\mathbf{Z} & \mathbf{E}_{[:,1]}  & \mathbf{E}_{[:,2]} & \dots  & \mathbf{E}_{[:,J]}
      \end{bmatrix}\rp> J + K$\label{conds:full-model-rank-big-mat}
    \end{enumerate}
  the model is locally identifiable.
  \end{theorem}
  The proof is in Appendix \ref{proof:full-model} and depends on showing that the model's Fisher information matrix $\mathcal{I}$ is positive definite. We use a technique employed in \cite{mukerjeePositiveDefinitenessInformation2002}, which establishes a lower bound for the positive definiteness of the Fisher Information matrix via a method of moments estimator. The idea rests on the derivation of the multivariate
Cram\'er-Rao lower bound in \cite{raoLinearStatisticalInference2002}. This partially establishes that the model is regular and shows that the model is locally identifiable \citep{watanabeAlgebraicGeometryStatistical2009,rothenbergIdentificationParametricModels1971}. 

Given \cref{subsec:prior-sens}, it is important to use prior information for minority groups when possible. 
To that end, the following priors can be employed:
\begin{align*}
  \lambda_{j} & \sim \text{LogNormal}(\mu_{\lambda_j}, s_\lambda^2) \quad \forall j \in [1, \dots, J],\\
  \eta_{j} & \sim  \text{Normal}(\mu_{\eta_j},  s_\eta^2)  \quad \forall j \in [1, \dots, J], \\
  \boldsymbol{\beta} & \sim \text{MultiNormal}(\boldsymbol{\mu}_\beta, \Sigma_\beta) \\
  \boldsymbol{\gamma} & \sim \text{MultiNormal}(\boldsymbol{\mu}_\gamma, \Sigma_\gamma) 
\end{align*}
where $\mu_{\lambda_j}, \mu_{\eta_j}, s_\lambda, s_\eta, \boldsymbol{\mu}_\beta, \Sigma_\beta, \boldsymbol{\mu}_\gamma$ and $\Sigma_\gamma$ are known hyperparameters.

\subsection{Modeling geographic heterogeneity in incidence and missingness}    

Suppose the case data are observed for more than one geographical area so we have an additional fixed categorical variable $L_n \in \{1, \dots, G\}$ encoding the geographic area to which each resident in the population is associated. 
We may expect that geographical heterogeneity in incidence and race/ethnicity missingness exists between
areas.
For instance, with respect to the COVID-19 pandemic, we might want to allow for geographic heterogeneity in population substrata incidence and missingness because we expect that areas have different contact patterns.
We can then further stratify the observations by area $g$ as:
$$
X_{igj} \,|\, \mathcal{U} = \textstyle\sum_{\{n \,|\, S_n = i, L_n = g, C_n = j, U_n = 1\}} R_n \,,\, M_{ig} \,|\, \mathcal{U} = \textstyle\sum_{j} \sum_{\{n \,|\, S_n = i, L_n = g, C_n = j, U_n = 1\}} (1 - R_n),
$$
and we can tabulate population counts $E_{igj}$ as
$E_{igj} = \sum_{n=1}^E \ind{S_n = i}\ind{C_n = j}\ind{L_n = g}$.

Model \cref{model:full} naturally extends to incorporate this structure.
Let $\mathbf{e}_{ig}$ be the $J$-vector with $j$-th element $E_{igj}$ for age-sex stratum $i$ and geographic area $g$. 
Similarly define the proportions of cases in stratum $i$ and geographic area $g$ with observed race/ethnic information as $\mathbf{p}_{ig}$, where $\mathbf{p}_{ig}$ is the $J$-vector with $j$-th element $p_{igj}$. 
We let the covariates for stratum $i$ vary by area $g$, $\mathbf{z}_{ig}$, vary by area $g$, and we also let the coefficients vary by $g$, so $\boldsymbol{\beta}_g, \boldsymbol{\gamma}_g \in \R^K$.
Let $\boldsymbol{\lambda}_g$ be the $J$-vector with $j$-th element $\lambda_{gj}$.
The observed-data model becomes
\begin{align}
  \begin{split}\label{model:full-geo}
    X_{igj} | \lambda_{gj}, \mathbf{z}_{ig}, \boldsymbol{\beta}_g, p_{igj}, E_{igj} & \sim \text{Poisson}(p_{igj} \lambda_{gj} \exp(\mathbf{z}_{ig}^T \boldsymbol{\beta}_g) E_{igj}),\\
    M_{ig} | \boldsymbol{\lambda}_{g}, \mathbf{z}_{ig}, \boldsymbol{\beta}_g, \mathbf{p}_{ig}, \mathbf{e}_{ig} & \sim \text{Poisson}(\exp(\mathbf{z}_{ig}^T \boldsymbol{\beta}_g) \textstyle \sum_j ((1 - p_{igj}) \lambda_{gj} E_{igj}), \\
    p_{igj} & = \lp 1 + \exp(-(\mathbf{z}_{ig}^T \boldsymbol{\gamma}_g + \eta_{gj}) \rp^{-1}.
  \end{split}
\end{align} 
See \Cref{app:model-diagram-hier} for a graphical depiction of the model with a table of model parameters.
We can draw on the results from \ref{full-model-id} to characterize the local identifiability of \ref{model:full-geo}.
By \ref{full-model-id}, within a geographic region $g$, the parameter set
$$
\boldsymbol{\theta}_g = \{\boldsymbol{\lambda}_{g},\boldsymbol{\eta}_{g},\boldsymbol{\beta}_g, \boldsymbol{\gamma}_g\}
$$
is locally identifiable provided the conditions in \ref{conds:full-model-id} hold.
However, when data are sparse, either because there is low incidence within an area or because there is a small minority group in geographic region $g$,  we would like to shrink our estimates for $\boldsymbol{\theta}_g$ to the global mean.
Ideally we would learn the degree of shrinkage for each dimension of $\boldsymbol{\theta}_g$. This motivates a hierarchical prior 
for elements of $\boldsymbol{\theta}_g$. 
\subsubsection{Hierarchical priors}
To that end, we may wish to incorporate area-level covariates, represented by a $D$-length vector $\mathbf{w}_g$, into the model for  $\boldsymbol{\theta}_g$.
Let $\boldsymbol{\Pi_\lambda}, \boldsymbol{\Pi_\eta}$ be in $\R^{J \times D}$  and let $\boldsymbol{\Pi_\beta}, \boldsymbol{\Pi_\gamma}$ be in $\R^{K \times D}$. 
A suitable model for the elements of $\boldsymbol{\theta}_g$ is:
\begin{align}
  \begin{split}\label{model:meta-model}
   \log(\boldsymbol{\lambda}_{g}) & \sim \text{MultiNormal}\lp \boldsymbol{\alpha}_{\boldsymbol{\lambda}} + \boldsymbol{\Pi}_{\boldsymbol{\lambda}} \mathbf{w}_g, \boldsymbol{\Sigma}_{\boldsymbol{\lambda}}\rp\\
   \boldsymbol{\eta}_g & \sim \text{MultiNormal}\lp \boldsymbol{\alpha}_{\boldsymbol{\eta}} + \boldsymbol{\Pi}_{\boldsymbol{\eta}} \mathbf{w}_g,\boldsymbol{\Sigma}_{\boldsymbol{\eta}}\rp \\
   \boldsymbol{\beta}_g & \sim \text{MultiNormal}\lp \boldsymbol{\alpha}_{\boldsymbol{\beta}} + \boldsymbol{\Pi}_{\boldsymbol{\beta}} \mathbf{w}_g,\boldsymbol{\Sigma}_{\boldsymbol{\beta}}\rp \\
   \boldsymbol{\gamma}_g & \sim \text{MultiNormal}\lp \boldsymbol{\alpha}_{\boldsymbol{\gamma}} + \boldsymbol{\Pi}_{\boldsymbol{\gamma}} \mathbf{w}_g,\boldsymbol{\Sigma}_{\boldsymbol{\gamma}}\rp \\
   \end{split}
   \end{align}
For a more detailed picture of how these parameters connect to \cref{model:full-geo}, see \cref{app:model-diagram-hier}.
Let the operation $\text{vec}(\mathbf{A}): \R^{M \times N} \to \R^{MN}$ via appending the $N$ $M$-length columns of $\mathbf{A}$ into an $NM$-length vector.
Then the vector of unknown hyperparameters can be represented as
\begin{align*}
\boldsymbol{\phi}  = (&\text{vec}(\boldsymbol{\Pi_\lambda}), \text{vec}(\boldsymbol{\Pi_\eta}), \text{vec}(\boldsymbol{\Pi_\beta}), \text{vec}(\boldsymbol{\Pi_\gamma}), \\
& 
\text{vec}(\boldsymbol{\Sigma_\lambda}), \text{vec}(\boldsymbol{\Sigma_\eta}), \text{vec}(\boldsymbol{\Sigma_\beta}), \text{vec}(\boldsymbol{\Sigma_\gamma}), \\
& \boldsymbol{\alpha_\lambda}, \boldsymbol{\alpha_\eta}, \boldsymbol{\alpha_\beta}, \boldsymbol{\alpha_\gamma})
\end{align*}
We can encode our prior knowledge about the geographic heterogeneity of parameters into a joint prior over $\boldsymbol{\phi}$.

While the hierarchical prior in \cref{model:meta-model} does not correspond to the set of priors in \cref{eq:simple-priors} the results in \cref{subsec:prior-sens} suggest that posterior inferences for incidence parameters for areas with small minority groups relative to the majority groups can be sensitive to the priors over $\boldsymbol{\Sigma_\lambda}, \boldsymbol{\Sigma_\eta}$, and $\boldsymbol{\alpha}_{\boldsymbol{\eta}}$.

\Cref{fig:asympt-rmse} in \Cref{app:post-mean-u} shows the large-population RMSE for the incidence of minority race/ethnicity cases that are missing race/ethnicity information, or $u_1$, under different prior scenarios.
The RMSE of the posterior mean estimators are minimized when the prior mean for $u_1$ is close to the true parameter, when the prior for $u_1$ excludes prior mass near zero and the prior mean underestimates the true parameter, or when the prior mean slightly overestimates the true parameter and the prior for $u_1$ does not put substantial prior mass near zero. 
The near-zero prior behavior for $u_1$ can be translated to priors on $\boldsymbol{\Sigma_\eta}$, and $\boldsymbol{\alpha}_{\boldsymbol{\eta}}$.
By limiting the amount of prior mass in the right tail of the distribution for $\boldsymbol{\alpha}_{\boldsymbol{\eta}}$ one can limit the amount of prior mass near zero for $u_1$; a normal distribution with substantial mass below 5 would suffice.
The prior over $\boldsymbol{\Sigma_\eta}$ will also affect the tails of the marginal prior for geographic-specific parameters, and can also adversely affect shrinkage.
If one uses a prior over population standard deviation with heavy tails, like a half-Cauchy, then the marginal prior for a geographic specific parameter will have substantial prior mass near zero. 
If, instead, the prior over the population standard deviation hews too closely to zero, like a half-normal with a standard deviation of 0.1, then the prior will shrink geographic-specific parameters too strongly towards the overall mean. 
Similar considerations about shrinkage should guide priors over $\boldsymbol{\Sigma_\lambda}$.
For more information on techniques for prior formulation in Bayesian models see \cite{gabry2019,gelmanPriorCanOften2017}.

See \Cref{sec:sim-study-sens} for more information on prior specification for population parameters.

\subsection{Inference}
We perform Bayesian inference in Stan \citep{stanmanual}. 
Stan is at once a domain-specific modeling language and a suite of inference algorithms, including dynamic Hamiltonian Monte Carlo (HMC), a descendant of the No-U-Turn-Sampler \citep{betancourtConceptualIntroductionHamiltonian2018,hoffman2014no}.
Stan's implementation of dynamic HMC adaptively sets the algorithm's tuning parameters (e.g. leapfrog integrator stepsize and mass matrix) during warmup iterations, which makes the sampler robust to many difficult-to-sample posteriors, such as those that arise from fitting hierarchical models like model \eqref{model:full-geo} \citep{betancourt2015hamiltonian}. 

We use Stan for inference because we are able to exactly marginalize over the discrete unknown cases as shown in \cref{app:like-deriv}.
While Stan does not directly allow inference over discrete parameters, as long as the target density can be expressed as a marginalization over the discrete unknowns, Stan can sample from the posterior over the continuous parameter space and subsequently draw discrete random variables conditional on the draws of the continuous parameters.

\section{Simulation study}

In this section, we present a simulation study designed to quantify the finite-sample properties of our model under varying degrees of missingness, as well as to compare the model's performance to alternative methods of inference commonly applied to datasets with missing covariates. 
We chose complete-case analysis, and two different multiple imputation approaches as the comparison methods because of their prevalence among researchers. 
The simulation study clarifies the potential pitfalls of using such methods when analyzing data with missing covariates.

\subsection{Population data}\label{subsec:sim-study-pop-data}

In our simulation study, we drew on georeferenced population data from  Wayne County, Michigan, which encompasses the City of Detroit and its surrounding suburbs.
The geographical areas of analysis were Public Use Microdata Areas (PUMAs), which are administrative areas defined by the Census Bureau such that they comprise at least $100,000$ people. 
We aggregated Census-tract-level data from IPUMS National Historic Geographic Information System into PUMA-level counts \citep{nhgis_data}.
In Wayne County, there are 13 PUMAs nested within the county borders. 
Within each PUMA, we stratified the population by age and sex, with age in years binned in $10$-year right-open intervals between $0$ and $80$: $[0,10), [10,20), \dots, [70,80)$ and used a single group to capture those $80$ and older. 
We used the $2010$ Decennial Census population counts as $E_{ij}$ for each PUMA. 
The use of U.S. Census data constrains our race and ethnicity classification because Hispanic/Latino ethnicity is treated as mutually exclusive with race. This prevents a more nuanced modeling of a separate effects of ethnicity and race.
Despite these limitations, for our simulation study we used the Census classifications to bucket the population into four groups: Black, Hispanic/Latino, Other, and White. 
\begin{table}[ht]
\caption{Population summary in Wayne County, Michigan as of the 2010 Decennial Census}
\label{table:population-summary-sim-study}
\centering
\begin{tabular}{@{}lrrrr@{}}
  \hline
 &  & Mean Age$\times$Sex$\times$Race/Eth. &  & $100\times$ Ratio\\ 
 Race/Ethnicity& Total Pop.  & $\times$PUMA Pop. & Std. dev. PUMA Pop. &  to White\\ 
  \hline
  Black & 732801 & 3132 & 3152 & 81 \\ 
  Hispanic/Latino & 95260 & 407 & 757 & 11 \\ 
  Other & 90343 & 386 & 397 & 10 \\ 
  White & 902180 & 3855 & 3225 & 100 \\ 
   \hline
\end{tabular}
\end{table}
The Black and White categories comprised people who identified as Black or White alone and not Hispanic or Latino, while the Hispanic/Latino category included anyone who identified as Hispanic and Latino. The Other category included Asians and Pacific Islanders, Native Americans and Alaska Natives, mixed race individuals, as well as people of Other races, all of whom did not identify as Hispanic or Latino.
From table \ref{table:population-summary-sim-study} we can see that in Wayne the majority of the population is White, though the Black population is of a similar order of magnitude. Hispanic/Latino people and people classified as Other are around $10\%$ of the White population.  

\subsection{Data generating process}\label{subsec:dgp}
 We simulated age-sex-stratum-specific incident cases of disease by PUMA from model \ref{model:full-geo}, with fixed hyperparameters $\boldsymbol{\phi}$ under four scenarios that varied the proportion of cases that had fully-observed covariates: $90\%$, $80\%$, $60\%$, and $20\%$. 
 The data were generated with two effects for sex, and nine effects for age, both with a sum to zero constraint in both the Poisson log-rate parameter and the Bernoulli log-odds parameter. 
 More explicitly, the $\boldsymbol{\beta}_g$ parameter was decomposed into $\beta^\text{sex}$, and $\boldsymbol{\beta}^{\text{age}}$; $\boldsymbol{\gamma}_g$, $\boldsymbol{\alpha}_\beta$, and $\boldsymbol{\alpha}_\lambda$ were similarly decomposed.
For the simulated datasets, the Poisson log-rate parameters for $\boldsymbol{\alpha}_\beta^\text{age}$ were fixed at values that mimicked the age pattern of relative risk of COVID-19 cumulative incidence in the first stage of the pandemic, roughly between March $1^\textrm{st}$, 2020 and July   $1^\textrm{st}$, 2020. 
The relative risk of COVID-19 for younger people was much lower compared to that of older people, especially those over 60, and we set the values of $\boldsymbol{\alpha}_\beta^\text{age}$ accordingly: $(-2.5,\allowbreak -2.0,\allowbreak  0.0,\allowbreak  0.0,\allowbreak  0.5,\allowbreak  0.5,\allowbreak  1.0,\allowbreak  1.0,\allowbreak  1.5$) \citep{zelnerRacialDisparitiesCoronavirus2021}. 
For the age pattern of the log-odds of missingness, older individuals were more likely to have race reported compared to younger ages and was thus reflected in our values for $\boldsymbol{\alpha}_\gamma^\text{age} = (-0.3,\allowbreak  -0.3,\allowbreak  -0.2,\allowbreak  -0.2,\allowbreak  -0.2,\allowbreak  -0.1,\allowbreak  0.1,\allowbreak  0.4,\allowbreak  0.8)$. 

In order to investigate hyperparameter inference as well as other functions of the parameters of epidemiological interest (like cumulative incidence per group at the county level or age-sex-standardized incidence) for majority and minority groups that was solely a function of missingness and not of rates of disease, we set each group's average log-rate of disease, or the elements of $\boldsymbol{\alpha}_\lambda$, to $-4$ for all simulations. 
We then set $\boldsymbol{\alpha}_\eta$, the group-wise population log-odds of observing race, to vary between scenarios according to the average proportion of cases observed with race. 
In order to set proportions of fully-observed cases for each race/ethnicity, we set ratios of the proportions relative to that of Whites and then varied the White proportion such that the population weighted average rate of cases with fully-observed covariates matched the population target rates of $90\%, 80\%, 60\%, 20\%$. 
Blacks-to-Whites was set to $\frac{0.75}{0.9}$, Hispanic/Latinos was set to $1$, Other was set to $\frac{0.6}{0.9}$. 
The generative model for the geography-specific parameters is:
\begin{align}
  \begin{split}\label{model:dgp-model-sim-study}
   \log\boldsymbol{\lambda}_g & \sim \text{MultiNormal}(\boldsymbol{\alpha}_\lambda, \text{diag}(\boldsymbol{\sigma}_\lambda)) \\
   \boldsymbol{\eta}_g & \sim \text{MultiNormal}(\boldsymbol{\alpha}_\eta, \text{diag}(\boldsymbol{\sigma}_\eta)) \\
   \boldsymbol{\beta}_g & \sim \text{MultiNormal}(\boldsymbol{\alpha}_\beta, \text{diag}(\boldsymbol{\sigma}_\beta)) \\
   \boldsymbol{\gamma}_g & \sim \text{MultiNormal}(\boldsymbol{\alpha}_\gamma, \text{diag}(\boldsymbol{\sigma}_\gamma)) 
  \end{split}
\end{align} 
with all elements of $\boldsymbol{\sigma}_\lambda$ and $\boldsymbol{\sigma}_\beta$ equal to $0.5$ and all elements of $\boldsymbol{\sigma}_\eta$ and $\boldsymbol{\sigma}_\gamma$ equal to $0.3$,.
The elements of the hierarchical scale parameters related to cumulative disease incidence, $\boldsymbol{\sigma}_\lambda$ and $\boldsymbol{\sigma}_\beta$, were set to larger values than the parameters related to the missingness process,  $\boldsymbol{\sigma}_\eta$ and $\boldsymbol{\sigma}_\gamma$, to reflect the fact that missingness of race data in Wayne County in the first wave of the pandemic was driven by local-level patient non-response and county-wide lab processing issues, while cumulative incidence was driven largely by local transmission.

Summaries of the simulated datasets are shown in Table \ref{table:sim-data-sum}. 
The differences between race in the true cumulative incidence were driven solely by the difference in age distributions between races within Wayne County. 
The table highlights the fact that, excluding random variation, the scenarios differ only in the observed incidence, as the disease process model as represented via hyperparameters $\boldsymbol{\alpha}_\lambda$ and $\boldsymbol{\alpha}_\beta$ remains fixed between scenarios.
The variance in incidence was a function of the variance of the realizations of the geography-specific parameters $\boldsymbol{\lambda}_g$ and $\boldsymbol{\beta}_g$ driven by the population scale parameters $\boldsymbol{\sigma}_\lambda$ and $\boldsymbol{\sigma}_\beta$.
\begin{table}[ht]
\caption{
The table summarises the simulation study by missingness scenario by race/ethnicity. 200 datasets were simulated in each scenario. 
The column ``Mean Obs.'' gives the average proportion of cases observed with race/ethnicity data across 200 simulated datasets. Similarly, ``Mean True Inc.'' is the mean true incidence by group, and ``Mean Obs. Inc.'' is the mean observed incidence by group.
} \label{table:sim-data-sum}
\centering
\begin{tabular}{@{}lccccccc@{}}
  \hline
Proportion cases &  & Mean &  & Mean &  & Mean &  \\ 
w/ race/ethnicity & Race/Ethnicity & Obs. & Std. dev.  & True Inc. & Std. dev. & Obs. Inc. & Std. dev. \\ 
  \hline
90\% & Black & 80.7\% & (2.4\%) & 3.4\% & (0.9\%) & 2.8\% & (0.8\%) \\ 
   & Hispanic/Latino & 96.7\% & (0.7\%) & 2.4\% & (0.7\%) & 2.3\% & (0.7\%) \\ 
   & Other & 63.9\% & (3.0\%) & 2.6\% & (0.6\%) & 1.7\% & (0.4\%) \\ 
   & White & 97.1\% & (0.5\%) & 4.4\% & (1.8\%) & 4.3\% & (1.7\%) \\ [6pt]
80\% & Black & 72.7\% & (3.0\%) & 3.4\% & (1.0\%) & 2.5\% & (0.8\%) \\ 
   & Hispanic/Latino & 85.0\% & (2.4\%) & 2.4\% & (0.6\%) & 2.1\% & (0.5\%) \\ 
   & Other & 57.4\% & (3.2\%) & 2.6\% & (0.6\%) & 1.5\% & (0.3\%) \\ 
   & White & 86.5\% & (2.1\%) & 4.2\% & (1.2\%) & 3.7\% & (1.0\%) \\ [6pt]
60\% & Black & 53.7\% & (4.6\%) & 3.5\% & (1.1\%) & 1.9\% & (0.7\%) \\ 
  & Hispanic/Latino & 60.3\% & (4.3\%) & 2.4\% & (0.6\%) & 1.5\% & (0.4\%) \\ 
   & Other & 42.3\% & (3.7\%) & 2.6\% & (0.5\%) & 1.1\% & (0.3\%) \\ 
   & White & 64.4\% & (4.4\%) & 4.3\% & (1.3\%) & 2.8\% & (1.0\%) \\ [6pt]
20\%  & Black & 17.2\% & (3.5\%) & 3.4\% & (0.8\%) & 0.6\% & (0.2\%) \\ 
   & Hispanic/Latino & 18.4\% & (3.2\%) & 2.4\% & (0.7\%) & 0.4\% & (0.2\%) \\ 
   & Other & 12.9\% & (2.3\%) & 2.7\% & (0.5\%) & 0.4\% & (0.1\%) \\ 
   & White & 21.7\% & (4.5\%) & 4.4\% & (1.5\%) & 1.0\% & (0.6\%) \\ 
   \hline
\end{tabular}
\end{table}

\subsection{Inferential models}\label{subsec:inf-models-sim-study}
We fitted four inferential models to the simulated datasets: model \eqref{model:full-geo}, which we will refer to as the ``joint'' model, the ``complete case'' model, defined in \Cref{model:comp-case}, in which cases with missing race/ethnicity are dropped, and two ``multiple imputation'' models in which we impute the missing/race ethnicity cases and subsequently fit the complete case model to the generated datasets.
The hierarchical prior structure of the joint model matched that of the data generating model in equation \ref{model:dgp-model-sim-study},
with priors over the hyperparameters:
\begin{equation}\label{eq:sim-study-priors}
    \begin{alignedat}{3}
   \boldsymbol{\alpha}_\lambda & \sim \text{MultiNormal}(-\mathbf{5}, \text{diag}(\mathbf{1})) \, , && \boldsymbol{\sigma}_\lambda & \sim \text{MultiNormal}^+(\mathbf{0}, \text{diag}(\mathbf{1}))  \\
   \boldsymbol{\alpha}_\eta & \sim \text{MultiNormal}(\mathbf{2}, \text{diag}(\mathbf{1})) \, , && \boldsymbol{\sigma}_\eta & \sim \text{MultiNormal}^+(\mathbf{0}, \text{diag}(\mathbf{1})) \\
   \boldsymbol{\alpha}_\beta & \sim \text{MultiNormal}(\mathbf{0}, \text{diag}(\mathbf{1})) \, , && \boldsymbol{\sigma}_\beta & \sim  \text{MultiNormal}^+(\mathbf{0}, \text{diag}(\mathbf{1}))\\
   \boldsymbol{\alpha}_\gamma & \sim \text{MultiNormal}(\mathbf{0}, \text{diag}(\mathbf{1})) \, , && \boldsymbol{\sigma}_\gamma & \sim  \text{MultiNormal}^+(\mathbf{0}, \text{diag}(\mathbf{0.25}))\\
    \end{alignedat}
\end{equation}
A noteworthy characteristic of the priors for the hyperparameters is that the priors over $\boldsymbol{\alpha}_\lambda$ and $\boldsymbol{\alpha}_\eta$ were misspecified compared to the data-generating parameters.
The true data-generating parameters fell one prior standard deviation above the prior means for $\boldsymbol{\alpha}_\lambda$, while the prior mean for $\boldsymbol{\alpha}_\eta$, which did not vary by scenario, was too large by 4 prior standard deviations in the 20\% observed scenario and was too small by 1.5 standard deviations in the 90\% observed scenario. 
This allowed us to examine the joint model's finite-sample properties for large groups and smaller groups.

\subsubsection{Complete case model definition}
The complete case model is 
\begin{align}
  \begin{split}\label{model:comp-case}
    X_{igj} | \lambda_{gj}, \mathbf{z}_{ig}, \boldsymbol{\beta}_g, p_{igj}, E_{igj} & \sim \text{Poisson}(\lambda_{gj} \exp(\mathbf{z}_{ig}^T \boldsymbol{\beta}_g) E_{igj}), \\
   \log\boldsymbol{\lambda}_g & \sim \text{MultiNormal}(\boldsymbol{\alpha}_\lambda, \text{diag}(\boldsymbol{\sigma}_\lambda)), \\
   \boldsymbol{\beta}_g & \sim \text{MultiNormal}(\boldsymbol{\alpha}_\beta, \text{diag}(\boldsymbol{\sigma}_\beta)), 
   \end{split}
\end{align}
which necessarily omits a model for the missing-race-data cases.
The priors for the hyperparameters matched those in \cref{eq:sim-study-priors} for the shared parameters between the joint model and the complete case model.
We used the results of Theorem \ref{full-model-id} to check that our PUMA-level models were locally identifiable. All 13 PUMAs satisfied the local identifiability criteria in \ref{full-model-id}.

\subsubsection{Multiple imputation method description}
\begin{enumerate}
    \item \textbf{Ad-hoc MI}:  The first multiple imputation model is an ad-hoc method which imputes missing cases using a multinomial distribution with a probability parameter equal to that of the population proportions. For example, suppose we observe $m_{ig}$ missing cases for a certain stratum $i$ in PUMA $g$, along with $\mathbf{x}_{ig}$ cases by race. In order to generate a single imputation draw, $\mathbf{y}_{ig}^{(s)}$, we draw the missing cases: $\boldsymbol{\epsilon}_{ig}^{(s)} \sim \text{Multinomial}(m_{ig}, \mathbf{e}_{ig} / \sum_{j} E_{igj})$ and add $\boldsymbol{\epsilon}_{ig}^{(s)}$ to $\mathbf{e}_{ig}$: $\mathbf{y}_{ig}^{(s)} = \boldsymbol{\epsilon}_{ig}^{(s)} + \mathbf{x}_{ig}$. 
    We loop through $i \in \{1, \dots, I\}$ to generate one complete dataset and repeat this step to generate multiple complete datasets.
    \item \textbf{Gibbs MI}:  The second multiple imputation model is described in Chapter 18 of \cite{gelman2013bayesian}: The method generates complete datasets using a Gibbs sampler that alternates between sampling missing cases $\boldsymbol{\epsilon}_{ig}^{(s)} | \boldsymbol{\theta}_{ig}^{(s-1)} \sim \text{Multinomial}(m_{ig}, \frac{\boldsymbol{\theta}_{ig}^{(s-1)}}{\sum_j \boldsymbol{\theta}_{ig}^{(s-1)}})$ and $\boldsymbol{\theta}^{(s)}  | \mathbf{y},\boldsymbol{\epsilon}^{(s)}  \sim 
    \text{Dirichlet}(\mathbf{1} + \mathbf{y} + \boldsymbol{\epsilon}^{(s)})$ where $\boldsymbol{\theta}^{(s)}$ is the concatenation of each $\boldsymbol{\theta}_{ig}^{(s)}$ for the Gibbs sampler iteration step $s$ into a single vector, and $\mathbf{y}, \boldsymbol{\epsilon}^{(s)}$ are also vectors formed by concatenating $\mathbf{y}_{ig}, \boldsymbol{\epsilon}_{ig}^{(s)}$ into single vectors appropriately matching the indexing of $\boldsymbol{\theta}^{(s)}$ and $\mathbf{1}$ is an appropriately sized vector of $1$s, representing the uniform prior over the simplex.
    We run the Gibbs sampler for 20 MCMC chains for 2,500 burn-in iterations and 2,500 samples, which we subsequently thin by 25 steps, resulting in 5,000 total posterior samples. 
    We then take a subset of these samples 5,000 as our completed datasets.
\end{enumerate}
We generate 100 imputed datasets from each method for each simulated dataset, fit model \eqref{model:comp-case} to each imputed dataset with Stan and combine the 100 sets of posterior draws into a single superset of posterior samples. 
We then compute posterior summary statistics including credible intervals for each method using the single superset of posterior samples, following advice in \cite{zhouNoteBayesianInference2010} which showed that proper Bayesian inference using multiple imputation must follow this procedure. 

\subsection{Estimands of interest}\label{subsec:estimands-simu-study}
In order to compare the models on a common subset of parameters, we limited our comparisons to those involving the data-generating disease process parameters  $\boldsymbol{\alpha}_{\lambda}$, and $\boldsymbol{\alpha}_\beta$.
The simplest estimands against which we measured each model's inferences were $\boldsymbol{\alpha}_\lambda$, and $\boldsymbol{\sigma}_\lambda$.
We were also interested in the following estimands:
$$
\left(\exp((\boldsymbol{\alpha}_\lambda)_{1} - (\boldsymbol{\alpha}_\lambda)_J), \, \dots\,
\exp((\boldsymbol{\alpha}_\lambda)_{J - 1} - (\boldsymbol{\alpha}_\lambda)_J)\right)
$$
which are Wayne-county-level group-specific rates of disease relative to the rate of disease in category $J$; in the simulation study category $J$ was Whites.
There are several more complex estimands which have epidemiological significance, which are similar to poststratification estimators \cite{gelman1997poststratification,gaoImprovingMultilevelRegression2021} that are functions of the PUMA-local parameters $\boldsymbol{\beta}_g$ and $\boldsymbol{\lambda}_{g}$, or the Poisson model coefficients for strata and rates of disease by race/ethnicity category in a geography $g$. 
\subsubsection{Modeled incidence}
The first will be total modeled incidence for a race/ethnicity category $j$, or $\mathbbm{I}_j$. 
Let 
$r_{igj} = \lambda_{gj} \exp(\mathbf{z}_i^T \boldsymbol{\beta}_g)$
be the rate of expected cases per person of disease in stratum $i$, geographical area $g$ for category $j$.
Then 
$$
\mathbbm{I}_j = \frac{\sum_{i=1}^I \sum_{g=1}^G E_{igj} r_{igj} }{\sum_{i=1}^I \sum_{g=1}^G E_{igj}}
$$
is the total incidence for category $j$.
Interest often lies in relative risk ratios, or 
$$
\mathbbm{I}_j / \mathbbm{I}_J.
$$
\subsubsection{Standardized incidence}
The second estimand is the standardized incidence or $\mathbbm{SI}_j$. 
Let 
$$
\psi_i = \frac{\sum_{j=1}^J \sum_{g=1}^G E_{igj} r_{igj}}{\sum_{j=1}^J \sum_{g=1}^G E_{igj}}
$$
be the population average incidence for a single stratum $i$.
Then the $\mathbbm{SI}_j$ for category $j$ is:
$$
\mathbbm{SI}_j = \frac{\sum_{g=1}^G\sum_{i=1}^I E_{igj} \psi_{i}}{\sum_{g=1}^G\sum_{i=1}^I E_{igj}}.
$$
The standardized incidence for race/ethnicity $j$ quantifies the cumulative incidence based solely on race $j$'s population distribution across strata.
\subsubsection{Standardized incidence ratio}
The third estimand is the standardized incidence ratio, denoted as the SIR in \cite{lashmodernepi}, though not to be confused with susceptible-infected-recovered models \citep{keeling2011modeling}, which is the ratio of the modeled incidence to standardized incidence:
$$
\text{SIR}_j = \frac{\mathbbm{I}_j}{\mathbbm{SI}_j}.
$$
The $\text{SIR}_j$ measures how modeled cumulative incidence for a race/ethnicity category $j$ deviates from the standardized incidence.
A ratio above one indicates that race/ethnicity category $j$ has experienced higher rates of disease than would be expected based on the population distribution across ages and sexes alone, while a ratio below one indicates the opposite.
We can then derive relative estimands from $\mathbbm{I}_j$, $\mathbbm{SI}_j$, and $\text{SIR}_j$ as we did using $\boldsymbol{\alpha}_\lambda$.

\subsection{Computation}
We ran Stan via  the \texttt{cmdstanr} interface in R \citep{stanmanual,cmdstanr,rcore} on University of Michigan's Great Lakes Slurm High Performance Computing Cluster.
For the exhaustive combination of models and datasets for the joint and complete-case models (1,600 in total), we ran four Markov chain Monte Carlo chains for 2,000 warmup iterations and 1,500 post-warmup iterations.
In order to ensure that the posteriors had been sufficiently explored, for each dataset/model combination we recorded the maximum of all parameters' rank-normalized $\hat{R}$s, and the minima of bulk effective sample size and tail effective sample size divided by the total post-warmup iterations, which was 6,000 (bulk ESS efficiency, and tail ESS efficiency, respectively) using the \texttt{posterior} package in \texttt{R} \citep{rposterior,rcore,vehtarirhat}. 

We generated 100 imputed datasets for each of the 800 simulated datasets for each imputation method, and subsequently ran \eqref{model:comp-case} for 500 warmup iterations and 1,000 post-warmup iterations with four MCMC chains, resulting in 160,000 fitted four-chain Stan models between both imputation methods.

Example \texttt{R} and Stan code, including models and code to verify identifiability \cref{conds:full-model-rank-big-mat}, can be found at \url{https://github.com/rtrangucci/epi-missing-data}.

\subsection{Results}\label{subsec:sim-study-results}

\subsubsection{Computation}

The joint and complete-case models ran with maximum rank-normalized $\hat{R}$s below 1.013.
All but one model ran with bulk ESS efficiency greater than 10.0\% (the 1 out of 1,600 model/data pair that violated the threshold ran with 9.7\% bulk ESS  efficiency) and all ran with minimum tail ESS efficiency greater than 10\%.
No divergent transitions were recorded, though 29 complete case models fitted to datasets generated in the 20\% observed scenario needed to be rerun with a warmup-iteration target Metropolis acceptance rate of 0.995, an increase compared to the 0.95 target acceptance rate that all models were run with initially.
No iterations were observed that hit maximum treedepth, which was set to 14 for all runs.

A small minority of the multiple imputation runs encountered treedepth issues, though all 160,000 model-by-imputed dataset combinations ran with bulk and tail ESS efficiencies greater than 10.0\%. 
The CPU time required to run the multiple imputation methods was, at a minimum, $\sim 42$  times greater than either the joint or the complete-case models which is a clear disadvantage to multiple imputation methods.
\citeauthor{zhouNoteBayesianInference2010} note that for Bayesian credible intervals to achieve nominal coverage with multiple imputation many more than the typically recommended 5-20 imputed datasets are required.

\subsubsection{Bias and root mean squared error}

\begin{figure}
    \centering
    \includegraphics[scale=0.5]{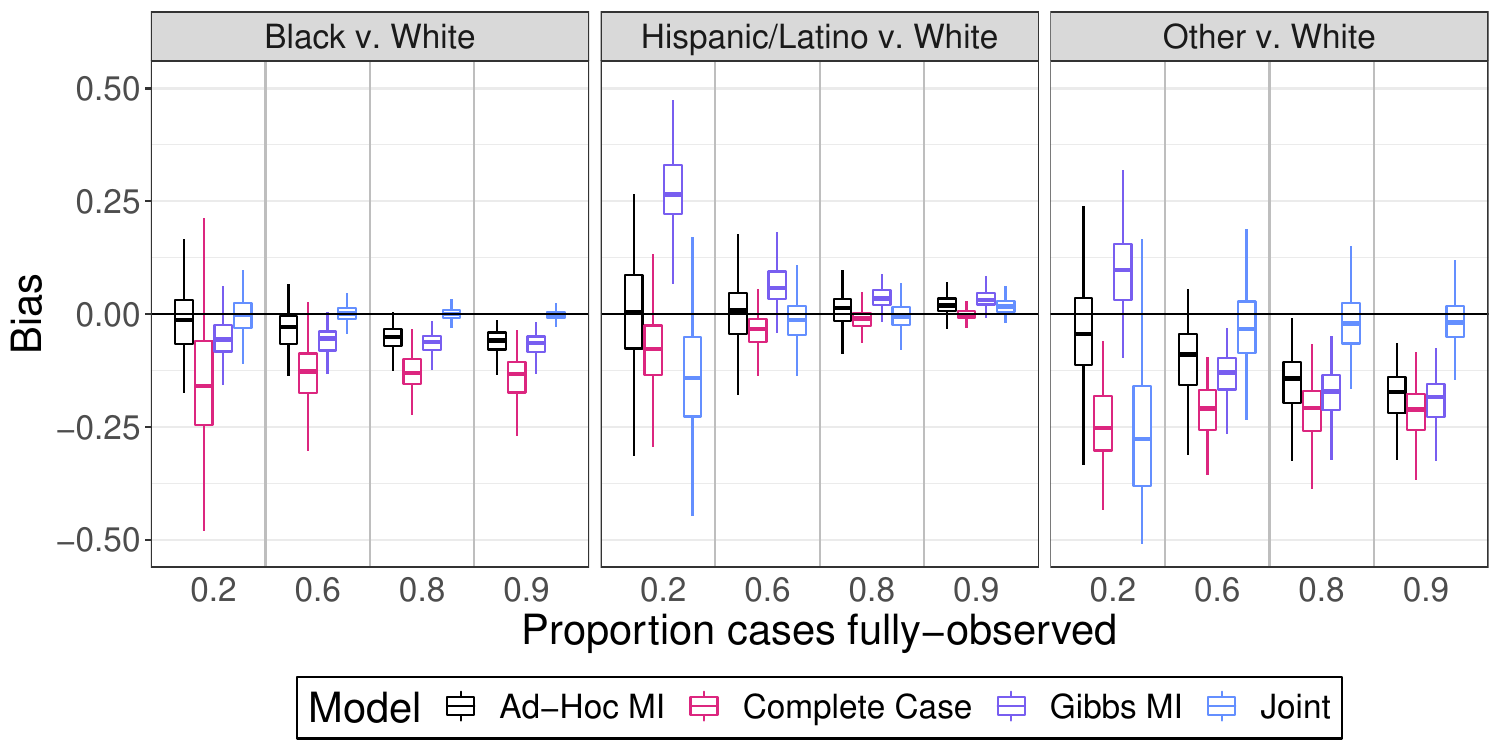}
    \caption{Bias across simulated datasets for the relative risk ratios, or $\mathbbm{I}_j / \mathbbm{I}_J$ for Blacks, Hispanic/Latinos, and Others relative to Whites plotted against the proportion of cases observed with race data.}
    \label{fig:bias-rr}
\end{figure}

We made boxplots of bias for each parameter across all simulation runs $S$. 
We used the posterior mean from each model as the estimator for each estimand $\theta$, or $\hat{\theta} = \Exp{\theta}{\theta | Y}$, and calculated bias for a simulation run $s$ as 
$$\text{bias}(\hat{\theta}_s, \theta_s) = \hat{\theta}_s - \theta_s.$$ 
Root mean squared error was calculated as 
$$
\text{RMSE}(\hat{\theta}, \theta) = \sqrt{\frac{1}{S}\textstyle \sum_{s=1}^S \text{bias}(\hat{\theta}_s, \theta_s)^2}.
$$
Asymptotic 95\% confidence intervals were calculated using the Delta method \citep{lehmannTheoryPointEstimation1998}.
\begin{figure}
    \centering
    \includegraphics[scale=0.5]{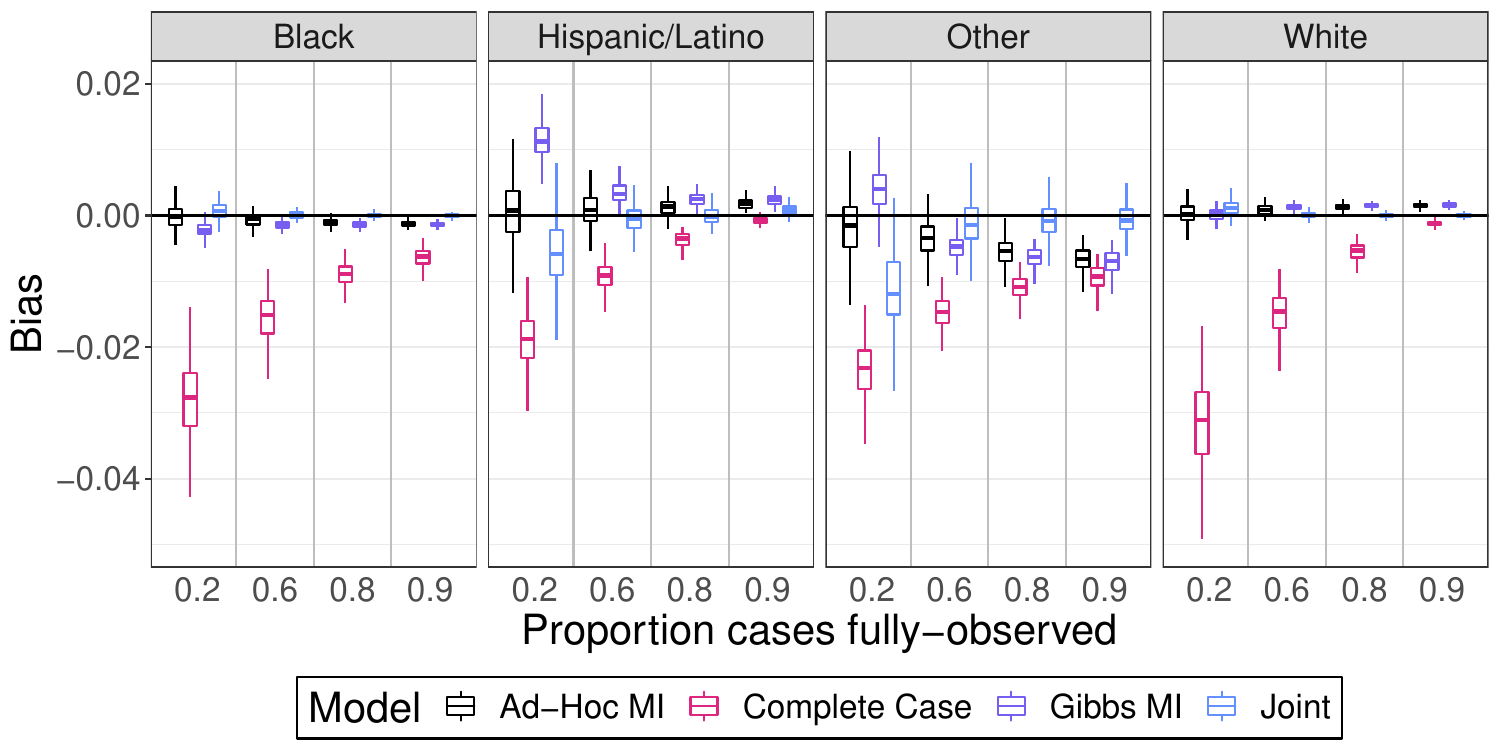}
    \caption{Bias across simulated datasets for the incidence, or $\mathbb{I}_j$ for Blacks, Hispanic/Latinos, Others, and Whites plotted against the proportion of cases observed with race data.}
    \label{fig:bias-inc}
\end{figure}
\paragraph{Bias in estimating incidence by race/ethnicity}
As can be seen in \Cref{fig:bias-inc}, for Blacks and Whites, which comprise 49\% and 40\% of the total population in Wayne County, the bias in the posterior mean incidence estimator generated by the joint model is small across all scenarios for most simulated datasets. 
For Whites, the average bias in the joint model posterior mean is not significantly different than zero in the 90\%, 80\% and 60\% scenarios, while for Blacks, there is statistically significant average bias for joint-model posterior mean incidence in all scenarios other than 80\%, but it is an order of magnitude smaller than the average bias of the posterior mean estimator from the imputation methods.
The complete case model, as expected, is significantly negatively biased in all scenarios.
The average bias from ad-hoc multiple imputation is smallest among all methods in the 20\% scenario because the data generating process, outlined in \Cref{subsec:dgp}, defines the true population rate of disease for each race/ethnicity group to be the same.
The distribution of missing cases by category conditional on the total missing cases is multinomial with parameter $\mathbf{e}_{ig} \odot (1 - \mathbf{p}_{ig}) / \sum_j E_{igj} (1 - p_{igj})$. 
When missingness is high, $(1 - p_{igj})$ is close to one, so the ad-hoc multinomial imputation procedure with parameter $\mathbf{e}_{ig} / \sum_j E_{igj}$ is approximately correct.
As missingness decreases, the ad-hoc imputation parameter diverges from the data generating process and the bias grows. 
This pattern can be seen in \Cref{fig:bias-rr} as well.
In sum, the averages of the joint model estimators' biases are sometimes more than two standard errors from zero, but the model's absolute bias is significantly smaller compared to the absolute bias of the competing estimators, with exceptions in the 20\% scenario compared to the Ad-Hoc MI method.
\paragraph{Bias in estimating relative risk by race/ethnicity}
In \Cref{fig:bias-rr} the joint model was able to estimate the relative risk of disease with mean bias that is not significantly different from zero for Blacks vs. Whites in the 80\%, 60\% and 20\% observed scenarios, while in the 90\% scenario the mean bias is significantly nonzero, but two orders of magnitude smaller than the mean bias incurred by the complete case model's estimators.
For Hispanic/Latinos and Others, there exists some mean bias in the 90\%, 60\%, and 20\% scenarios, though in the 80\% and 60\% scenarios the mean bias is an order of magnitude smaller than that of the complete case analysis.
Complete case analysis does yield estimators with average bias that is not significantly different from zero for the relative risk of disease for Hispanics/Latinos to Whites in the 90\% observed scenario and has smaller average bias compared to the joint model's estimators. 
This is due to the fact that in the simulated datasets the log-odds of observing race data was equal for Whites and Hispanics/Latinos, all else being equal. 
The average bias from multiple imputation using Gibbs sampling is consistently nonzero across all missingness scenarios for all groups in  \Cref{fig:bias-rr}.
The Gibbs multiple imputation procedure assumes the data are MAR, when the DGP is NMAR for all scenarios. 
This highlights the danger of using a MAR procedure when the data are NMAR.
The pattern of bias is similar for $\exp((\boldsymbol{\alpha}_\lambda)_j - (\boldsymbol{\alpha}_\lambda)_J)$: the complete-case estimators are comparable in terms of mean bias to that of the joint-model estimators in the Hispanic/Latino group, while the complete-case estimators underperform in Blacks and Others.
For $\boldsymbol{\sigma}_\lambda$, the complete-case posterior mean estimators are positively biased compared to the joint-model's estimators, likely due to the fact that the complete case analysis attributes all variance in local area estimates of $\boldsymbol{\lambda}_g$ to variation in disease incidence while the joint model attributes some of the variation to variation in the observational process. 
The estimators from the joint model are, however, negatively biased, likely due to the fact that we have only 13 PUMAs and relatively strong $\text{Normal}^+(0, 0.5^2)$ priors that shrink towards zero on the population scale parameters $\boldsymbol{\sigma}_\lambda$.  

\paragraph{Root mean squared error}
The RMSEs are shown on in \Cref{app:mse-plots}. That of the joint-model estimators for $\text{SIR}_j$ are significantly smaller (as measured shown by nonoverlapping 95\% confidence intervals) than the RMSEs for the complete-case estimators in the 90\%, 80\%, and 60\% scenarios for nearly all groups (the exception is for Hispanics/Latinos in the 60\% scenario, where the RMSEs are not significantly different).
In the 20\% observed scenario, the RMSEs of the joint-model estimators for Blacks and Whites are smaller than those of the complete case model, but the RMSEs of the joint-model estimators for Hispanic/Latinos and Others are larger than the complete-case estimators. 
This is due to the fact that Hispanic/Latinos and Others are smaller populations in Wayne County, and the parameter space for the is $2\times$ as large as the complete-case model's parameter space. 
We also present the RMSE comparisons for the relative risk ratios and relative county-level rates, shown in figures \ref{fig:mse-rr} and \ref{fig:mse-exp-alpha}, respectively. 
The relative risk ratio plots show a similar pattern to that of the $\text{SIR}_j$ estimates, with the exception of relative risk ratios for Hispanics/Latinos, for which the RMSEs of the complete-case estimators are smaller than those of the joint model's. 
This is due to the fact that White and Hispanics/Latinos case-patients are observed at similar relative rates across simulations because the observation ratio, or $\text{inv\_logit}((\boldsymbol{\alpha}_\eta)_j)/ \text{inv\_logit}((\boldsymbol{\alpha}_\eta)_J) = 1$ for these two groups and the complete case analysis model implicitly assumes the observation ratios for all races to be exactly 1.

On the contrary, figure \ref{fig:mse-exp-alpha} shows that the RMSEs for the joint-model's estimators are similar in magnitude or larger in all scenarios.
While the joint-model's estimators show smaller mean biases, the variance for the estimators is much larger compared to the complete case analysis. 
This is again due to the fact that there are only 13 PUMAs included in the simulation study, and the fact that the dimension of the parameter space is twice as large for the joint model as that of the complete case model. 
The RMSEs for the ad-hoc imputation approach are small in the 20\% scenario for the same reason the bias is small in the 20\% scenario, but the RMSE increases as the missingness decreases.
This is a clear indication that the data generating process does not agree with the imputation procedure.
The RMSEs for the Gibbs imputation approach are large for the 20\% scenario, likely owing to the fact that as the number of missing cases increases, the variance of the imputed datasets increased due to increased posterior uncertainty for the imputation model.
This could be an indication that more than 100 imputed datasets are necessary for the imputation procedure when missingness is high, which would accord with the observations in \cite{zhouNoteBayesianInference2010}, though we were constrained by computational budget to use only 100 imputed datasets per simulated dataset.

\subsubsection{Coverage and interval length}

Table \ref{table:50pct-intervals} summarizes the interval coverage for the complete-case model, the joint model, and the multiple imputation procedures. All intervals that follow are central $p\%$ posterior credible intervals. In the event the joint distribution of the simulated parameters and data matches the prior and the likelihood of the inferential model and we can properly draw samples from the posterior, the central $p\%$ posterior credible intervals (and any other posterior intervals, for that matter) will contain the parameter that generated the data with exactly $p\%$ probability \citep{cook2006validation}. 
As expected, the complete-case model's 50\% intervals severely under cover for all but the county-level relative rates of disease for Hispanics and Latinos compared to the rate for Whites. 
The ad-hoc imputation method's intervals over-covered for the population-level relative risk comparisons (as seen in \ref{table:full-50pct-intervals}), while they undercovered for the standardized incidence and relative risk measures, while the Gibbs sampler imputation's intervals severely undercovered in all scenarios for all the parameters of interest.
Despite the ad-hoc methods near-match to the data generating process in the 20\% scenario, the intervals for incidence under-cover more than the joint model's credible intervals. 
The joint model's intervals are near the nominal coverage probabilities, i.e. the 50\% intervals cover the true parameter value in 50\% of simulations, though they do under-cover for sparsely populated groups like Others and Hispanic/Latinos, especially so with significant numbers of missing cases.

The same pattern is exhibited in the figure \ref{fig:breakdown-intervals}, which shows boxplots of the average coverage across all parameters related to the disease process for each simulated dataset for all models.
The complete-case model's 50\% and 80\% interval coverage is about 25\% and 35\%, respectively, while the joint model's intervals achieve the nominal coverage probability on average.
The multiple imputation methods' intervals fare a bit better though they still under-cover: the rates are near 30\%-35\% and 60\% to 65\% on average.

In appendix section \ref{app:post-intervals} we present table \ref{table:80pct-intervals}, which mirrors table \ref{table:50pct-intervals} but for 80\% intervals.
The pattern of performance is similar. 

\begin{table}[ht]
\caption{Table shows 50\% posterior credible interval coverages and lengths for estimands of interest from the simulation study. Coverage proportion is calculated across 200 simulated datasets for each model/simulation scenario. Column headers for percentages (e.g. 20\%) indicate the missing-data simulation scenario which corresponds to the statistic calculated in the table column; the simulation scenario corresponds to the proportion of cases observed with completely observed race covariates.}
\label{table:50pct-intervals}
\tabcolsep=0.10cm
\centering
\begin{tabular}{@{}lccccccccc@{}}
  \hline
  &  & \multicolumn{4}{c}{50\% interval coverage} & \multicolumn{4}{c}{50\% mean interval length} \\ 
 Parameter & Model  & 20\% & 60\% & 80\% & 90\% & 20\% & 60\% & 80\% & 90\% \\ 
    \hline
$\mathbbm{I}_\text{Blacks}$ 
   & Complete Case & 0.00 & 0.00 & 0.00 & 0.00 & 1e-04 & 2e-04 & 2e-04 & 3e-04 \\ 
   & Joint & 0.37 & 0.48 & 0.46 & 0.51 & 2e-03 & 7e-04 & 5e-04 & 4e-04 \\ 
   & Ad-Hoc MI& 0.05 & 0.13 & 0.03 & 0.00 & 3e-04 & 3e-04 & 3e-04 & 3e-04 \\ 
   & Gibbs MI & 0.01 & 0.01 & 0.00 & 0.00 & 5e-04 & 3e-04 & 3e-04 & 3e-04 \\ [6pt]
  $\mathbbm{I}_\text{Hispanics/Latinos}$ 
   & Complete Case & 0.00 & 0.00 & 0.00 & 0.20 & 3e-04 & 5e-04 & 6e-04 & 7e-04 \\ 
   & Joint & 0.26 & 0.47 & 0.48 & 0.42 & 6e-03 & 3e-03 & 2e-03 & 1e-03 \\ 
   & Ad-Hoc MI& 0.07 & 0.15 & 0.18 & 0.00 & 9e-04 & 8e-04 & 8e-04 & 7e-04 \\ 
   & Gibbs MI & 0.00 & 0.01 & 0.01 & 0.00 & 2e-03 & 1e-03 & 8e-04 & 7e-04 \\ [6pt]
  $\mathbbm{I}_\text{Others}$ 
   & Complete Case & 0.00 & 0.00 & 0.00 & 0.00 & 3e-04 & 5e-04 & 5e-04 & 6e-04 \\ 
   & Joint & 0.12 & 0.44 & 0.48 & 0.41 & 6e-03 & 5e-03 & 3e-03 & 3e-03 \\ 
   & Ad-Hoc MI& 0.07 & 0.07 & 0.01 & 0.00 & 1e-03 & 8e-04 & 7e-04 & 7e-04 \\ 
   & Gibbs MI & 0.09 & 0.01 & 0.00 & 0.00 & 2e-03 & 9e-04 & 7e-04 & 7e-04 \\ [6pt]
  $\mathbbm{I}_\text{Whites}$ 
   & Complete Case & 0.00 & 0.00 & 0.00 & 0.00 & 1e-04 & 2e-04 & 3e-04 & 3e-04 \\ 
   & Joint & 0.30 & 0.54 & 0.49 & 0.50 & 1e-03 & 7e-04 & 5e-04 & 4e-04 \\ 
   & Ad-Hoc MI& 0.13 & 0.12 & 0.00 & 0.00 & 3e-04 & 3e-04 & 3e-04 & 3e-04 \\ 
   & Gibbs MI & 0.14 & 0.01 & 0.00 & 0.00 & 4e-04 & 3e-04 & 3e-04 & 3e-04 \\ [6pt]
  $\mathbbm{I}_\text{Blacks}/\mathbbm{I}_\text{Whites}$ &
  Complete Case & 0.03 & 0.01 & 0.00 & 0.00 & 0.02 & 0.01 & 0.01 & 0.01 \\ 
   & Joint & 0.48 & 0.54 & 0.52 & 0.48 & 0.06 & 0.03 & 0.02 & 0.01 \\ 
   & Ad-Hoc MI& 0.04 & 0.10 & 0.01 & 0.00 & 0.01 & 0.01 & 0.01 & 0.01 \\ 
   & Gibbs MI & 0.07 & 0.01 & 0.00 & 0.00 & 0.02 & 0.01 & 0.01 & 0.01 \\ [6pt]
  $\mathbbm{I}_\text{Hispanics/Latinos}/$ &
  Complete Case & 0.09 & 0.08 & 0.33 & 0.53 & 0.03 & 0.02 & 0.02 & 0.02 \\ 
$\mathbbm{I}_\text{Whites}$  & Joint & 0.24 & 0.46 & 0.51 & 0.45 & 0.14 & 0.07 & 0.05 & 0.03 \\ 
    & Ad-Hoc MI& 0.09 & 0.12 & 0.17 & 0.24 & 0.02 & 0.02 & 0.02 & 0.02 \\ 
   & Gibbs MI & 0.00 & 0.09 & 0.07 & 0.05 & 0.05 & 0.02 & 0.02 & 0.02 \\ [6pt]
  $\mathbbm{I}_\text{Others}/\mathbbm{I}_\text{Whites}$ &
  Complete Case & 0.00 & 0.00 & 0.00 & 0.00 & 0.03 & 0.02 & 0.02 & 0.01 \\ 
   & Joint & 0.12 & 0.43 & 0.47 & 0.41 & 0.16 & 0.12 & 0.09 & 0.07 \\ 
     & Ad-Hoc MI& 0.09 & 0.06 & 0.00 & 0.00 & 0.02 & 0.02 & 0.02 & 0.02 \\ 
   & Gibbs MI & 0.09 & 0.00 & 0.00 & 0.00 & 0.05 & 0.02 & 0.02 & 0.02 \\ [6pt]
   \hline
\end{tabular}
\end{table}

\subsubsection{Breakdown analysis}

\begin{figure}
    \centering
    \includegraphics[scale=0.5]{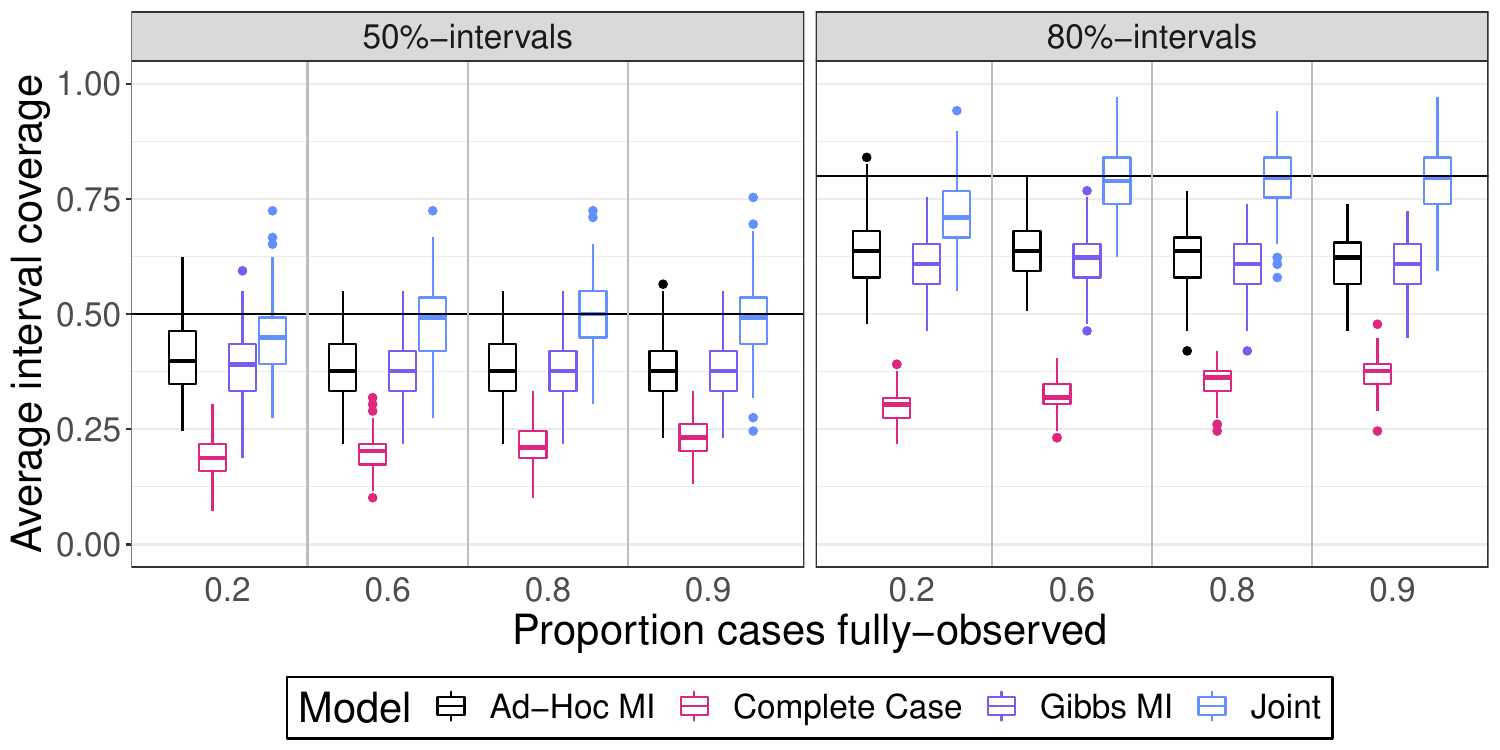}
    \caption{Boxplots of simulation-wise mean 50\% and 80\% interval coverage by observed data proportion scenario for the joint model, the complete-case model, and the multiple imputation methods. Horizontal black lines indicate the nominal coverage probability rates.}
    \label{fig:breakdown-intervals}
\end{figure}

The joint model performs well under the 90\%-, 80\%- and 60\%-observed scenarios, but when there is a significant proportion of cases that are missing race data, like in the 20\%-observed scenario, the model's posterior intervals begin to undercover compared to the nominal coverage probabilities.
One can see this in figure \ref{fig:breakdown-intervals}, as the interval coverage in the 60\% observed scenario begin to undercover slightly as measured by the median across the 200 simulated datasets. 
In the 20\% observed scenario, the $75^\textrm{th}$ quantiles of the mean parameter coverage for the full model for both the 50\% and 80\% intervals lie below the nominal coverage rates.
This leads us to conclude that informative priors are necessary when the model is fitted to datasets that have significant numbers of cases that are missing race data. If the likelihood and prior conflict, however, these priors may have an outsized influence on the posterior estimands.

\subsection{Prior sensitivity results} \label{sec:sim-study-sens}

In order to test the sensitivity of model inferences to priors over population hyperparameters such as the population mean log-incidence ($\boldsymbol{\alpha}_{\boldsymbol{\lambda}}$), or population mean log-odds of observing a specific race/ethnicity category ($\boldsymbol{\alpha}_{\boldsymbol{\eta}}$), we used a subset of 100 simulated datasets from the 20\% observed scenario. We varied the parameters of the priors over the population hyperparameters over a grid and re-estimated the quantities of interest for each prior specification.
We varied one prior parameter at a time while holding the other prior parameters fixed at the values shown in \cref{eq:sim-study-priors}.
The parameter values are shown in \Cref{tab:sim-study-settings}.
\begin{table}[H]
    \centering
    \begin{tabular}{c|c|c}
       Population parameter & Prior parameter & Values \\
       \hline \\
       ${\boldsymbol{\alpha}_{\boldsymbol{\eta}}}_j$ & $\Exp{\boldsymbol{\alpha}_{\boldsymbol{\eta}}}_j$ & $\{0.5,1,\mathbf{2},3\} \forall j$  \\
       ${\boldsymbol{\alpha}_{\boldsymbol{\lambda}}}_j$ & $\Exp{\boldsymbol{\alpha}_{\boldsymbol{\lambda}}}_j$ & $\{-3.5,-4,-4.5,\mathbf{-5}\} \forall j$ \\
       ${\boldsymbol{\alpha}_{\boldsymbol{\eta}}}_j$ & $\mathrm{SD}(\boldsymbol{\alpha}_{\boldsymbol{\eta}})_j$ & $\{0.3, 0.5,\mathbf{1},2,3\} \forall j$  \\
       ${\boldsymbol{\alpha}_{\boldsymbol{\lambda}}}_j$ & $\mathrm{SD}(\boldsymbol{\alpha}_{\boldsymbol{\lambda}})_j$ & $\{0.3, 0.5,\mathbf{1},2,3\} \forall j$ \\
       ${\boldsymbol{\sigma}_{\boldsymbol{\eta}}}_j$ & $\Exp{\boldsymbol{\sigma}_{\boldsymbol{\eta}}}_j/\sqrt{2/\pi}$ & $\{0.25,0.5,\mathbf{1},2\} \forall j$ \\
       ${\boldsymbol{\sigma}_{\boldsymbol{\lambda}}}_j$ & $\Exp{\boldsymbol{\sigma}_{\boldsymbol{\lambda}}}_j/\sqrt{2/\pi}$ & $\{0.25,0.5,\mathbf{1},2\} \forall j$ 
    \end{tabular}
    \caption{Prior sensitivity simulation study prior settings.\\ Bold values correspond to settings used for results presented in \ref{subsec:sim-study-results}. Prior parameter for $\boldsymbol{\sigma}_{\boldsymbol{\lambda}}$ and $\boldsymbol{\sigma}_{\boldsymbol{\eta}}$ is the standard deviation parameter for a half-normal distribution.}
    \label{tab:sim-study-settings}
\end{table}
We measured 1) the sensitivity of the estimated posterior mean incidence by race/ethnic group, or $\mathbb{I}_j$, and 2) its bias.
Our measure of posterior mean sensitivity to the prior mean was the change in posterior mean against a reference mean scaled by a reference standard deviation, where the reference mean and standard deviation were those obtained using the prior settings set out in \Cref{eq:sim-study-priors}.
Specifically, for an estimand $g(\boldsymbol{\theta})$, with a posterior over $\boldsymbol{\theta}$ $\pi_{\mathrm{b}}(\boldsymbol{\theta} | \mathrm{Data})$ under a prior with reference parameters $\mathrm{b}$ and a posterior $\pi_{\mathrm{a}}(\boldsymbol{\theta} | \mathrm{Data})$ under a prior with alternative parameters $\mathrm{a}$:
\begin{align}\label{eq:post-mean-z}
\text{Posterior Z-score}\,= \frac{\mathbb{E}_{\pi_{\mathrm{a}}(\boldsymbol{\theta} | \mathrm{Data})}[g(\boldsymbol{\theta})] - \mathbb{E}_{\pi_{\mathrm{b}}(\boldsymbol{\theta} | \mathrm{Data})}[g(\boldsymbol{\theta})]}{\sqrt{\mathrm{Var}_{\pi_{\mathrm{b}}(\boldsymbol{\theta} | \mathrm{Data})}(g(\boldsymbol{\theta}))}}
\end{align}
The measure of bias for a true estimand $g(\boldsymbol{\theta}^\dagger)$ is
\begin{align}\label{eq:post-mean-bias}
\frac{\mathbb{E}_{\pi_{\mathrm{a}}(\boldsymbol{\theta} | \mathrm{Data})}[g(\boldsymbol{\theta})]-g(\boldsymbol{\theta}^\dagger)}{g(\boldsymbol{\theta}^\dagger)}
\end{align}
\Cref{fig:prior-sens-z-score-alpha} shows that the posterior incidence estimate is somewhat sensitive to the priors over log-population mean incidence and log-odds of observing race/ethnicity information.
The right-hand column in \Cref{fig:prior-sens-z-score-alpha} shows that as the prior mean for $\alpha_{\lambda}$ for the Other group differs from the true data-generating mean by 3 prior standard deviations, the posterior mean can change by roughly half a posterior standard deviation from the baseline prior.

Meanwhile, the left-hand column of \Cref{fig:prior-sens-z-score-alpha} shows the sensitivity of the posterior mean for incidence by race/ethnicity to the prior for $\alpha_{\lambda}$. 
Of interest is the posterior mean for the Other group because it is the minority group.
In the $20$\%-observed scenario, the true $\alpha_\lambda$ for the Other group is approximately $0.3$, while the prior mean for $\alpha_\lambda$ is $2$. When the prior standard deviation is decreased to 0.5 from 1, the prior mean is approximately 3 prior standard deviations away from the true data generating parameter, and the posterior mean decreases by about half a posterior standard deviation. 
Despite the fact that the posterior means can shift due to changes in the prior, however, the posterior mean never exceeds 2 posterior standard deviations, implying that the inferences do not appreciably change.

Digging deeper into the upper-left-hand plot in \Cref{fig:prior-sens-bias-alpha} shows that when the prior for $\boldsymbol{\alpha}_{\boldsymbol{\eta}}$ is centered on missing-at-random missingness and the prior mean is too large compared to the true proportion of cases with observed race/ethnicity, the model over-allocates missing cases to majority groups while it under-allocates cases to minority groups.
If we instead center the prior too low then we may over-allocate cases to minority groups.

The lower-left-hand plot in \Cref{fig:prior-sens-z-score-alpha} shows a similar phenomenon when the prior reflects too-strong certainty that the data-generating process is nearly missing-at-random. 
When too much prior weight is allocated to near-missing-at-random $\boldsymbol{\alpha}_{\boldsymbol{\eta}}$, the model deflates incidence for groups with higher-than-average missingness and inflates incidence for groups with lower-than-average incidence.

\Cref{fig:prior-sens-bias-alpha} shows that the bias is not appreciable for incidence, with the exception of the Other group when the prior for $\alpha_\lambda$ is about 3 standard deviations or more too large.

\Cref{fig:prior-sens-sigma} shows posterior Z-score and bias plots for changes to the prior for population inter-geography standard deviation parameters for $\lambda_{gj}$ and $\eta_{gj}$, or $\boldsymbol{\sigma}_{\boldsymbol{\lambda}}$ and $\boldsymbol{\sigma}_{\boldsymbol{\eta}}$. The posterior for incidence is not especially sensitive to the prior over these parameters.

The results of the prior sensitivity simulation study show that the model inferences for incidence are relatively robust to misspecification of priors for population hyperparameters, but that care should be taken with the prior mass apportioned to data generating processes that are centered on missing-at-random scenarios.

\section{Application to COVID-19 case data in Wayne County, Michigan}

In this section we will apply both the complete-case model and the joint model to COVID-19 case data in Wayne County from the first wave of the pandemic.

\subsection{Data}

\begin{table}[ht]\label{table:population-summary-real-data}
\caption{Population summary in Wayne County, Michigan as of the 2010 Decennial Census}
\centering
\begin{tabular}{@{}lrrrr@{}}
  \hline
 &  & Mean Age$\times$Sex$\times$Race/Eth. &  & $100\times$ Ratio\\ 
Race/Ethnicity & Total Pop. & $\times$PUMA Pop. & Std. dev. PUMA Pop.  &  to White\\ 
  \hline
  Asian/Pacific Islander & 45894 & 196 & 315 & 5 \\ 
  Black & 732801 & 3132 & 3152 & 81 \\ 
  Hispanic/Latino & 95260 & 407 & 757 & 11 \\ 
  Other & 44449 & 190 & 150 & 5 \\ 
  White & 902180 & 3855 & 3225 & 100 \\ 
   \hline
\end{tabular}
\end{table}

\begin{table}[ht]\label{table:case-summary-real-data}
\caption{Cumulative incidence of PCR-confirmed COVID-19 infections in Wayne County, MI from March 1, 2020 through June 30, 2020. Mean and variance for Total uses only observed-race/ethnicity cases. Mean and Variance columns rounded to zero digits.}
\centering
\begin{tabular}{@{}lcccccc@{}}
  \hline
&  & Cumulative &  Risk Relative & & & Prop. zero \\ 
Race/Ethnicity  & Total Cases & Incidence & to Whites & Mean & Variance & counts\\ 
  \hline
  Asian/Pacific Islander &  229 & 0.005 & 1.0 & 1 & 3 & 0.55   \\ 
  Black &  9,577 & 0.013 & 2.6 & 41 & 1904 & 0.02 \\ 
  Hispanic/Latino &    708 & 0.007 & 1.5 & 3 & 34 & 0.37  \\ 
  Other &    834 & 0.019 & 3.8 & 4 & 13 & 0.18  \\ 
  White &  4,476 & 0.005 & 1.0 & 19 & 389 & 0.08  \\ 
  Missing &  3,464 & NA & NA & 15 & 204 & 0.06  \\ 
  Total & 19,288 & 0.011 & 2.1 & 14 & 697 & 0.24  \\ 
   \hline
\end{tabular}
\end{table}

The source of our case data is the Michigan Disease Surveillance System (MDSS) maintained by the Michigan Department of Health and Human Services (MDHHS).
MDHHS's guidelines for the collection of probable COVID-19 cases is set out in \cite{mdhhssop} as outlined in \cite{zelnerRacialDisparitiesCoronavirus2021}.
We included all reported PCR-confirmed COVID-19 cases for individuals outside of state prisons with that were entered into MDSS between $2020-03-01$ through $2020-06-30$. 
This comprises 22,141 cases of COVID-19. 
We then filtered out 1,374 cases, or about 6\% of the total cases, that could not be geocoded to a unique address in Wayne County. 
We filtered a further 74 cases for which the case patients' sex at birth was unknown, as well as 7 cases for which age was unknown.
Finally, we dropped 1,398 cases which were matched to the address of a licensed nursing homes or long-term care facility (LTCF).
We excluded these cases for two reasons: 1) the populations of nursing homes and LTCFs are likely not well-represented by the 2010 Census denominators and 2) the high incidence among nursing home and LTCF residents does not accord with our assumption of a Poisson process for disease cases.
This results in a final dataset of 19,288 PCR-confirmed COVID-19 cases.

In total, approximately 18\% of the 19,288 cases, or 3,464 cases, are missing race data.
For cases that do include the race of the respondent and are not identified as Hispanic or Latino, we classify those who are identified as Asian or Hawaiian or Pacific Islander as Asian, those identified as Black/African American or Black/African American/Unknown as Black, and Caucasian and Caucasian/Unknown as White. 
We classify cases as Hispanic or Latino if the data field for patient ethnicity is equal to Hispanic or Latino. 
We classify those who identify as Native American or Alaska Native, mixed race, or other race as Other.
Cases that are not missing race info but are missing patient ethnicity information are classified as the indicated race and are treated as not Hispanic or Latino. 

We again have 13 PUMAs that comprise Wayne county, and 18 age by sex-at-birth strata per PUMA. 
\subsubsection{Aggregation to PUMAs} This yields 234 observations of the counts of PCR-confirmed COVID-19 cases within each race/ethnicity category, or 1,170 total observations of PUMA by age by sex-at-birth by race/ethnicity.
The mean count is 13.5 while the variance is 696.9. 
As for observations of total counts of cases missing race and ethnicity information by PUMA by age by sex-at-birth, 6\% of the 234 PUMAs have zero observed cases with missing race and ethnicity.

\subsubsection{Population data}
We added the Asian/Pacific Islander group as an additional race/ethnic category, because such individuals make up a significant fraction of the population in Wayne County, though in all other respects the PUMA-level population data is the same as in the simulation study in Subsection \ref{subsec:sim-study-pop-data}.

\subsection{Models and priors}

We fitted four of the models presented in Section \ref{subsec:inf-models-sim-study}: the joint model, the complete-case model, and the ad-hoc and Gibbs multiple imputation models. 
The full specification for the joint model is:
\begin{align}
  \begin{split}\label{model:full-geo-applied}
    X_{igj} | \lambda_{gj}, \mathbf{z}_i, \boldsymbol{\beta}_g, p_{igj}, E_{igj} & \sim \text{Poisson}(p_{igj} \lambda_{gj} \exp(\mathbf{z}_i^T \boldsymbol{\beta}_g) E_{igj}),\\
    M_{ig} | \boldsymbol{\lambda}_{g}, \mathbf{z}_i, \boldsymbol{\beta}_g, \mathbf{p}_{ig}, \mathbf{e}_{ig} & \sim \text{Poisson}(\exp(\mathbf{z}_i^T \boldsymbol{\beta}_g) \textstyle \sum_j ((1 - p_{igj}) \lambda_{gj} E_{igj}), \\
    p_{igj} & = \lp 1 + \exp(-(\mathbf{z}_i^T \boldsymbol{\gamma}_g + \eta_{gj}) \rp^{-1}, \\
   \log\boldsymbol{\lambda}_g | \boldsymbol{\alpha}_\lambda, \boldsymbol{\sigma}_\lambda & \sim \text{MultiNormal}(\boldsymbol{\alpha}_\lambda, \text{diag}(\boldsymbol{\sigma}_\lambda^2)), \\
   \boldsymbol{\eta}_g | \boldsymbol{\alpha}_\eta, \boldsymbol{\sigma}_\eta & \sim \text{MultiNormal}(\boldsymbol{\alpha}_\eta, \text{diag}(\boldsymbol{\sigma}_\eta^2)), \\
   \boldsymbol{\beta}_g | \boldsymbol{\alpha}_\beta, \boldsymbol{\sigma}_\beta  & \sim \text{MultiNormal}(\boldsymbol{\alpha}_\beta, \text{diag}(\boldsymbol{\sigma}_\beta^2)), \\
   \boldsymbol{\gamma}_g | \boldsymbol{\alpha}_\gamma, \boldsymbol{\sigma}_\gamma & \sim \text{MultiNormal}(\boldsymbol{\alpha}_\gamma, \text{diag}(\boldsymbol{\sigma}_\gamma^2)), 
  \end{split}
\end{align} 
with the same priors over the hyperparameters as in \cref{eq:sim-study-priors} with the exception of the prior scale for $\boldsymbol{\sigma}_\gamma$ set to $1$ instead of $0.5$.

The full specification for the complete-case model is 
\begin{align}
  \begin{split}\label{model:comp-case-applied}
    X_{igj} | \lambda_{gj}, \mathbf{z}_i, \boldsymbol{\beta}_g, p_{igj}, E_{igj} & \sim \text{Poisson}(\lambda_{gj} \exp(\mathbf{z}_i^T \boldsymbol{\beta}_g) E_{igj}), \\
   \log\boldsymbol{\lambda}_g | \boldsymbol{\alpha}_\lambda, \boldsymbol{\sigma}_\lambda & \sim \text{MultiNormal}(\boldsymbol{\alpha}_\lambda, \text{diag}(\boldsymbol{\sigma}_\lambda^2)), \\
   \boldsymbol{\beta}_g | \boldsymbol{\alpha}_\beta, \boldsymbol{\sigma}_\beta  & \sim \text{MultiNormal}(\boldsymbol{\alpha}_\beta, \text{diag}(\boldsymbol{\sigma}_\beta^2)), 
   \end{split}
\end{align}
with the same priors as the joint model over the shared hyperparameters $\boldsymbol{\alpha}_\lambda, \boldsymbol{\alpha}_\beta, \boldsymbol{\sigma}_\lambda$ and $\boldsymbol{\sigma}_\beta$.

$\mathbf{z}_i$ was $9$-dimensional, with the first element encoding male vs. female and the next eight elements encoding the age stratum from $[0,10)$ to $[70,80)$. 
We used a sum contrast for age and a scaled sum contrast for male vs. female. 
We used the results of Theorem \ref{full-model-id} to check that our model as defined is locally identifiable for each PUMA. All 13 PUMAs meet our criteria for the model to be locally identifiable.
We needed to rerun the identifiability analysis because we expanded our race/ethnicity categories by one to include Asians/Pacific Islanders as a separate group.
Our construction of the $\boldsymbol{\beta}_g$ and $\boldsymbol{\gamma}_g$ is the same as in the simulation study.

\subsubsection{Computational results}

We again used \texttt{cmdstanr} as the Stan interface via \texttt{R} \citep{cmdstanr,rcore}.
Each model was run with 8 MCMC chains with 3,000 warmup iterations, and 2,000 post-warmup iterations with a target Metropolis acceptance rate of 0.99.
For the joint model, all $\hat{R}$s were less than 1.01, while the minimum bulk and tail ESS efficiencies were 0.098 and 0.200 rounded, respectively.
For the complete-case model, all $\hat{R}$s were less than 1.01, while the minimum bulk and tail ESS efficiencies were 0.156 and 0.238, respectively. All ESS efficiency numbers are rounded to three digits.

The multiple imputation methods were run for 1,000 warmup, and 2,000 post-warmup iterations for each of the 100 imputed datasets.
All $\hat{R}$s were below 1.01 for each imputed datasets MCMC run, and minimum bulk and tail efficiencies exceeded 10\% for the Gibbs imputation scheme while minimum bulk and tail efficiencies exceed 9\% and and 10\%, respectively for the ad-hoc imputation scheme.
Note that the $\hat{R}$ statistics for the combined chains are typically larger than 1.01 for many parameters of interest, which can be seen in \cref{tab:sampling-eff}. This is due to the between-imputed-dataset variance. 

\subsection{Results and Model Comparison}

\subsubsection{Comparison of model results on completely-observed cases}

Following \cite{gelman2020bayesian} and \cite{gabry2019}, we performed a series of graphical posterior predictive checks, or PPCs, using the \texttt{bayesplot} package \citep{bayesplot}. These involved simulating PUMA by age by sex by race case counts from the fitted models and comparing these outputs to the observed data. Along this dimension, the joint model and the complete-case model were indistinguishable in terms of errors, squared errors, and 50\%, 80\% and 95\% interval coverage for the observed data. 

These checks also revealed that the observational variance, or $\frac{1}{IJ - 1}\sum_{i,j}(x_{ij} - \bar{x})^2$, $\bar{x} = \frac{1}{IJ} \sum_{i,j} x_{ij}$ and the proportion of zeros, or $\frac{1}{IJ} \sum_{i,j} \ind{x_{ij} = 0}$, fell near the $50^\textrm{th}$ percentile for each model's posterior over the two statistics, which indicates that the Poisson distribution is a suitable outcome distribution for this dataset.

We also used graphical PPCs to gauge whether the model assumption that there is no interaction between race and age is reasonable. 
The plots are included in Appendix Section \ref{app:ppc-age}, and show that while there were deviations from the model's posterior distribution for age by race cumulative incidence, they are small compared to the total cumulative incidence. 
Moreover, our interest lies in quantifying cumulative incidence by race for Wayne county instead of capturing all sources of variation in the observed data.

\subsubsection{Posterior predictive checks on missing cases}

We can compare the observed statistics for the missing cases to the joint model's posterior predictive distribution for the same statistics. 
The mean, variance, and proportion of age/race/sex strata with zero cases observed all fell well within the joint model's central 50\% posterior intervals.
A posterior predictive rootogram shown in Appendix Section \ref{app:root} that the tail is a bit thicker than the joint model expects, but the deviation is not extreme enough to warrant modifying the model.  

\subsubsection{Inference on epidemiological estimands}

Following the results of our simulation study, the models' inferences differed for the estimands introduced in Subsection \ref{subsec:estimands-simu-study}, like modeled incidence, standardized incidence, standardized incidence ratios, and functions of these estimands.

\begin{figure}
    \centering
    \includegraphics[scale=0.5]{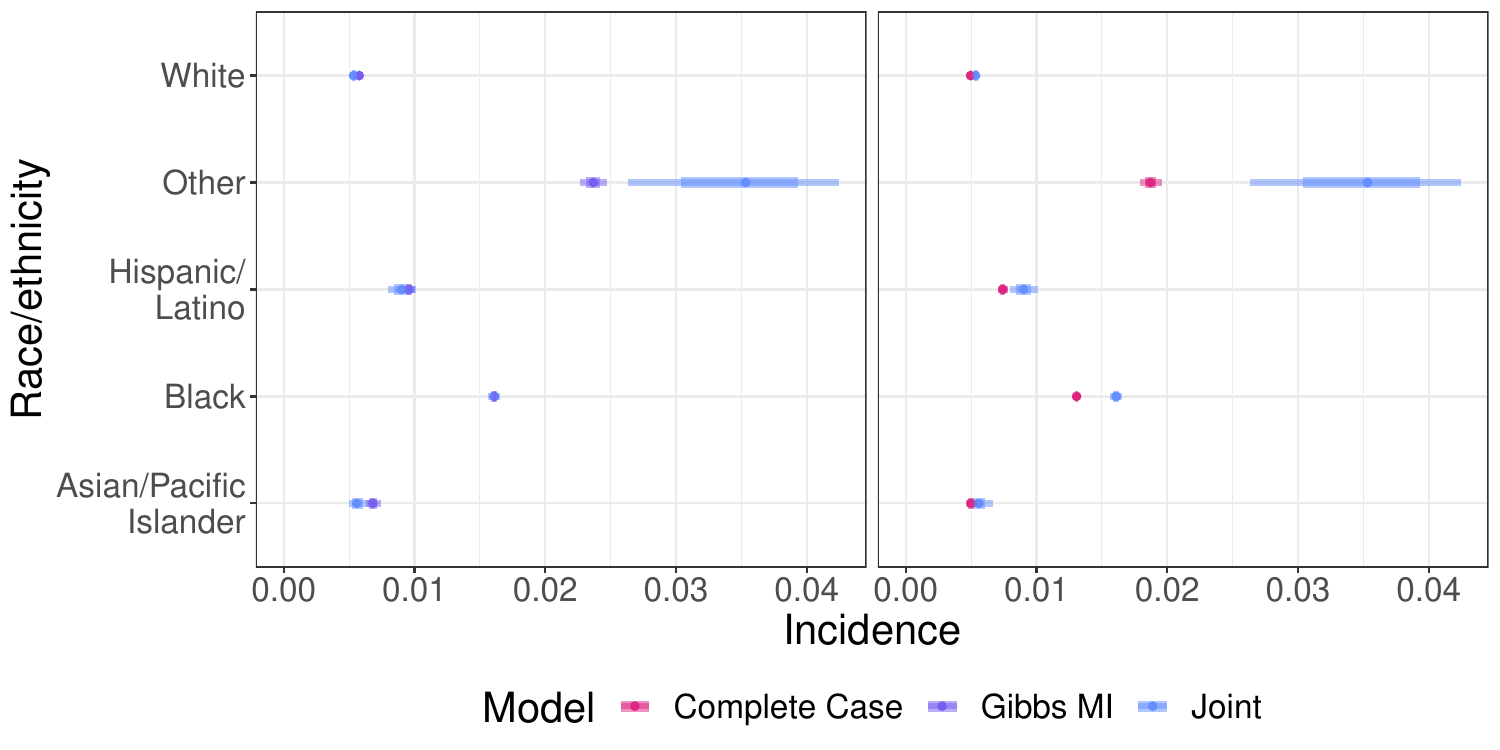}
    \caption{Race/ethnicity category-specific modeled incidence by model. The inner intervals are 50\% and the outer intervals are 80\%.}
    \label{fig:inc-applied}
\end{figure}

A comparison of the modeled incidence inferences for the joint model, the complete-case model, and the Gibbs-sampler-imputation method is shown in Figure \ref{fig:inc-applied}. 
The most striking aspect of the figure is the elevated incidence in the Other race category across all methods.
The complete case model infers uniformly lower incidence than does the joint model, which makes sense as the complete case model omits cases that are missing race/ethnicity information. 
The left-hand panel shows the Gibbs-imputation method imputes higher incidence for Whites, Asians/Pacific Islanders, and Hispanics/Latinos compared to the joint model.
This mirrors the Gibbs performance in the simulation study as shown in \Cref{fig:sim-ias-20pct-miss}. The plot shows that the standardized difference in posterior means between the Gibbs imputation and the joint model is systematically greater than zero for Hispanic/Latinos and Whites, while it is systematically lower than zero for Others in the 80\% observed data scenario.
Visually, we can see that the understatement for incidence is more extreme for Blacks and Others than it is for Hispanics or Latinos and for Whites.
Both the Gibbs and complete case intervals are shorter than the joint-model intervals.

\begin{figure}
    \centering
    \includegraphics[scale=0.5]{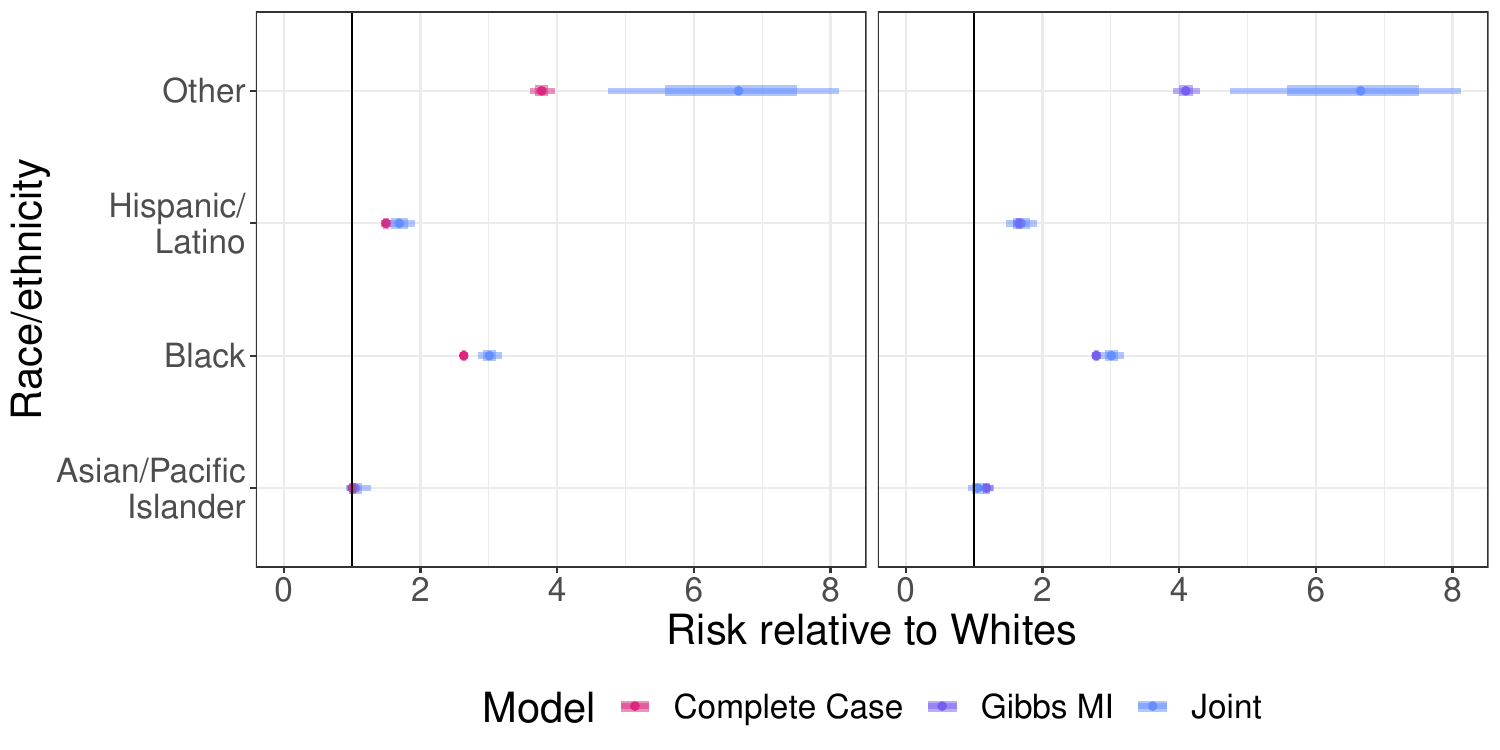}
    \caption{Relative risk of COVID-19. The inner intervals are 50\% and the outer intervals are 80\%.}
    \label{fig:rr-applied}
\end{figure}

Figure \ref{fig:rr-applied} shows the relative modeled incidence, or $\mathbb{I}_j / \mathbb{I}_J$, where $J$ is the category for Whites. 
The plot shows that relative risks for all nonwhites are smaller when using complete case analysis compared to that of the joint model.
The increase is most substantial for the Other race/ethnicity category, but both Blacks and Hispanic and Latinos have significant increases in relative risk. 

Table \ref{tab:estimands-applied} in Appendix Subsection \ref{app:post-table-applied} shows the exhaustive comparison between the two models for all of the estimands. 
Despite the models showing statistically significant differences for posterior means among the standardized incidence ratios, the practical differences are small for Blacks, Hispanics and Latinos and Asians and Pacific Islanders.
The 80\% posterior credible intervals, on the contrary, are larger on average for the joint models' inferences on the standardized incidence ratios.
Whites and people of Other races are seen to have statistically significant and practically significant differences in the models' posterior mean estimators. 
The models' inferences differed most significantly in terms of relative incidence, as can also be seen in Figures \ref{fig:rr-applied}, where for Blacks vs. Whites and Others vs. Whites the posterior 80\% credible intervals do not overlap, even after taking into account Monte Carlo standard error.
The joint model's posterior intervals for the epidemiological parameters of interest were wider on average, in agreement with the simulation study results.
\subsubsection{Inference on missingness parameters}

\begin{figure}
    \centering
    \includegraphics[scale=0.5]{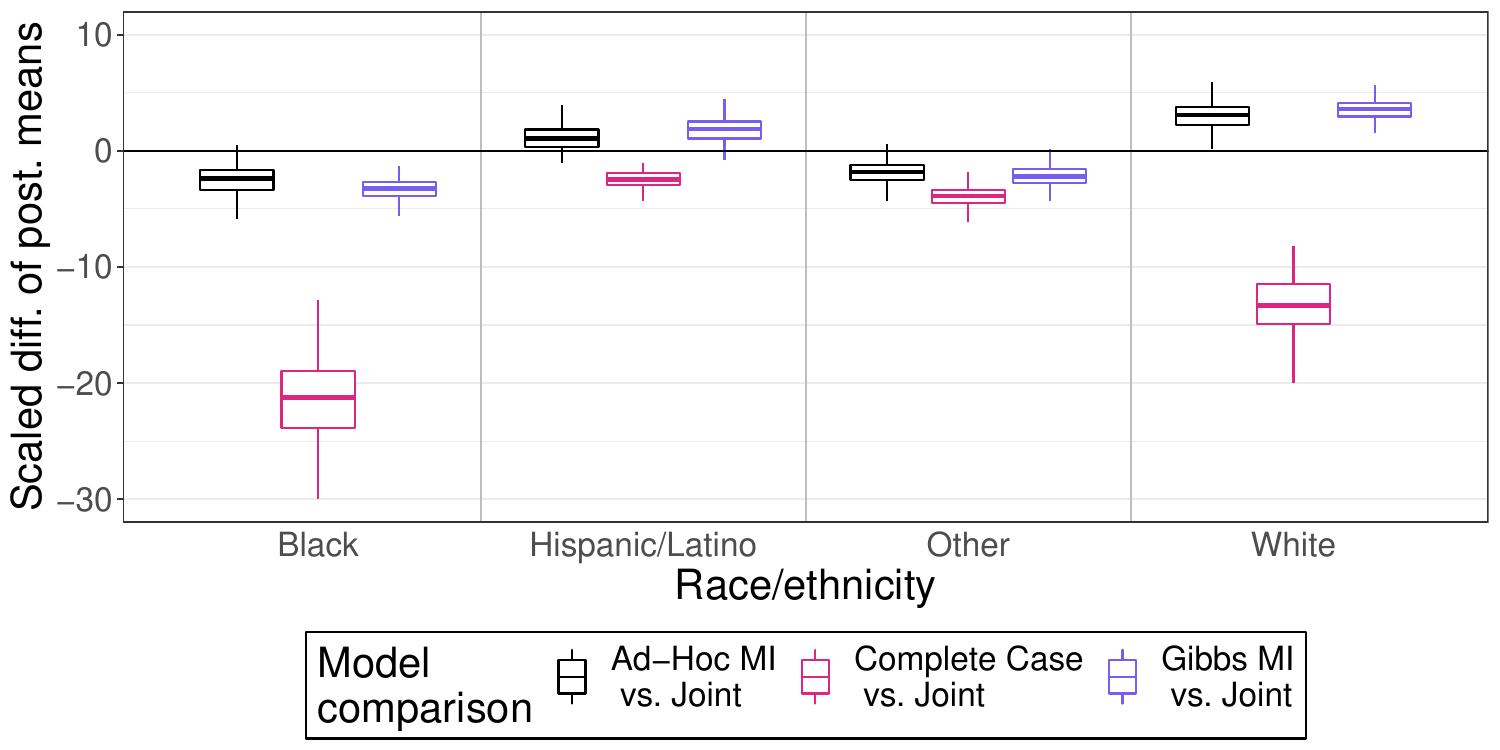}
    \caption{Boxplots of differences in posterior means between indicated methods and joint model scaled by pooled posterior standard deviation by race/ethnicity category-specific modeled incidence by simulated dataset for the 80\% observed data scenario. }
    \label{fig:sim-ias-20pct-miss}
\end{figure}

\begin{figure}[!htb]
    \centering
        \centering
        \includegraphics[scale=0.5]{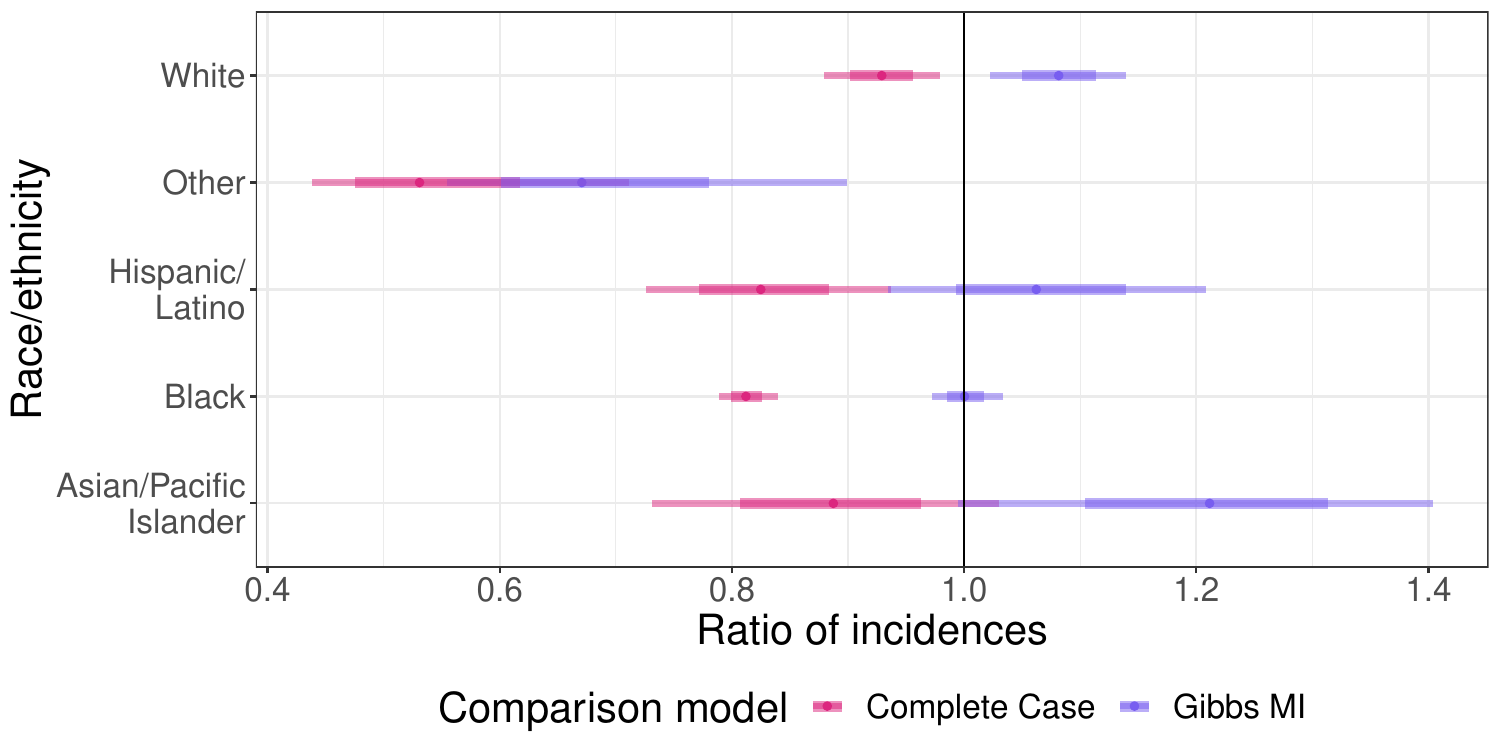}
        \caption{Posterior credible intervals for the ratio of modeled incidences by race/ethnicity, or $\mathbb{I}^\text{CC}_j / \mathbb{I}^\text{J}_j$ where $\text{CC}$ stands for complete case model and $J$ stands for the joint model. The inner and outer intervals are 50\% and 80\% respectively.}
        \label{fig:inc-ratio}
\end{figure}

\begin{figure}[!htb]
        \centering
        \includegraphics[scale=0.5]{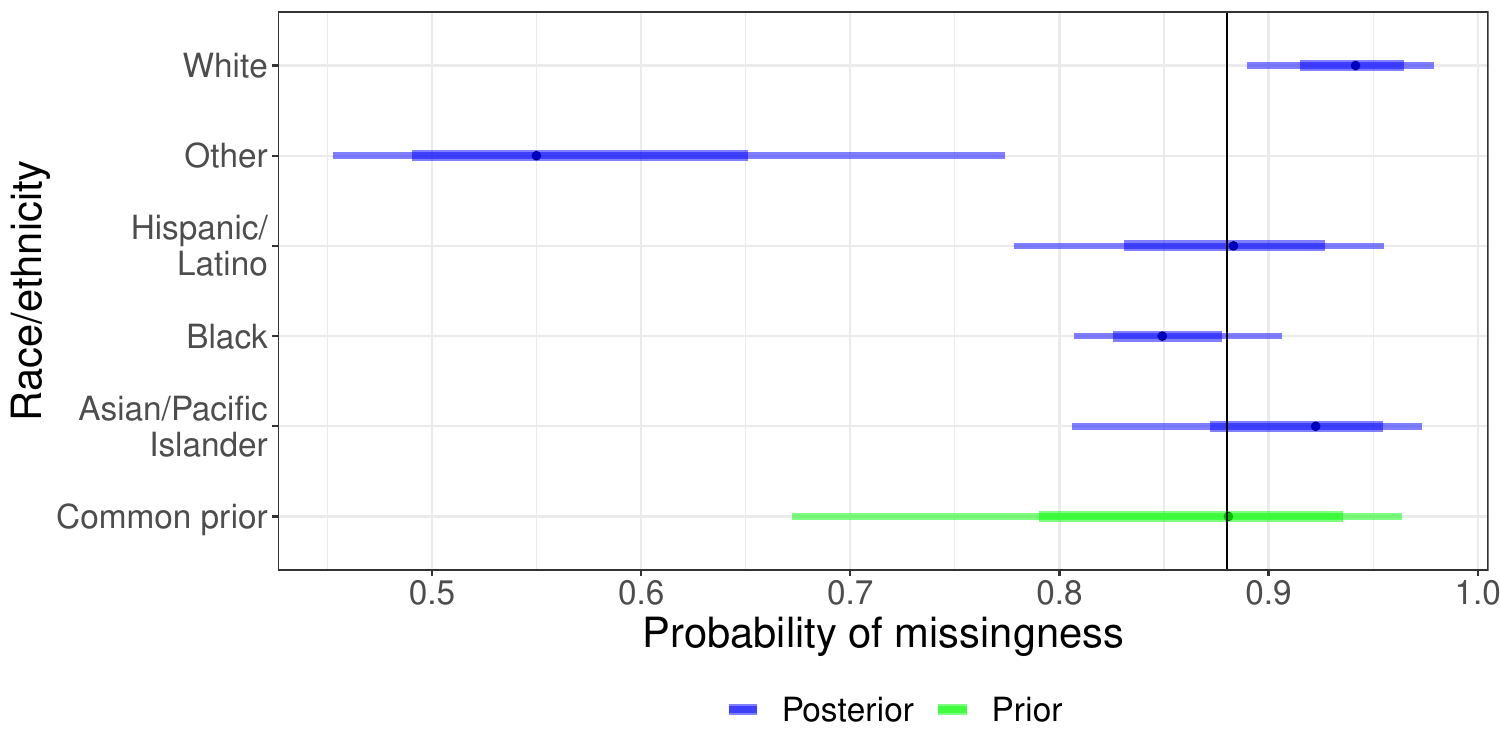}
        \caption{Posterior credible intervals for the population proportion of cases with fully-observed race data, all else being equal, by race/ethnicity, or $\text{inv\_logit}((\boldsymbol{\alpha}_\eta)_j)$. The inner and outer intervals are 50\% and 80\% respectively.}
        \label{fig:population-prob-obs}
\end{figure}

We cannot directly compare the inferences for the missingness parameters for the complete-case model to the joint model. 
We can, however, examine how the ratio of modeled incidences by race differs between races. 
In Figure \ref{fig:inc-ratio}, one can see that the 80\% posterior credible intervals for the ratio of the complete-case model's incidence to the joint model's incidence do not include 1 for all races other than Asians and Pacific Islanders. 
The only groups for which the ratio of Gibbs-to-joint-model incidences exclude zero are Whites and Others, which again mirrors the pattern in \Cref{fig:sim-ias-20pct-miss}, though the difference is less extreme for Hispanics/Latinos and Asians/Pacific Islanders for the real-world data. 
The Gibbs imputation method's inference for Blacks nearly matched that of the joint model. 
This is not surprising when we consider the fact that between-group comparisons of the ratio of incidences reveals that non-White residents, excluding Others had missingness proportions that were near equal between groups. 
This can be seen from \ref{fig:inc-ratio} as the posterior intervals for the Complete Case comparison overlap for Hispanics/Latinos, Blacks and Asian/Pacific Islanders.
It can also be seen that the posterior intervals for Others do not overlap with any other category, and that Whites and Blacks are also do not overlap. 
Figure \ref{fig:population-prob-obs} shows the supporting evidence for NMAR missingness of race; the plot shows the Wayne-County-wide population inferences for the probability that an individual with COVID-19 of a certain race will have race reported in their case line-listing, all else being equal.
This estimand is a transformation of the $\boldsymbol{\alpha}_\eta$ parameter, namely, $\text{inv\_logit}(\boldsymbol{\alpha}_\eta)$.
The strongest evidence for NMAR missingness exists for the Other category, whose 80\% posterior credible intervals do not intersect any other category's intervals. 
There is also some evidence for NMAR missingness for Blacks with respect to Whites, as the 80\% posterior credible intervals for the probability of completely observing race are $(0.81,0.91)$ vs. $(0.89,0.98)$, respectively, as shown in Table \ref{tab:miss-estimands} in Appendix Subsection \ref{app:post-table-applied}.

\subsubsection{Summary} Our results largely align with those of prior analyses of racial disparities in COVID-19 incidence in the U.S. For example, \citeauthor{labgoldEstimatingUnknownGreater2021} found a similarly large incidence among case-patients of Other race.
The authors find a bias-adjusted PCR-confirmed COVID-19 rate of nearly 14\% among Other race case-patients compared to rates of at most nearly 4\% in Hispanic/Latino case-patients, who had the next-largest incidence among the races included in \citeauthor{labgoldEstimatingUnknownGreater2021}'s study.
The relative incidence between Others and Whites is nearly 14, which puts our 80\% posterior credible interval of $(4.77, 8.11)$ in context.

Several explanations are plausible for the elevated incidence among people of Other races; Wayne county has a large Middle Eastern population and these individuals may identify themselves as not being Black, Hispanic or Latino, Asian or Pacific Islander or White. 
The case data does include a field for Arab ethnicity, but the 2010 Census did not include a Middle Eastern or North African category for ethnicity.
Another explanation may be our treatment of missing Hispanic/Latino ethnicity information. 
If many people who are identified as Other but do not have a recorded Hispanic/Latino ethnicity are truly Hispanic/Latino then our model would inflate the incidence in the Other category at the expense of the Hispanic/Latino category; given that the Other group is so small a small inflation in counts would result in a large inflation of risk.  

There is strong evidence for nonignorable missingness driven by not-missing-at-random race covariates.
The evidence is strongest for people of Other races.
This means that omitting cases that are missing race and calculating relative risk between any race and Other would yield a biased estimate. 
Moreover, the size of the bias would be large because the probability of observing race for Others is low compared to the other categories; the 80\% posterior credible interval is $(0.45,077)$.
There is also some evidence for NMAR missingness for Blacks with respect to Whites.
Given the small number of PUMAs we modeled, there would likely be stronger evidence in favor of NMAR missingness for other race/ethnicity categories if we were to model a larger geographical area, like all of Southeastern Michigan instead of just modeling Wayne County.

While the Complete Case inferences are predictably different from the joint model's, the multiple imputation using Gibbs sampling also produced significantly different inferences.
The coherence between the simulated data example and the applied data analysis suggest that multiple imputation procedures that assume MAR missingness when data are NMAR can exacerbate biases in the data by over-imputing cases for groups that are over-represented in the data because of NMAR missingness. 
This suggests that care must be taken when choosing an imputation procedure for missing demographic data.


Overall, our case study illustrates the large risk of bias associated with ignoring NMAR categorical data when inferring relative risks from real-world data. 

\section{Discussion}\label{sec:discussion}

Non-random missingness of race/ethnicity covariate data is a critical challenge for the analysis of public heath data during the COVID-19 pandemic.
Multiple imputation methods, which have been adopted broadly in the analysis of survey data in which the assumption of ignorability is typically reasonable \citep{audigier2018multiple}, may not be appropriate for the analysis of missing race/ethnicity covariates in public health surveillance data in which the possibility of not-missing-at-random (NMAR) missingness is greater.

In order to meet the needs of public health researchers to model disease data that are missing important covariates, we developed a method to jointly model the missingness process along with the disease process. 
Most importantly, the model can learn the extent to which the missingness process is NMAR, so our method is broadly applicable to scenarios where missingness could plausibly be NMAR, like that of missing race data. 

We use a selection model formulation that combines a Poisson sampling model for the counts of disease by stratum and a conditional binomial sampling model for cases with completely observed race/ethnicity with a probability of success parameter that depends on the race/ethnicity category. 
Through the incorporation of known population counts from census data, the model parameters can be identified. The model can be extended to incorporate a log-linear model for incidence, a logistic model for missingness, and a hierarchy to allow for geographic heterogeneity in local parameters. 

Our use case is focused on missing race data in COVID-19 cases in Wayne County, Michigan from March 2020 through June 2020, which we suspect may have been NMAR.
Wayne county saw the largest share of PCR-confirmed COVID-19 cases in the first wave of the pandemic, and also had a large share of cases that were missing race data, so it makes for an appropriate test bed for our method. 

We ran a simulation study using Wayne county as the setting where we varied the proportion of cases with observed race from as high as 90\% to as low as 20\% to quantify the joint model's finite sample performance and to compare its performance against a complete-case analysis and two multiple imputation methods.
The results showed that the joint model performed well in the 90\%- through 40\%-observed scenarios compared to the competing methods though its performance suffered in the 20\% observed-data scenario.
This leads us to conclude that in order to use the joint model effectively in sparse data scenarios, better priors will be needed; prior formulation for the model is an area of active research.

We then applied the models to a dataset comprising PCR-confirmed COVID-19 cases with incomplete race data from Wayne County between March 2020 through June 2020.
The differences between the joint-model inferences and the multiple-imputation inferences suggest that the missingness process for race may be NMAR and that care must be taken when applying methods that assume data are MAR.
Model results also suggest that cases in the Other category, which comprises those of mixed race, Native Americans, and Other races, are being undercounted in Wayne County.

\subsection{Limitations}

The biggest limitation of our analysis is the result of the joint model's dependence on census data for identifiability.  
This required the use of 2010 Decennial Census data, which is 10 years old, and may be systematically different than the true population distribution in Wayne County in 2020.
The 2010 Decennial Census, however, is the most up-to-date source of spatially detailed  population information, reflecting a broader limitation of any analysis that is dependent on decennial census data to estimate infection rates and relative risks.
Because of this dependence on census population data, we were unable to model risk for race/ethnic categories that were potentially important in the Wayne County COVID-19 dataset, but for which census data were not available.   
MDSS collected information on Hispanic/Latino ethnicity separately from race, which resulted in missing covariate information in both categorical variables. 
Ideally we would have applied our method to multiple categorical covariates with missingness, but we were prevented from doing so due to the Census' coarse race and ethnicity categories.
As stated in Section \ref{subsec:sim-study-pop-data}, if the Census recorded ethnicity and race separately, we would be able to model the effect of ethnicity separately from that of race and we could treat the missing ethnicity data separately from that of missing race data.
Instead, we set race/ethnicity as being equal to the observed race if Hispanic/Latino ethnicity was missing, which can understate uncertainty in our posterior and could result in understating incidence for Hispanics and Latinos and overstating incidence in all other categories.
We ran a separate analysis where we treated these observations as missing race; the incidence results largely agreed with the model we presented in the main text for Blacks, Latinos, and Asians, though we observed significant differences in the White and Other incidences.
This analysis overstates uncertainty, because individuals for whom we observe race but not ethnicity can be only one of two categories, but our model in its current formulation treats these cases as potentially arising from any of the race/ethnicity categories. 
Until we have detailed 2020 Decennial Census results, we cannot model ethnicity and race separately.

We are also constrained by the mismatch of the 2010 Census question about sex and our dataset's definition of sex at birth. 
As \cite{kennedy2020using} argues, responses to the U.S. Census' question of sex may not correspond to sexes at birth. 
This mismatch can lead to bias in our parameter estimates and an understatement of uncertainty.

Another limitation of our model is that it assumes a Poisson sampling distribution for incident cases of disease. 
When cumulative incidence increases over time, as has occurred with COVID-19, a binomial sampling model may be more appropriate\footnote{See \cref{app:dynamic} for an extension to a binomial likelihood}.
Similarly, our model assumes conditional independence between disease counts, which may not be appropriate as cumulative incidence grows\footnote{See \cref{subsec:future} for more discussion}.
Both of these reasons are why we decided to focus on the first wave of the pandemic, which is when the disease was relatively rare among the population of Wayne County and for which the violations of conditional independence assumption could be reasonably assumed to not lead to too much understatement of uncertainty.

\subsection{Conclusion}\label{sec:conclusion}

Public health surveillance systems will always have to contend with missing data. Because  the nature and causes of this missingness are likely to change over time and across disease systems, it is important that the methods used to address missingness are flexible and able to account for both MAR and NMAR covariates. In Michigan, missingness of categorical demographic data among COVID-19 cases has varied over the course of the pandemic. For example, some localities in our data reported as much as 40\% of PCR-confirmed COVID-19 cases having missing data on race/ethnicity for the period of rapidly-increasing incidence from October 2020 to February 2021.

Our simulation study shows that complete-case analysis or na\"ive multiple imputation can yield uncertainty intervals that are too 
short to be useful and point estimators that can over- or under-state between-group relative risks.
Our method represents a computationally tractable and analytically transparent alternative that performs well in many scenarios, as evidenced in our simulation studies as well as analysis of data from Wayne County, Michigan.Given the need for public health authorities to characterize risks of disease among different population groups in as close to real-time as possible, flexible, efficient methods such as ours, are urgently needed.
\subsubsection{Extensions and future work}\label{subsec:future}
This work can serve as a foundation on which to build new joint-disease-missing-covariate models targeted to specific applications. Although the model presented here can give useful inferences in its own right in a variety of settings, despite its relative simplicity,  domain-specific modifications may be appropriate. For example, future models could incorporate multi-level information on the public health and healthcare systems generating surveillance data to account more explicitly for contextual drivers of missingness.

The joint model can also be extended to account for infectious disease transmission dynamics and other sources of temporal and spatial autocorrelation. For example, the one-period Poisson sampling model can be extended to a time-series susceptible-infected-recovered (TSIR) model\footnote{TSIR models use a negative binomial likelihood; the code in \cref{app:dynamic} is easily extensible from binomial to negative binomial} or an endemic/epidemic model, both of which are discrete time analogues to classical susceptible-infected-recovered models \citep{,heldModelingSeasonalitySpacetime2012,meyerPowerlawModelsInfectious2014,wakefieldhandbook,bauerStratifiedSpaceTime2018,keeling2011modeling}. 
When the disease becomes more widespread, potentially requiring a binomial likelihood, modeling the data a finer spatial resolution would make integrating over non-Poisson random variables more computationally efficient, particularly when combined with parallel computation of the likelihood. 
A dynamic programming implementation of the likelihood using binomial- instead of Poisson-distributed disease counts is included in \cref{app:dynamic}. 
In order to regularize the model's inferences as the parameter space dimension increases in step with the spatial resolution, one can use a computationally-efficient log-Gaussian Cox process as a prior for the spatially-dependent parameters \citep{liLogGaussianCox2012,simpsonGoingGridComputationally2016}.
Furthermore, relaxing the conditional independence between categories is possible through a latent Poisson model, which we leave for future work.

The dependence of the joint model on the availability of sufficiently detailed and recent census data can also be mitigated. For example, uncertainty in group-specific population denominators can be  accounted using   frequently updated population datasets, such as the American Community Survey, even if these data are not available at the same level of spatial granularity as decennial census data.
An alternative route is to perform a  ``tipping point'' sensitivity analysis \citep{liublinskaSensitivityAnalysisPartially2014} to flag changes in census data that would lead to a substantive change in conclusions (e.g. a reversal of the sign for log relative-risk measures).
Given the many degrees-of-freedom of census population data, and the critical role played by such data in population-based analyses of health and illness, this presents an interesting and important challenge that should be explored in future work.

\begin{appendix} 

\section{Selection model derivation}\label{app:selection-model}

Following \cite{littleConditionsIgnoringMissingData2017} and \cite{gelman2013bayesian}, we wish to model the joint distribution for the data:
$$
\prod_{(i,j)} f((y_{ij}, x_{ij}) | \mu_{ij}, p_{ij})
$$
which we have factorized according to a selection model paradigm: $f(y_{ij} | \mu_{ij})$, and $f(x_{ij} | y_{ij}, p_{ij})$, which follows from the conditional independence across $i$ and $j$ assumed in the generative model above.
Let the vectors $\boldsymbol{\mu}_i$, $\mathbf{p}_i$ be the $J$-vectors with respective $j^\textrm{th}$ elements $\mu_{ij}$ and $p_{ij}$.  
Let $x_{ij}$, $w_{ij}$, $m_i$ be a specific realizations of $X_{ij}$, $W_{ij}$, $M_i$  and let $\mathbf{x}_i$, $\mathbf{w}_i$ be defined as $\boldsymbol{\mu}$ was defined, and where $W_{ij} \coloneqq Y_{ij} - X_{ij}$.
Let $\mathbf{m}$ be the $I$ vector with $i^\textrm{th}$ element $m_i$.
Then the complete data likelihood is defined as:
\begin{align*}
L((\boldsymbol{\mu}_1,\mathbf{p}_1), \dots, (\boldsymbol{\mu}_I, \mathbf{p}_I) & | (\mathbf{x}_1,\mathbf{w}_1), \dots, (\mathbf{x}_I, \mathbf{w}_I)) \\ &= \int \prod_{(i,j)} e^{-\mu_{ij}}\frac{\mu_{ij}^{y_{ij}}}{y_{ij}!} \frac{y_{ij}!}{x_{ij}!w_{ij}!} p_{ij}^{x_{ij}}(1 - p_{ij})^{w_{ij}} d\mathbf{w}_1 \dots d\mathbf{w}_I
\end{align*}
which is shown in \cref{app:like-deriv} to be
\begin{align*}
L((\boldsymbol{\mu}_1,\mathbf{p}_1), \dots, (\boldsymbol{\mu}_I, \mathbf{p}_I) & | (\mathbf{x}_1,m_1), \dots, (\mathbf{x}_I, m_I)) \\ &= \lp\prod_{(i,j)} e^{-p_{ij}\mu_{ij}}\frac{(p_{ij}\mu_{ij})^{x_{ij}}}{x_{ij}!}\rp\prod_{i} e^{-\sum_j (1 - p_{ij})\mu_{ij}}\frac{(\sum_j (1 - p_{ij})\mu_{ij})^{m_{i}}}{m_{i}!}
\end{align*}
By the properties in \cite{littleConditionsIgnoringMissingData2017} and \cite{gelman2013bayesian} if $p_{ij} \neq p_i \forall (i,j)$ the complete data likelihood does not factorize into a term governing the observational process in $Y$ and the missinginess process in $R$, viz.
\begin{align*}
    L((& \boldsymbol{\mu}_1, \mathbf{p}_1), \dots, (\boldsymbol{\mu}_I, \mathbf{p}_I) | (\mathbf{x}_1,m_1), \dots, (\mathbf{x}_I, m_I)) \neq \\ 
    & L(\boldsymbol{\mu}_1, \dots, \boldsymbol{\mu}_I | (\mathbf{x}_1,m_1), \dots, (\mathbf{x}_I, m_I)) L(\mathbf{p}_1, \dots, \mathbf{p}_I | (\mathbf{x}_1,m_1), \dots, (\mathbf{x}_I, m_I))
\end{align*}
Given that equality does not hold when $p_{ij}$ vary by $j$, we can say that in this case the data are not missing at random (NMAR), and thus we must model the joint distribution of observed data  and missing data.

The observed data likelihood above is equivalent to the following generative model for the observed random variables $X_{ij}$ and $M_i$:
\begin{align}\label{model:general}
  \begin{split}
    X_{ij} | p_{ij} \mu_{ij} & \sim \text{Poisson}(p_{ij} \mu_{ij}) \\
    M_{i} | \boldsymbol{\mu}_i, \mathbf{p}_i & \sim \text{Poisson}(\textstyle \sum_{j} \mu_{ij} (1 - p_{ij}) )
  \end{split}
\end{align}

If we observe data for more than one geographic area, say for $g \in \{1, \dots, G\}$, we might expect our parameters to vary across locations. 
For example, geographic heterogeneity in cumulative incidence has been  a fundamental characteristic of the COVID-19 pandemic and many other infectious disease outbreaks and epidemics \citep{bilal2021spatial,wakefieldhandbook}.
We can extend our generative model to capture geographic variation if we index our parameters with $g$ and model them as jointly distributed under $F_{\boldsymbol{\phi}}$, with $\boldsymbol{\phi}$ as a vector of unknown hyperparameters:
\begin{align}\label{model:geo-hierarchical}
((\boldsymbol{\mu}_{1g},\mathbf{p}_{1g}),  \dots, (\boldsymbol{\mu}_{Ig},\mathbf{p}_{Ig})) | \boldsymbol{\phi} \sim F_{\boldsymbol{\phi}} \ , \forall g,
\end{align}
where $\boldsymbol{\mu}_{ig}$ and $\mathbf{p}_{ig}$ are $J$-vectors where the $j^\textrm{th}$ elements are equal to $\mu_{igj}$ and $p_{igj}$, respectively.
The observed data model becomes
\begin{align}\label{model:general-geo}
  \begin{split}
    X_{igj} | p_{igj} \mu_{igj} & \sim \text{Poisson}(p_{igj} \mu_{igj}), \\
    M_{ig} | \boldsymbol{\mu}_{ig}, \mathbf{p}_{ig} & \sim \text{Poisson}(\textstyle \sum_{j} \mu_{igj} (1 - p_{igj}) ),
  \end{split}
\end{align}

By extension, the joint hierarchical likelihood is:
\begin{align} \label{mod:hier-general-like}
\begin{split}
\prod_g \big(& L((\boldsymbol{\mu}_{1g},\mathbf{p}_{1g}), \dots, (\boldsymbol{\mu}_{Ig}, \mathbf{p}_{Ig}) | (\mathbf{x}_{1g},m_{1g}), \dots, (\mathbf{x}_{Ig}, m_{Ig})) \\ &f((\boldsymbol{\mu}_{1g},\mathbf{p}_{1g}),  \dots, (\boldsymbol{\mu}_{Ig},\mathbf{p}_{Ig}) | \boldsymbol{\phi})\big)
\end{split}
\end{align}
where $f((\boldsymbol{\mu}_{1g},\mathbf{p}_{1g}),  \dots, (\boldsymbol{\mu}_{Ig},\mathbf{p}_{Ig}) | \boldsymbol{\phi})$ is the density associated with $F_{\boldsymbol{\phi}}$.

In the context of the COVID-19 case data, one might focus their analysis on a single county comprised of many smaller spatial units, with county-level parameters the target of inference, as in  \eqref{mod:hier-general-like} and \eqref{model:geo-hierarchical} as $\boldsymbol{\phi}$.
We will typically have prior information about the hyperparameters from data in other states or in nearby counties, so we opt to use Bayesian inference. If we represent the prior density for $\boldsymbol{\phi}$ as $h(\boldsymbol{\phi} | \boldsymbol{\tau})$ and $\boldsymbol{\tau}$ are known, the joint posterior is:
\begin{align} \label{mod:hier-general-post}
\begin{split}
\pi(& (\boldsymbol{\mu}_{1g},\mathbf{p}_{1g}),  \dots, (\boldsymbol{\mu}_{Ig},\mathbf{p}_{Ig}), \boldsymbol{\phi} \big| (\mathbf{x}_{1g},m_{1g}), \dots, (\mathbf{x}_{Ig}, m_{Ig})) \propto \\ 
\big(\prod_g \big(& L((\boldsymbol{\mu}_{1g},\mathbf{p}_{1g}), \dots, (\boldsymbol{\mu}_{Ig}, \mathbf{p}_{Ig}) | (\mathbf{x}_{1g},m_{1g}), \dots, (\mathbf{x}_{Ig}, m_{Ig})) \\ &f((\boldsymbol{\mu}_{1g},\mathbf{p}_{1g}),  \dots, (\boldsymbol{\mu}_{Ig},\mathbf{p}_{Ig}) | \boldsymbol{\phi})\big)\big)h(\boldsymbol{\phi} | \boldsymbol{\tau})
\end{split}
\end{align}

Given the structure of the model, the marginal posterior for $\boldsymbol{\phi}$ is informed by the data via the terms $L((\boldsymbol{\mu}_{1g},\mathbf{p}_{1g}), \dots, (\boldsymbol{\mu}_{Ig}, \mathbf{p}_{Ig}) | (\mathbf{x}_{1g},m_{1g}), \dots, (\mathbf{x}_{Ig}, m_{Ig}))$, so it is important to understand the characteristics of the likelihood. 

It can be seen that neither \eqref{model:general} nor \eqref{model:general-geo} is identifiable as written without further assumptions.

\section{Derivation of likelihood in Section~\ref{sec:method}}\label{app:like-deriv}

Here we give proof of the following property used in Section~\ref{sec:method}: the data generating model given by the Poisson process and Binomial selection process at the beginning of Section~\ref{sec:method} results in model~\eqref{model:general}.

Consider two groups, $j \in [1,2]$. As above, our fully observed likelihood gives the density for the vector of random variables, $(X_{i1}, X_{i2} , W_{i1}, W_{i2}) \forall i$, while we observe only  $(X_{i1}, X_{i2} , M_{i}) \forall i$. Thus, we must integrate over the set of all $\{(W_{i1}, W_{i2} | W_{i1} + W_{i2} = M_i\}$.
\begin{align}
 &\Prob{X_{i1} = x_{i1}, X_{i2} = x_{i2}, M_{i} = m_{i}}  = \\
 & \Prob{(X_{i1} = x_{i1}, W_{i1} = 0), (X_{i2} = x_{i2}, W_{i2} = 0)}\mathbbm{1}(m_i = 0) \\
 & + \sum_{e = 0}^{m_i} \Prob{(X_{i1} = x_{i1}, W_{i1} = e), (X_{i2} = x_{i2}, W_{i2} = (m_i - e)}\mathbbm{1}(m_i > 0).
\end{align}
Given that $W_{ij} = Y_{ij} - X_{ij}$, this expression is equivalent to
\begin{align}
 & \Prob{(X_{i1} = x_{i1}, Y_{i1} = x_{i1}), (X_{i2} = x_{i2}, Y_{i2} = x_{i2})}\mathbbm{1}(m_i = 0) \\
 & + \sum_{e = 0}^{m_i} \Prob{(X_{i1} = x_{i1}, Y_{i1} = x_{i1} + e), (X_{i2} = x_{i2}, Y_{i2} = x_{i2} + (m_i - e)}\mathbbm{1}(m_i > 0).
\end{align}
Each term
$$
\Prob{(X_{i1} = x_{i1}, Y_{i1} = y_{i1}), (X_{i2} = x_{i2}, Y_{i2} = y_{i2})}
$$
decomposes to
$$
\Prob{X_{i1} = x_{i1} | Y_{i1} = y_{i1}}\Prob{Y_{i1} = y_{i1}}
\Prob{X_{i2} = x_{i2} | Y_{i2} = y_{i2}}\Prob{Y_{i2} = y_{i2}}
$$
given the independence between $Y_{i1}$ and $Y_{i2}$ and the conditional independence of $X_{i1} | Y_{i1}$ and $X_{i2} | Y_{i2}$. 

\begin{align*}
    &\prod_{i=1}^I \lp\frac{\lambda_{i1}^{x_{i1}} e^{-\lambda_{i1}}}{x_{i1}!} p_{i1}^{x_{i1}}
    \frac{\lambda_{i2}^{x_{i2}} e^{-\lambda_{i2}}}{x_{i2}!}p_{i2}^{x_{i2}}\rp^{\mathbbm{1}(m_i = 0)}\\
    &\lp\sum_{e=0}^{m_i}\frac{\lambda_{i1}^{x_{i1} + (m_i - e)} e^{-\lambda_{i1}}}{(x_{i1} + (m_i - e))!} \binom{x_{i1} + m_i - e}{x_{i1}} p_{i1}^{x_{i1}} (1 - p_{i1})^{m_i - e} \frac{\lambda_{i2}^{x_{i2} + e} e^{-\lambda_{i2}}}{(x_{i2} + e)!} \binom{x_{i2} + e}{x_{i2}} p_{i2}^{x_{i2}} (1 - p_{i2})^e \rp^{1 - \mathbbm{1}(m_i = 0)}
\end{align*}
This simplifies to
\begin{align*}
    &\prod_{i=1}^I \lp\frac{\mu_{i1}^{x_{i1}} e^{-\mu_{i1}}}{x_{i1}!} p_{i1}^{x_{i1}}
    \frac{\mu_{i2}^{x_{i2}} e^{-\mu_{i2}}}{x_{i2}!}p_{i2}^{x_{i2}}\rp\\
    &\lp\sum_{e=0}^{m_i}\frac{((1 - p_{i1})\mu_{i1})^{(m_i - e)}}{(m_i - e)!} \frac{((1 - p_{i2})\mu_{i2})^{e}}{e!} \rp
\end{align*}
which, multiplying by $\frac{m_i!}{m_i!}$ and using the binomial theorem, further simplifies to
\begin{align*}
    &\prod_{i=1}^I \lp\frac{\mu_{i1}^{x_{i1}} e^{-\mu_{i1}}}{x_{i1}!} p_{i1}^{x_{i1}}
    \frac{\mu_{i2}^{x_{i2}} e^{-\mu_{i2}}}{x_{i2}!}p_{i2}^{x_{i2}}\rp\\
    &\frac{\lp (1 - p_{i1})\mu_{i1} + (1 - p_{i2})\mu_{i2}\rp^{m_i}}{m_i!}
\end{align*}
Finally we multiply by $e^{-((1 - p_{i1})\mu_{i1} + (1 - p_{i2})\mu_{i2})}e^{(1 - p_{i1})\mu_{i1} + (1 - p_{i2})\mu_{i2}}$ to yield
\begin{align*}
    &\prod_{i=1}^I \lp\frac{(\mu_{i1} p_{i1})^{x_{i1}} e^{-p_{i1}\mu_{i1}}}{x_{i1}!} 
    \frac{(p_{i2} \mu_{i2})^{x_{i2}} e^{-p_{i2}\mu_{i2}}}{x_{i2}!}\rp \frac{e^{-((1 - p_{i1})\mu_{i1} + (1 - p_{i2})\mu_{i2})}\lp (1 - p_{i1})\mu_{i1} + (1 - p_{i2})\mu_{i2}\rp^{m_i}}{m_i!}
\end{align*}
which we recognise as the product of filtered Poisson random variables, and the marginally Poisson distributed cases missing stratum
information.

The proof of the generalization to $J$ groups, which can be show with induction, has been omitted. 

\section{Graphical model depictions}

\subsection{Graphical model of model with covariates} \label{app:model-diagram}
\begin{figure}[H]
    \centering
\begin{minipage}{0.7\textwidth}
\begin{minipage}{0.35\textwidth}
\centering
 \tikz{
     \node[obs] (x) {$x_{ij}$};%
     \node[latent,above=of x] (y) {$y_{ij}$}; %
     \node[obs,left=of y] (e) {$E_{ij}$}; %
     \node[obs,below=of x,xshift=-1.7cm] (z) {$z_{i}$}; %
     \node[latent,above=of y,xshift=-2cm] (lambda) {$\lambda_j$}; %
     \node[latent,above=of x,xshift=2cm] (eta) {$\eta_j$}; %
     \node[latent,above=of eta,xshift=1cm] (gamma) {$\boldsymbol{\gamma}$}; %
     \node[latent,above=of gamma] (beta) {$\boldsymbol{\beta}$}; %
     \plate  {plate1} {(z)(x)(y)(e)} {$I$}; %
     \plate [yshift=-0.1cm] {plate2} {(x)(y)(e)(lambda)(eta)} {$J$}; %
     \edge {y,z}{x}
     \edge {eta,gamma}{x}
     \edge {e,z,lambda,beta}{y}
     } 
\end{minipage}
\vspace{.5in}
\hspace{-1.0in}
\begin{minipage}{0.49\textwidth}
{\footnotesize
\begin{align*}
  i & \in \{1, \dots, I\} \text{ : Stratum} \\
 j  & \in \{1, \dots, J\} \text{ : Category}  \\
  \tikzcircle[fill=gray!40]{5pt} & \text{ : Observed} \\
  \tikzcircle[fill=white]{5pt} & \text{ : Latent} 
\end{align*}
}
\end{minipage}
\begin{minipage}{\textwidth}
\centering
\begin{tabular}{c|c|c}
Variable & Domain & Description\\
\hline 
  $y_{ij}$  & $\mathbb{N}_0$  & Total cases  \\
  $x_{ij}$  & $\mathbb{N}_0$ & Observed cases  \\
  $E_{ij}$  & $\mathbb{N}_0$ & Observed population \\
  $\mathbf{z}_{i}$ & $\mathbb{R}^K$  & Observed covariates  \\
  $\lambda_j$  & $\mathbb{R}^+$ & Per-capita rate of disease  \\
  $\eta_j$  & $\mathbb{R}$  & Log-odds of observing category info.  \\
  $\boldsymbol{\beta}$& $\mathbb{R}^K$ & Log-relative rates of disease \\
  $\boldsymbol{\gamma}$& $\mathbb{R}^K$ & Log-odds of observing category info.
\end{tabular}
\captionof{table}{Table of generative model variables for \Cref{model:full}}\label{tab:model-params}
\end{minipage}
\end{minipage}
\end{figure}
The parameters of interest in \Cref{tab:model-params} are $\lambda_j$, which give the category-specific, per-capita rates of disease, and transformations of the parameters like those enumerated in \Cref{subsec:estimands-simu-study}.

\subsection{Graphical model of hierarchical model with covariates} \label{app:model-diagram-hier}
\begin{figure}[H]
    \centering
    \begin{minipage}{\textwidth}
\begin{minipage}{0.59\textwidth}
\centering
 \tikz{
     \node[obs] (x) {$x_{igj}$};%
     \node[latent,above=of x] (y) {$y_{igj}$}; %
     \node[obs,below=of x,xshift=-1.7cm] (z) {$\mathbf{z}_{ig}$}; %
     \node[obs,left=of y] (e) {$E_{igj}$}; %
     \node[latent,above=of y,xshift=-2cm] (lambda) {$\lambda_{gj}$}; %
     \node[latent,left=of lambda,yshift=1.0cm] (alpha_lambda) {$\boldsymbol{\alpha}_{\boldsymbol{\lambda}}$}; %
     \node[latent,left=of lambda] (pi_lambda) {$\boldsymbol{\Pi}_{\boldsymbol{\lambda}}$}; %
     \node[latent,left=of lambda,yshift=-1.0cm] (sigma_lambda) {$\boldsymbol{\Sigma}_{\boldsymbol{\lambda}}$}; %
     \node[latent,above=of x,xshift=2cm,yshift=-0.5cm] (eta) {$\eta_{gj}$}; %
     \node[latent,right=of eta,xshift=0.5cm,yshift=0.5cm] (alpha_eta) {$\boldsymbol{\alpha}_{\boldsymbol{\eta}}$}; %
     \node[latent,right=of eta,xshift=0.5cm,yshift=-0.5cm] (pi_eta) {$\boldsymbol{\Pi}_{\boldsymbol{\eta}}$}; %
     \node[latent,right=of eta,xshift=0.5cm,yshift=-1.5cm] (sigma_eta) {$\boldsymbol{\Sigma}_{\boldsymbol{\eta}}$}; %
     \node[latent,above=of eta,xshift=1cm] (gamma) {$\boldsymbol{\gamma}_g$}; %
     \node[latent,above=of gamma] (beta) {$\boldsymbol{\beta}_g$}; %
     \node[latent,above=of beta,xshift=-0.5cm] (alpha_beta) {$\boldsymbol{\alpha}_{\boldsymbol{\beta}}$}; %
     \node[latent,above=of beta,xshift=0.5cm] (pi_beta) {$\boldsymbol{\Pi}_{\boldsymbol{\beta}}$}; %
     \node[latent,above=of beta,xshift=1.5cm] (sigma_beta) {$\boldsymbol{\Sigma}_{\boldsymbol{\beta}}$}; %
     \node[obs,left=of pi_beta,xshift=-1cm] (w) {$\mathbf{w}_g$}; %
     \node[latent,right=of gamma,yshift=2.0cm] (alpha_gamma) {$\boldsymbol{\alpha}_{\boldsymbol{\gamma}}$}; %
     \node[latent,right=of gamma,yshift=1.0cm] (pi_gamma) {$\boldsymbol{\Pi}_{\boldsymbol{\gamma}}$}; %
     \node[latent,right=of gamma] (sigma_gamma) {$\boldsymbol{\Sigma}_{\boldsymbol{\gamma}}$}; %
     \plate  {plate1} {(z)(x)(y)(e)} {$I$}; %
     \plate [yshift=-0.1cm] {plate2} {(x)(y)(e)(lambda)(eta)} {$J$}; %
     \plate [yshift=-0.1cm] {plate3} {(z)(x)(y)(e)(lambda)(eta)(beta)(gamma)} {$G$}; %
     \edge {y,z,eta,gamma}{x}
     \edge {z,e,lambda,beta}{y}
     \edge {w}{gamma,beta,lambda,eta}
     \edge {alpha_lambda,sigma_lambda,pi_lambda}{lambda}
     \edge {alpha_eta,pi_eta,sigma_eta}{eta}
     \edge {alpha_beta,pi_beta,sigma_beta}{beta}
     \edge {alpha_gamma,pi_gamma,sigma_gamma}{gamma}
     } 
\end{minipage}
\vspace{0.5in}
\begin{minipage}{0.7\textwidth}
{\footnotesize
\begin{align*}
  i & \in \{1, \dots, I\} \text{ : Stratum} \\
 g & \in \{1, \dots, G\} \text{ : Geographic area}\\
 j &  \in \{1, \dots, J\} \text{ : Category} \\
 \text{For } \boldsymbol{\nu} & \in  \{\boldsymbol{\beta}, \boldsymbol{\gamma}, \boldsymbol{\eta}, \boldsymbol{\lambda} \} \text{ : }\\
  \boldsymbol{\alpha}_{\boldsymbol{\nu}} & \text{ : Inter-geography mean for $\nu$} \\
  \boldsymbol{\Pi}_{\boldsymbol{\nu}} & \text{ : Coefs. on $\mathbf{w}_g$ for $\nu$}\\
  \boldsymbol{\Sigma}_{\boldsymbol{\nu}} &  \text{ : Inter-geography cov. matrix for $\nu$} \\
  \tikzcircle[fill=gray!40]{5pt} & \text{ : Observed} \\
  \tikzcircle[fill=white]{5pt} & \text{ : Latent} 
\end{align*}
}
\end{minipage}
\begin{minipage}{\textwidth}
\centering
\begin{tabular}{c|c|l}
Variable & Domain & Description\\
\hline 
  $y_{igj}$  & $\mathbb{N}_0$  & Total cases  \\
  $x_{igj}$  & $\mathbb{N}_0$ & Observed cases  \\
  $E_{igj}$  & $\mathbb{N}_0$ & Observed population  \\
  $\mathbf{z}_{ig}$ & $\mathbb{R}^K$  & Observed covariates \\
  $\mathbf{w}_{g}$ & $\mathbb{R}^D$  & Observed geographic-specific covariates \\
  $\lambda_{gj}$  & $\mathbb{R}^+$ & Per-capita rate of disease  \\
  $\eta_{gj}$  & $\mathbb{R}$  & Log-odds of observing category \\
  $\boldsymbol{\beta}_g$ & $\mathbb{R}^K$ & Log-relative rates of disease\\
  $\boldsymbol{\gamma}_g$& $\mathbb{R}^K$ & Log-odds of observing category\\
\end{tabular}
\captionof{table}{Table of generative model variables for \ref{model:full-geo}}\label{tab:model-params-geo}
\end{minipage}
\end{minipage}
\end{figure}
The parameters of interest in \Cref{tab:model-params-geo} are $\lambda_{gj}$, which give the category-specific geographical-area-specific, per-capita rates of disease, and transformations of the parameters like those enumerated in \Cref{subsec:estimands-simu-study}.
Interest may also lie in the across-geography mean category-specific log per-capita rates of disease for category $j$, $\boldsymbol{\alpha}_{\boldsymbol{\lambda}}$ the coefficients on $\mathbf{w}_g$ or $\boldsymbol{\Pi}_{\boldsymbol{\lambda}}$, and the inter-geography covariance between log-per-capita, category-specific rates of disease, $\boldsymbol{\Sigma}_{\boldsymbol{\lambda}}$

\clearpage

\section{Lemmas and Proofs for Model Identifiability Properties}

\subsection{Fisher information positive definiteness of simple model}\label{app:id-simple}

The following is a derivation of the Fisher information matrix $\mathcal{I}$ for model \ref{model:no-cov-obs}.
The likelihood for the model is
\begin{align*}
   \ell(\lambda_1, p_1, \dots, \lambda_J, p_J) & = \sum_{i=1}^I \sum_{j=1}^{J} -E_{ij}\lambda_j + x_{ij}\log \lp E_{ij} \lambda_j p_j \rp - \log x_{ij}! \\
    & + \sum_{i=1}^I m_i \log\lp\sum_{j=1}^{J}E_{ij}\lambda_j(1-p_j)\rp - \log m_i! 
\end{align*}
We reparameterize as we do in \ref{thm:glob-id-no-cov}:
\begin{align}
    (\lambda_j, p_j) \to (v_j, u_j) \forall j \in [1,\dots,J]
\end{align}
where $v_j = \lambda_j p_j$  $u_j = \lambda_j (1 - p_j)$. Then the likelihood is
\begin{align*}
   \ell(v_1, u_1, \dots, v_j,  u_J) & = \sum_{i=1}^I \sum_{j=1}^{J} -E_{ij}(u_j + v_j) + x_{ij}\log \lp E_{ij} v_j \rp - \log x_{ij}! \\
    & + \sum_{i=1}^I m_i \log\lp\sum_{j=1}^{J}E_{ij}u_j\rp - \log m_i! 
\end{align*}

Let $\delta_{i,j}$ be the Kronecker delta function.
The Fisher information matrix for the reparameterized log-likelihood is:
\begin{align}
   -\Exp{\frac{\partial \ell}{\partial v_j v_k}} & = \frac{\sum_{i=1}^IE_{ij}}{v_j}\delta_{j,k} \\
   -\Exp{\frac{\partial \ell}{\partial u_j u_k}} & =  
   \sum_{i=1}^I\frac{E_{ij} E_{ik}}{\sum_{m=1}^J E_{im} u_m} \\
   -\Exp{\frac{\partial \ell}{\partial u_j v_k}} & =  0
\end{align}
We can arrange the Fisher information in block matrix form for all $I$ observations:
\begin{align}
\mathcal{I} = \begin{bmatrix}
\mathbf{V} & 0 \\
0 & \mathbf{U} 
\end{bmatrix}
\end{align}
with 
\begin{align}
\mathbf{U}_{jk} & =  \sum_{i=1}^I\frac{E_{ij} E_{ik}}{\sum_{m=1}^J E_{im} u_m}\\
\mathbf{V}_{jk} & =    \frac{\sum_{i=1}^IE_{ij}}{v_j}\delta_{j,k} 
\end{align}
$\mathcal{I}$ is positive definite if $\mathbf{U} \succ 0$ and $\mathbf{V} \succ 0$.
$\mathbf{V}$ is a diagonal matrix and is positive definite as long as all elements along the diagonal are strictly positive. 
The parameter constraints on $u_j$ and $v_j$ yield $u_j, v_j > 0$ for each $j$ so as long as $\sum_{i=1}^I E_{ij} > 0$ for all $j$ $\mathbf{V} \succ 0$ . 

The matrix $\mathbf{U}$ can be represented as the matrix product of three matrices.
Let $\mathbf{E}$ be the $I \times J$ matrix with its $(i, j)^\textrm{th}$ entry as $E_{ij}$, and let $\mathbf{S}$ be a diagonal matrix defined as
\begin{align}
\mathbf{S}_{jk} & =  \frac{1}{\sqrt{\sum_{m=1}^J E_{im} u_m}} \delta_{j,k}
\end{align}
Then $\mathbf{U} = \mathbf{E}^T \mathbf{S}^2 \mathbf{E}$.
In order for $\mathbf{U} \succ 0$  $\mathbf{S} \mathbf{E}$ must be rank $J$.
This will be so if $\mathbf{S}$ is invertible and if $\mathbf{E}$ is rank $J$.
If $\det{\mathbf{S}} \neq 0$ $\mathbf{S}$ is invertible: 
\begin{align}
    \det(\mathbf{S}) = (\prod_{i=1}^I\sum_{m=1}^J E_{im} u_m)^{-1/2}
\end{align}
$\mathbf{S}$ has a nonzero determinant if at least one $E_{im}u_ > 0$ for all $i \in [1, \dots, I]$. 
Given the constraints on the parameter space, $p_j \in (0,1)$ and $\lambda_j > 0$
for all $j$ ensures that $u_j > 0$ for all $j$. 
The minimal conditions for the positive definiteness of $\mathcal{I}$ are as follows:
\begin{itemize}
    \item $\mathbf{E}$ is rank $J$
    \item $p_j \in (0,1)$
    \item $\lambda_j \in (0,\infty)$
    \item $\sum_{m=1}^J E_{im} > 0 \forall i$
\end{itemize}

Given estimators $\hat{v}_j = \sum_{i=1}^I X_{ij} / \sum_{i=1}^I E_{ij}$ for $v_j$ and $\hat{\mathbf{u}} = (\mathbf{E}^T \mathbf{E})^{-1} \mathbf{E}^T\mathbf{m}$,
the observed Fisher information matrix $\hat{\mathcal{I}}$ is:
\begin{align}
\hat{\mathcal{I}} = \begin{bmatrix}
\hat{\mathbf{V}} & 0  \\
0 & \hat{\mathbf{U}}
\end{bmatrix}
\end{align}
where
$$
\hat{\mathbf{V}}_{mn} = \delta_{m,n} \frac{(\sum_{i=1}^I E_{ij})^2}{\sum_{i=1}^I X_{ij}}
$$
and 
$$
\hat{\mathbf{U}} = \mathbf{E}^T \text{diag}(\mathbf{E}(\mathbf{E}^T \mathbf{E})^{-1} \mathbf{E}^T\mathbf{m})^{-1}\mathbf{E} 
$$
The inverse of $\hat{\mathcal{I}}$ is just the block diagonal matrix with $\hat{\mathbf{V}}^{-1}$ and $\hat{\mathbf{U}}^{-1}$ along the diagonal:

\subsection{Derivation of posterior mean of $\lambda_1$ for minority group}\label{app:post-mean-u}

Let $J = 2$ and let the following priors be implemented for $p_j, \lambda_j$
\begin{align*}
    p_j & \overset{\text{iid}}{\sim} \text{Beta}(\alpha_j, \beta_j) \\
    \lambda_j &\overset{\text{iid}}{\sim} \text{Gamma}(\alpha_j + \beta_j, r_j) 
\end{align*}
then $v_j = p_j \lambda_j \indy u_j = (1 - p_j) \lambda_j$ with $v_j \sim \text{Gamma}(\alpha_j, r_j)$ and $u_j \sim \text{Gamma}(\beta_j, r_j)$
We can write the observed-data likelihood in terms of $u_j$ and $v_j$ as follows:
\begin{align*}
    \exp \lp-\textstyle\sum_j u_j\textstyle\sum_i E_{ij} \rp \prod_{i=1}^I \frac{\lp E_{i1} u_1 + E_{i2} u_2 \rp^{m_i}}{m_i!}
    \prod_{j=1}^J \frac{\exp \lp-v_j\textstyle\sum_i E_{ij} \rp \lp\prod_{i=1}^I (E_{ij})^{x_{ij}}\rp v_j^{ \sum_i x_{ij}}}{\prod_{i=1}^I x_{ij}!}
\end{align*}
If we assume that, WLOG, group $2$ is the majority group for all $i$, or, in other words, $E_{i1} \ll E_{i2}$ for all $i$, then using the approximation 
\begin{align*}
\lp E_{i1} u_1 + E_{i2} u_2 \rp^{m_i} & = (E_{i2} u_2)^{m_i} \exp m_i \log \lp 1 + \frac{E_{i1}}{E_{i2} u_2} u_1\rp \\
& \approxeq (E_{i2} u_2)^{m_i} \exp m_i \lp u_1 \frac{E_{i1}}{E_{i2} u_2} - u_1^2\frac{E_{i1}^2}{2 E_{i2}^2 u_2^2} \rp
\end{align*}
Leading to the approximate likelihood:
\begin{align*}
    & \exp \lp-\textstyle\sum_j u_j\textstyle\sum_i E_{ij} \rp \lp\prod_{i=1}^I E_{i2} \rp u_2^{\sum_i m_i} \frac{\prod_{i=1}^I \exp\lp u_1 \frac{E_{i1}}{E_{i2} u_2} - u_1^2\frac{E_{i1}^2}{2 E_{i2}^2 u_2^2}\rp^{m_i}}{m_i!}\\
    \times
    & \prod_{j=1}^2 \frac{\exp \lp-v_j\textstyle\sum_i E_{ij} \rp \lp\prod_{i=1}^I (E_{ij})^{x_{ij}}\rp v_j^{ \sum_i x_{ij}}}{\prod_{i=1}^I x_{ij}!}
\end{align*}
When we multiply by the priors and collect terms, we get an expression separable in $u_j$ and $v_j$:
\begin{align*}
    & \lp\prod_{i=1}^I E_{i2} \rp \exp \lp-u_2\textstyle\sum_i E_{i2} \rp u_2^{\beta_2 + \sum_i m_i - 1} / m_i! \\ 
    & u_1^{\beta_1 - 1} \exp\lp -u_1^2 u_2^{-2}\textstyle\sum_i \frac{m_i E_{i1}^2}{2 E_{i2}^2} + u_1 \lp u_2^{-1}\textstyle\sum_i \frac{m_i E_{i1}}{E_{i2}} - r_1 - \textstyle \sum_i E_{i1}\rp\rp\\
    \times
    & \prod_{j=1}^2 \frac{\exp\lp-v_j\lp r_j + \textstyle\sum_i E_{ij} \rp\rp \lp\prod_{i=1}^I (E_{ij})^{x_{ij}}\rp v_j^{\alpha_j + \sum_i x_{ij}} - 1}{\prod_{i=1}^I x_{ij}!} 
\end{align*}
The approximate posterior distribution, $\pi(u_1 | u_2, m)$, is a modified half-normal distribution, as introduced in \cite{sunModifiedHalfNormalDistributionProperties2021}. 
While the functional form of the posterior mean is complicated for general values of $\beta_1$, we compute the posterior mean, variance, and local sensitivity of the posterior mean to $r_1$ when $\beta_1 = 1$ and $\beta_1 = 2$, which correspond to implied priors for $u_1$ of an exponential distribution with rate $r_1$, and a gamma distribution with shape $2$ and rate $r$, respectively.
Let $s_1 = \textstyle\sum_i \frac{m_i E_{i1}}{E_{i2}}, s_2 = \textstyle\sum_i \frac{m_i E_{i1}^2}{E_{i2}^2}$, and $E_{+1} = \textstyle \sum_i E_{i1}$. Further, let $\phi$ and $\Phi$ be the density and distribution function of the standard normal distribution, respectively, and $z(u_2, r_1) = \frac{s_1 - u_2(r_1  + E_{+1})}{\sqrt{s_2}}$.
If $\beta_1 = 1$, the posterior mean of $u_1 | u_2, m$ is
\begin{align*}
 \Exp{u_1 | u_2, r_1, \beta_1 = 1} = \frac{u_2}{\sqrt{s_2}}\lp z(u_2, r_1) + \phi(z(u_2, r_1))\Phi(z(u_2, r_1))^{-1}\rp
\end{align*}
while the posterior variance is
\begin{align*}
  \Var{u_1 | u_2, r_1, \beta_1 = 1} =  \frac{u_2^2}{s_2}\lp 1 - z(u_2, r_1)\phi(z(u_2, r_1))\Phi(z(u_2, r_1))^{-1} - \phi(z(u_2, r_1))^2\Phi(z(u_2, r_1))^{-2})\rp
\end{align*}
The posterior mean for $\lambda_1$ is then
\begin{align*}
\frac{\alpha_1 + \sum_i x_{i1}}{r_1 + \sum_i E_{ij}} + 
   \frac{u_2}{\sqrt{s_2}}(z(u_2, r_1) + \phi(z(u_2, r_1))\Phi(z(u_2, r_1))^{-1})
\end{align*}
with variance:
\begin{align*}
\frac{\alpha_1 + \sum_i x_{i1}}{(r_1 + \sum_i E_{i1})^2} + 
   \frac{u_2^2}{s_2}\lp 1 - z(u_2, r_1)\phi(z(u_2, r_1))\Phi(z(u_2, r_1))^{-1} - \phi(z(u_2, r_1))^2\Phi(z(u_2, r_1))^{-2})\rp
\end{align*}
The partial derivative of $\Exp{u_1 | u_2, r_1}$ with respect to $r_1$ is
\begin{align*}
    \frac{-u_2^2}{s_2}\bigg(1 - & \frac{1}{\sqrt{s_2}} \phi(z(u_2, r_1))\Phi(z(u_2, r_1))^{-2}\\
    & \times\bigg(\sqrt{s_2} \phi(z(u_2, r_1)) + (s_1 - u_2 (E_{+j} + r_1))\Phi(z(u_2, r_1)) \bigg)\bigg)
\end{align*}
which simplifies to
\begin{align*}
    \frac{\partial \Exp{u_1 | u_2, r_1}}{\partial r_1} &= \frac{-u_2^2}{s_2} \lp 1 - z(u_2, r_1)\phi(z(u_2, r_1))\Phi(z(u_2, r_1))^{-1} - \phi(z(u_2, r_1))^2\Phi(z(u_2, r_1))^{-2}\rp \\
    & = -\Var{u_1 | u_2, r_1}
\end{align*}

If $\beta_1 = 2$, the posterior mean of $u_1 | u_2, m$ is
\begin{align*}
   \frac{u_2}{\sqrt{s_2}} \lp z(u_2, r_1) + \frac{\Phi(z(u_2, r_1))}{\phi(z(u_2, r_1)) + z(u_2, r_1)\Phi(z(u_2, r_1))} \rp
\end{align*}
while the posterior variance is
\begin{align*}
   \frac{u_2^2}{s_2}\lp 2 - \lp z(u_2, r_1) + \frac{\Phi(z(u_2, r_1))}{\phi(z(u_2, r_1)) + z(u_2, r_1)\Phi(z(u_2, r_1))} \rp\rp
\end{align*}
The posterior mean for $\lambda_1$ is then
\begin{align*}
\frac{\alpha_1 + \sum_i x_{i1}}{r_1 + \sum_i E_{ij}} + 
\frac{u_2}{\sqrt{s_2}} \lp z(u_2, r_1) + \frac{\Phi(z(u_2, r_1))}{\phi(z(u_2, r_1)) + z(u_2, r_1)\Phi(z(u_2, r_1))} \rp
\end{align*}
with variance:
\begin{align*}
\frac{\alpha_1 + \sum_i x_{i1}}{(r_1 + \sum_i E_{i1})^2} + 
   \frac{u_2^2}{s_2}\lp 2 - \lp z(u_2, r_1) + \frac{\Phi(z(u_2, r_1))}{\phi(z(u_2, r_1)) + z(u_2, r_1)\Phi(z(u_2, r_1))} \rp\rp
\end{align*}
Taking the difference between $\Exp{u_1 | u_2, r_1, \beta_1 = 2}$ and $\Exp{u_1 | u_2, r_1, \beta_1 = 1}$ yields
\begin{align*}
   \frac{u_2}{\sqrt{s_2}} \Bigg( z(u_2, r_1) + \frac{\Phi(z(u_2, r_1))}{\phi(z(u_2, r_1)) + z(u_2, r_1)\Phi(z(u_2, r_1))}& \\
   - \lp z(u_2, r_1) + \phi(z(u_2, r_1))\Phi(z(u_2, r_1))^{-1}\rp \Bigg)&
\end{align*}
Algebra reveals that $\Exp{u_1 | u_2, r_1, \beta_1 = 2} - \Exp{u_1 | u_2, r_1, \beta_1 = 1}$ is
\begin{align*}
   \frac{u_2}{\sqrt{s_2}} \frac{1 - z(u_2, r_1)\phi(z(u_2, r_1))\Phi(z(u_2, r_1))^{-1} - \phi(z(u_2, r_1))^2\Phi(z(u_2, r_1))^{-2})}{z(u_2, r_1) + \phi(z(u_2, r_1))\Phi(z(u_2, r_1))^{-1}}
\end{align*}
or
\begin{align*}
   \frac{\Var{u_1 | u_2, r_1, \beta_1 = 1}}{\Exp{u_1 | u_2, r_1, \beta_1 = 1}}
\end{align*}
Dividing this by the standard deviation:
\begin{align*}
   & \frac{\sqrt{\Var{u_1 | u_2, r_1, \beta_1 = 1}}}{\Exp{u_1 | u_2, r_1, \beta_1 = 1}} \\
   & = \frac{\sqrt{1 - z(u_2, r_1)\phi(z(u_2, r_1))\Phi(z(u_2, r_1))^{-1} - \phi(z(u_2, r_1))^2\Phi(z(u_2, r_1))^{-2})}}{z(u_2, r_1) + \phi(z(u_2, r_1))\Phi(z(u_2, r_1))^{-1}}
\end{align*}
If we further assume that $E_{i2}, E_{i1} \to \infty$ such that $E_{i1} / E_{i2} \to 0$ and $E_{i1}^2 / E_{i2} \to K < \infty$ for all $i$, the posterior for $u_2$ converges to a point mass at $u_2^\star$, the true data generating parameter. 

This can be seen from the fact that the MLE for $u_2$ converges to $u_2^\star$.
The gradient of the log-likelihood $\ell$ with respect to $u_2$ and $u_1$ is:
\begin{align*}
  \frac{\partial \ell}{\partial u_1}  & = -E_{+1} - \frac{s_2 u_1}{u_2^2} + \frac{s_1}{u_2} \\ 
  \frac{\partial \ell}{\partial u_2}  & = -E_{+2} + \frac{s_2 u_1^2}{u_2^3} - \frac{s_1 u_1}{u_2^2} + \frac{m_{+}}{u_2} 
\end{align*}
Setting these equal to zero yields the following two solutions:
\begin{align*}
  \hat{u}_1(u_2)  & = u_2 \frac{s_1-u_2 E_{+1}}{s_2} \\ 
  \hat{u}_2 & = \frac{E_{+1} s_1 + E_{+2} s_2 \pm \sqrt{-4 E_{+1}^2 m_+ s_2 + (-E_{+1} s_1 - E_{+2} s_2)^2}}{2 E_{+1}^2}
\end{align*}
Recall that $M_i \sim \mathrm{Poisson}(u_1^\star E_{i1} + u_2^\star E_{i2})$. For $E_{i1},E_{i2} \to \infty$ with $u_1^\star$ and $u_2^\star$ bounded away from zero and $< \infty$, $\frac{M_i - (u_1^\star E_{i1} + u_2^\star E_{i2})}{\sqrt{u_1^\star E_{i1} + u_2^\star E_{i2}}} \overset{d}{\to} \mathrm{N}(0, 1)$ by the CLT. 
Let $\mathcal{Z} \sim \mathrm{N}(0,1)$, then $s_1 - u_2^\star E_{+1} \overset{d}{\to} u_1^\star I K + \sqrt{u_2^\star I K}\mathcal{Z}, s_2 \overset{p}{\to} u_2^\star I K$. 
$$
\frac{\sqrt{-4 E_{+1}^2 m_+ s_2 + (-E_{+1} s_1 - E_{+2} s_2)^2}}{2E_{+1}^2}
=\sqrt{\frac{(E_{+1} I K u_1 - E_{+1}^2 u_2 + E_{+2} I K u_2)^2}{4 E_{+1}^4} +  O_p(\frac{1}{\sqrt{E_{+2}}})}
$$
so 
$$
\frac{E_{+1} s_1 + E_{+2} s_2 \pm \sqrt{-4 E_{+1}^2 m_+ s_2 + (-E_{+1} s_1 - E_{+2} s_2)^2}}{2 E_{+1}^2}\overset{p}{\to} u_2^\star
$$
Finally, by Slutsky's theorem 
$$
\hat{u}_1 \overset{d}{\to} u_1^\star + \lp\sqrt{\frac{u_2^\star}{IK}} \rp \mathcal{Z}
$$
We can calculate the asymptotic MSE of $\hat{u}_1$, assuming that when $\hat{u}_1 \leq 0$ we set $\hat{u}_1$ to be $0$. 

Let $\alpha = -u_1^\star\sqrt{IK/u_2^\star}$. Then bias of $\hat{u}_1$ is
$$
\sqrt{u_2^\star/IK} \phi(\alpha) -u_1^\star \Phi(\alpha) 
$$
and the variance is
$$
\sqrt{u_2^\star/IK}(1 +\alpha \phi(\alpha) - \phi(\alpha)^2 - \Phi(\alpha)) + 2 \phi(\alpha) \Phi(\alpha) \sqrt{u_2/IK} u_1 +\Phi(-\alpha)\Phi(\alpha) u_1^2
$$
The asymptotics above also imply:
$z(u_2, r_1) \overset{d}{\to} \frac{u_1^\star I K - u_2^\star r_1}{\sqrt{u_2^\star I K}} + \mathcal{Z} = z^\star$, and the posterior mean
for $u_1$ given $\beta=1$ is 
\begin{align*}
 \sqrt{\frac{u_2}{IK}}\lp z^\star + \phi(z^\star)\Phi(z^\star)^{-1}\rp =
u_1^\star - \frac{r_1 u_2^\star}{I K}
+ \frac{\sqrt{u_2^\star} \phi(z^\star)\Phi(z^\star)^{-1}}{\sqrt{I K}} + \sqrt{\frac{u_2^\star}{IK}}\,\mathcal{Z}
\end{align*}
The expression $z + \phi(z)/\Phi(z) \geq 0 \forall z$, so the Bayesian posterior mean for $u_1$ will be positive a.s. whereas the MLE may be 0 with positive probability depending on $I$, $K$ and $u_1^\star,u_2^\star$.
The asymptotic posterior mean for $\beta = 2$ is 
\begin{align*}
 \sqrt{\frac{u_2}{IK}}\lp z^\star + \frac{1}{z^\star + \phi(z^\star)\Phi(z^\star)^{-1}}\rp 
\end{align*}

We can compare the asymptotic root mean-squared error for the MLE and the posterior mean under an exponential prior for $u_1$, or $\mathrm{Exp}(r_1)$ and under a $\mathrm{Gamma}(2, r_1)$ prior for $u_1$ for a range of values of $u_1^\star$ for $I=15$ and $K=1$. Note that $u_1 \sim \mathrm{Exp}(r_1)$ corresponds to $p_1 \sim \mathrm{Beta}(\alpha_1, 1)$ and $u_1 \sim \mathrm{Gamma}(2, r_1)$ corresponds to $p_1 \sim \mathrm{Beta}(\alpha_1, 2)$.
We use the square root of the exact asymptotic MSE for the MLE, while we use a Monte Carlo approximation to the RMSE for the two Bayesian estimators.
We assume that $u_i^\star = (1 - p_i^\star) \lambda_i^\star$ and fix $p_1^\star = 0.6$ and $p_2^\star = 0.9$, which represents a high race/ethnicity reporting rate for the majority group, and low race/ethnicity reporting rate for the minority group.
We assume $\lambda_2^\star \in \{0.001,0.009,0.02\}$ while we examine $\lambda_1^\star$ from $0.001$ to $0.05$.
We fix the posterior mean for $u_1$ at $0.01$, which implies $r_1 = 100$ for the exponential prior and $r_1 = 200$ for the gamma prior.
\begin{figure}[H]
    \centering
    \includegraphics[width=\textwidth]{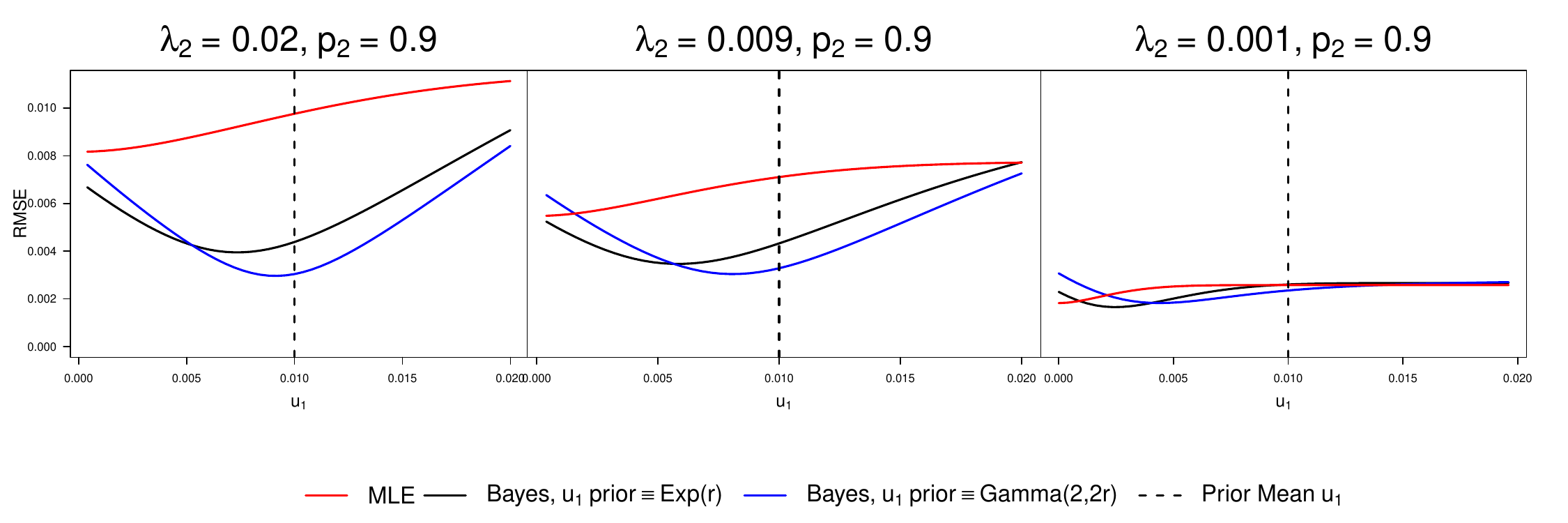}\hfill
    \caption{Asymptotic root mean-squared error (RMSE) of posterior mean for two Bayes estimators vs. MLE. Monte Carlo approximation to RMSE for posterior means, with standard error on the order of $10^{-6}$ for all $u_1$.
    Note an exponential prior puts prior mass near zero while the $\mathrm{gamma}(2,r_1)$ prior puts vanishing prior mass as $u_1 \to 0$.
    The $y$-axis represents the RMSE of the a given point estimator for certain data-generating values of $u_1$ and $u_2$. The panels of the graphs represent different true values of $u_2$, corresponding to $u_2 = \lambda_2 p_2$, while the $x$-axes represent a continuum of true values for $u_1$.
    The dashed vertical line represents the prior mean for $u_1$. Thus each panel of the graph shows how RMSE of each point estimator varies as $u_1$ increases from $4\times10^{-6}$ to $1.9\times10^{-2}$ given a certain value of $u_2$.
    The RMSE of the MLE, shown as the solid red line, slowly increases as $u_1$ increases as the variance of the MLE increases faster than the squared bias decreases. The Bayes estimators show decreasing RMSE as the prior mean for $u_1$ approaches the true $u_1$. Two conclusions can be drawn from the graphs: Both Bayes solutions dominate the MLE for reasonable values of $u_1$ and $u_2$. The exception is for small $u_2$ and when the prior for $u_1$ is several orders of magnitude too large. The second conclusion is that the Bayes estimator with gamma prior dominates the exponential-prior estimator when the prior mean for $u_1$ is moderately larger than the true $u_1$ and when the prior mean underestimates the true $u_1$.}
    \label{fig:asympt-rmse}
\end{figure}
\Cref{fig:asympt-rmse} shows that Bayes estimators yield gains over the MLE for minority groups. Within the class of Bayes estimators, estimators derived from models with priors that put too much support near zero (e.g. an exponential prior) can shrink too the posterior mean too strongly towards zero even when the prior mean over-estimates the true parameter.
Given that the near-zero behavior is driven by the prior over $p_1$, limiting prior mass near $1$ for $p_1$ can yield point estimators with lower MSEs for a broad range of values for $u_1$.

\subsection{DCT lemma}

\begin{lemma}\label{lemma:dct-pois}
  Let $p_{\eta(\theta)}(X = k)$ be defined
  $$
  \frac{1}{k!} \exp\lp\eta(\theta) k - e^{\eta(\theta)}\rp,
  $$
  where $\eta(\theta)$ is a univariate differentiable function of $\theta$, $\theta \in \R^d$.
  Let $g(\eta(\theta)) = \int f(x) p_{\eta(\theta)}(x) d\mu(x)$ where $\mu$ is
  the counting measure on $[0, 1, 2, \dots]$. If we define the set $\theta \in \Theta_f$ 
  as the set for which $\int \abs{f(x)} p_{\eta(\theta)}(x) d\mu(x) < \infty$,
  then
  $$
  \frac{\partial}{\partial \theta_j} g(\eta(\theta)) = \int f(x) \frac{\partial}{\partial \theta_j} p_{\eta(\theta)}(x) d\mu(x)
  $$
\end{lemma}
\begin{proof}
  By the chain rule, the $\frac{\partial}{\partial \theta_j} g(\eta(\theta)) = \frac{d g(\eta(\theta))}{d \eta(\theta)}\frac{\partial \eta(\theta)}{\partial \theta_j}$.
  By Theorem 2.4 in \cite{keenerTheoreticalStatisticsTopics2010}, $\frac{d g(\eta(\theta))}{d \eta(\theta)}$ exists and can be
  obtained via differentiating under the integral sign:
  $$
  \frac{d}{d \eta(\theta)} g(\eta(\theta)) = \int f(x) \frac{dp_{\eta(\theta)}(x)}{d \eta(\theta)} d\mu(x).
  $$
  Using this result and the chain rule again yields
  \begin{align}
    \frac{\partial}{\partial \theta_j} g(\eta(\theta)) & = \frac{d g(\eta(\theta))}{d \eta(\theta)}\frac{\partial \eta(\theta)}{\partial \theta_j} \\
                                                       & = \int f(x) \frac{dp_{\eta(\theta)}(x)}{d \eta(\theta)} \frac{\partial \eta(\theta)}{\partial \theta_j}d\mu(x) \\
                                                       & = \int f(x) \frac{\partial}{\partial \theta_j} dp_{\eta(\theta)}(x)
  \end{align}
\end{proof}

\subsection{Lemmas and theorems in service of Fisher Info}

Our local identifiability result is based on the following theorem that is referenced, though not proven, in \cite{mukerjeePositiveDefinitenessInformation2002}.
To our knowledge an explicit proof has not been given, though it follows directly from the proof of the Cram\'er-Rao lower bound in \cite{raoLinearStatisticalInference2002}.
This proof provides a slightly different route to showing local identifiability compared to that of \cite{catchpoleDetectingParameterRedundancy1997}.
We give a proof below where we use the same notation as used in \cite{raoLinearStatisticalInference2002} for clarity's sake.  
\begin{theorem} \label{thm:pos-def-info}
Suppose we have observations $X_n$ and let $\mathbf{x} \in \R^{N}$ be the collection of all observations, with the $n$-th element equal to $X_n$, where $\pi(\mathbf{x}, \boldsymbol{\theta})$, parameterized by $\boldsymbol{\theta} \in \R^d$ with $i^\textrm{th}$ element $\theta_i$ is the joint density of the observations.  
Let $f_1(\mathbf{x}), \dots, f_r(\mathbf{x})$ be $r$ statistics for which $\Exp{f_i(\mathbf{x})} = g_i(\boldsymbol{\theta})$.
Further, assume that $\frac{\partial}{\partial \theta_j} \int f_i(\mathbf{x}) \pi(\mathbf{x}, \boldsymbol{\theta}) d\mathbf{x} = \int f_i(\mathbf{x}) \frac{\partial}{\partial \theta_j} \pi(\mathbf{x}, \boldsymbol{\theta}) d\mathbf{x} = \frac{\partial g_i(\boldsymbol{\theta})}{\partial \theta_j}$. Let $\Delta$ be a matrix in $\R^{r \times d}$ with elements $\Delta_{ij} = \frac{\partial g_i(\boldsymbol{\theta})}{\partial \theta_j}$.
Let $\mathcal{I}$ be a matrix in $\R^{d \times d}$ where the $(i,j)$-th element is defined as $\mathcal{I}_{ij} = \Cov{\frac{\partial \log P(\mathbf{x}, \boldsymbol{\theta})}{\partial \theta_i}}{\frac{\partial \log P(\mathbf{x}, \boldsymbol{\theta})}{\partial \theta_j}}$.
Let $\mathbf{V}$ be the matrix in $\R^{r \times r}$ with $(i,j)$ elements $\mathbf{V}_{ij} = \Cov{f_i(\mathbf{x})}{f_j(\mathbf{x})}$. 
If $\mathbf{V}$ is positive definite, and $\Delta$ is full-rank, then the Fisher information matrix $\mathcal{I}$
is positive definite.
\end{theorem}
\begin{proof}
Let $\mathbf{f}$ be the ordered collection of elements $f_i(\mathbf{x})$, and let 
$\nabla \log \pi(\mathbf{x}, \boldsymbol{\theta})$ be the score vector. Then let the random vector $\rho = (\mathbf{f}, \nabla \log \pi(\mathbf{x},\boldsymbol{\theta}))$. The covariance 
matrix associated with $\rho$, $\Sigma$, is a block matrix. Under 
$\pi(\mathbf{x}, \boldsymbol{\theta})$, $\Cov{f_i}{\frac{\partial \log \pi(\mathbf{x}, \boldsymbol{\theta})}{\partial \theta_j}} = \frac{\partial g_i(\boldsymbol{\theta})}{\partial \theta_j}$, which
is element $(i,j)$ of the matrix $\Delta$. The block covariance matrix for $\rho$ is:
\begin{align*}
\Sigma = \begin{bmatrix}
    \mathbf{V} & \Delta \\
    \Delta^T & \mathcal{I}
    \end{bmatrix}
\end{align*}
Suppose $\mathbf{V}$ is positive definite. We know $\mathcal{I} - \Delta^T \mathbf{V}^{-1} \Delta \succeq 0$,
because $\Sigma$ is a covariance matrix which ensures it is positive semi definite. 
Furthermore, $\mathcal{I} \succeq \Delta^T \mathbf{V}^{-1} \Delta$. If $\Delta$ is full-rank then $\mathcal{I} \succ 0$.
\end{proof}

We need the following two lemmas to prove that the Fisher information is positive definite.
First, the following lemma is stated in \cite{tian2004rank}:
\begin{lemma} \label{lemma:block-rank}
Let $\mathbf{A}^\dagger$ be the Moore-Penrose inverse of a matrix $\mathbf{A} \in \R^{L \times M}$. Let $\mathbf{B} \in \R^{L \times T}$.
The rank of a block matrix $[\mathbf{A}\,\, \mathbf{B}]$ is the rank of $\mathbf{A}$ plus the rank of $\mathbf{B} - \mathbf{A}\mathbf{A}^\dagger \mathbf{B}$.
\end{lemma}

Next, we will need this lemma later on:
\begin{lemma}\label{lemma:model-block-rank}
Suppose $\mathbf{W} \in \R^{I \times K}$ and 
$\mathbf{E} \in \R^{I \times J}$. Let $\mathbf{E}_{[:,j]}$ be an $I$-vector of the $j^\textrm{th}$ column of matrix $\mathbf{E}$, let $\mathbf{F}_j = {\rm diag}(\mathbf{E}_{[:,j]})$.
\begin{enumerate}[label=(L.\alph*)]\label{conds:rank-lemma}
\item $I \geq J + K$\label{conds:rank-lemma-dim}
\item ${\rm rank}\lp\mathbf{E}\rp = J$\label{conds:rank-lemma-rank-E}
\item ${\rm rank}\lp\mathbf{W}\rp = K$\label{conds:rank-lemma-rank-W}
\item $\forall i \in 1, \dots, I$ $\sum_{j=1}^J E_{ij} > 0$\label{conds:rank-lemma-pos-E}
\item $\lambda_j > 0 \forall j \in 1, \dots, J$\label{conds:rank-lemma-pos-lambda}
    \item 
    ${\rm rank}\lp \begin{bmatrix}
    \mathbf{F}_1\mathbf{W} & \dots & \mathbf{F}_J\mathbf{W} & \mathbf{E}_{[:,1]}  & \mathbf{E}_{[:,2]} & \dots & \mathbf{E}_{[:,J-1]}  & \mathbf{E}_{[:,J]}
  \end{bmatrix}\rp > J + K$\label{conds:rank-lemma-rank-big-mat}
\end{enumerate}
Then the matrix
$$
\begin{bmatrix}
    \lp \sum_{j} \lambda_j\mathbf{F}_j \rp \mathbf{W} & \mathbf{E}_{[:,1]}  & \mathbf{E}_{[:,2]} & \dots & \mathbf{E}_{[:,J-1]}  & \mathbf{E}_{[:,J]}
    \end{bmatrix}
$$
is rank $J + K$.
\end{lemma}
\begin{proof}
  Given \crefrange{conds:rank-lemma-dim}{conds:rank-lemma-pos-lambda} above , the matrix 
$\lp \sum_{j} \lambda_j\mathbf{F}_j \rp \mathbf{W}$ is rank $K$. Using lemma
\ref{lemma:block-rank} we proceed sequentially, showing first that 

$$\text{rank}\lp
\begin{bmatrix}
    \lp \sum_{j} \lambda_j\mathbf{F}_j \rp \mathbf{W} & \mathbf{E}_{[:,1]}
\end{bmatrix}\rp = K + 1
$$
By \ref{lemma:block-rank}, 
\begin{align*}
\text{rank} \lp
\begin{bmatrix}
 \lp \sum_{j} \lambda_j\mathbf{F}_j \rp \mathbf{W} & \mathbf{E}_{[:,1]}
\end{bmatrix}
\rp = &
\text{rank} \lp
 \lp \sum_{j} \lambda_j\mathbf{F}_j \rp \mathbf{W} 
\rp \\
 & + \text{rank} \lp\mathbf{E}_{[:,1]} -
 \lp \sum_{j} \lambda_j\mathbf{F}_j \rp \mathbf{W} \lp\lp \sum_{j} \lambda_j\mathbf{F}_j \rp \mathbf{W}\rp^\dagger\mathbf{E}_{[:,1]}
\rp
\end{align*}
We can show that
$\text{rank} \lp\mathbf{E}_{[:,1]} - \lp \sum_{j} \lambda_j\mathbf{F}_j \rp \mathbf{W} \lp\lp \sum_{j} \lambda_j\mathbf{F}_j \rp \mathbf{W}\rp^\dagger\mathbf{E}_{[:,1]}\rp = 1$.
 We show by contradiction that given conditions on the coefficient matrix, no 
 solution to the equation in $a_k$
\begin{align}\label{eqn:base-case-sys}
  \mathbf{E}_{[:,1]} = \sum_{k=1}^K a_k \lp \sum_{j=1}^J \lambda_j \mathbf{E}_{[:,j]} \odot \mathbf{W}_k\rp
\end{align}
can be found.
This follows from examining the matrix of coefficients of the equation, shown
here in block form:
\begin{align}\label{eqn:base-case-coef-mat}
\begin{bmatrix}
\mathbf{E}_{[:,1]} & \lambda_1 \mathbf{F}_1 \mathbf{W} & \dots & \lambda_J \mathbf{F}_J \mathbf{W}
\end{bmatrix}
\end{align}
Given \cref{conds:rank-lemma-pos-lambda},  $\lambda_j > 0$ for all $j$, for a fixed set of parameters $\lambda_j$ the matrix in \cref{eqn:base-case-coef-mat} has the same column space as
\begin{align}\label{eqn:base-case-col-space-coef-mat}
\begin{bmatrix}
\mathbf{E}_{[:,1]} & \mathbf{F}_1 \mathbf{W} & \dots & \mathbf{F}_J \mathbf{W}
\end{bmatrix}
\end{align}
Given \cref{conds:rank-lemma-rank-big-mat}, matrix \ref{eqn:base-case-col-space-coef-mat} has rank greater than $K$, so the system of equations in \ref{eqn:base-case-sys} will not have a solution in $a_k, k \in [1, \dots, K]$. Thus the rank of matrix 
\begin{align}
  \mathbf{E}_{[:,1]} - \lp \sum_{j} \lambda_j\mathbf{F}_j \rp \mathbf{W} \lp\lp \sum_{j} \lambda_j\mathbf{F}_j \rp \mathbf{W}\rp^\dagger\mathbf{E}_{[:,1]}
  \end{align}
  is $1$ so the rank of .
  $$
  \begin{bmatrix}
    \lp \sum_{j} \lambda_j\mathbf{F}_j \rp \mathbf{W} & \mathbf{E}_{[:,1]}
  \end{bmatrix}
  $$
  is $K + 1$.

Now for the induction step:
Suppose that 
$$
\begin{bmatrix}
    \lp \sum_{j} \lambda_j\mathbf{F}_j \rp \mathbf{W} & \mathbf{E}_{[:,1:M]}
    \end{bmatrix}
$$
is rank $M + K$
and suppose we want to determine the rank of
$$
\begin{bmatrix}
    \lp \sum_{j} \lambda_j\mathbf{F}_j \rp \mathbf{W} & \mathbf{E}_{[:,1:M]} & \mathbf{E}_{[:,M+1]}
    \end{bmatrix}.
$$
By lemma \ref{lemma:block-rank}, the rank of the above matrix is 
\begin{align}
&\text{rank}\begin{bmatrix}
    \lp \sum_{j} \lambda_j\mathbf{F}_j \rp \mathbf{W} & \mathbf{E}_{[:,1:M]}
    \end{bmatrix}\label{eqn:ind-step-induction-hypo} \\
    & +
   \text{rank}\lp\mathbf{E}_{[:,M+1]} - \begin{bmatrix}
    \lp \sum_{j} \lambda_j\mathbf{F}_j \rp \mathbf{W} & \mathbf{E}_{[:,1:M]}
    \end{bmatrix}\begin{bmatrix}
    \lp \sum_{j} \lambda_j\mathbf{F}_j \rp \mathbf{W} & \mathbf{E}_{[:,1:M]}
  \end{bmatrix}^\dagger\mathbf{E}_{[:,M+1]} \rp \label{eqn:ind-step-to-prove}
\end{align}
By the induction hypothesis, we know that \cref{eqn:ind-step-induction-hypo} equals $K + M$ and we 
look to see whether \cref{eqn:ind-step-to-prove} equals 1 by determining whether the system of equations below: 
\begin{align}\label{eqn:ind-step-sys}
   \mathbf{E}_{[:,M+1]} = \sum_{k=1}^K a_k \lp \sum_{j=1}^J \lambda_j \mathbf{E}_j \odot \mathbf{W}_k\rp
   + \sum_{m=1}^M d_m \mathbf{E}_{[:,m]}
\end{align}
has a solution in the variables $a_k, k \in [1,\dots,K]$, $d_m, m \in [1,\dots,M], M < J$.
Then for a fixed $\{\lambda_j, j \in [1, \dots, J]\}$ the coefficient matrix for the system of equations in \cref{eqn:ind-step-sys} has the same column space as
\begin{align}\label{eqn:ind-step-coef-mat}
\begin{bmatrix}
\mathbf{E}_{[:,M+1]}  & \mathbf{F}_1 \mathbf{W} & \dots & \mathbf{F}_J \mathbf{W} & \mathbf{E}_{[:,1:M]}
\end{bmatrix}
\end{align}
By \cref{conds:rank-lemma-rank-big-mat}, the rank of matrix \eqref{eqn:ind-step-coef-mat} is greater than $M + K$, which precludes $\mathbf{E}_{[:,M+1]}$ from lying in the column space of
$$
\begin{bmatrix}\lp \sum_{j} \lambda_j\mathbf{F}_j \rp \mathbf{W} & \mathbf{E}_{[:,1:M]}\end{bmatrix}
$$. Thus the line \eqref{eqn:ind-step-to-prove} is equal to $1$ and summing with line \eqref{eqn:ind-step-induction-hypo} shows that the rank of 
$$
\begin{bmatrix}
  \lp \sum_{j} \lambda_j\mathbf{F}_j \rp \mathbf{W} & \mathbf{E}_{[:,1:M]} & \mathbf{E}_{[:,M+1]}
\end{bmatrix}
$$
is $M + 1 + K$.
Therefore by induction,  
$$
\begin{bmatrix}
  \lp \sum_{j} \lambda_j\mathbf{F}_j \rp \mathbf{W} & \mathbf{E}_{[:,1]}  & \mathbf{E}_{[:,2]} & \dots & \mathbf{E}_{[:,J-1]}  & \mathbf{E}_{[:,J]}
\end{bmatrix}
$$
is rank $J + K$.
\end{proof}

\section{Full model local identifiability proof}\label{proof:full-model}
\begin{proof}
  In order to draw on the results of \ref{thm:pos-def-info}, we must ensure $\frac{\partial}{\partial \theta_j} \int f_i \pi(\mathbf{x}, \theta) d\nu = \int f_i \frac{\partial}{\partial \theta_j} \pi(\mathbf{x}, \theta) d\nu$ holds for model \eqref{model:full}.
  This condition does indeed hold by lemma \ref{lemma:dct-pois} because our observational density is an exponential family density.
  Now we look for moment estimators, $\mathbf{f} = (f_1, \dots, f_r)$ with full-rank $\Delta$ and positive definite covariance matrices. 
    As shown in section \ref{app:like-deriv}, conditionally on unknown parameters, known covariates $\mathbf{z}_i$ and population counts $\mathbf{e}_i$, $X_{ij}$ is independent of $M_i$ for all $i$ and $j$.
    Let $\mathbf{x}_j = (X_{1j}, X_{2j},\dots, X_{Ij})$, 
    and set
    $$\mathbf{f} = (\mathbf{x}_{1},\mathbf{x}_{2}, \dots, \mathbf{x}_{J}, M_1, M_2, \dots, M_I).$$ 
    From the independence of the elements of the vector $(X_{i1},X_{i2}, \dots,X_{iJ},,  M_i)^T$, and the conditional independence between observations $i$, $\text{Cov}(\mathbf{f})$ is diagonal, and is positive definite because \crefrange{cond:full-model-p}{cond:full-model-E} are assumed to hold.
The final condition for \cref{thm:pos-def-info} to hold, which establishes a lower bound on the positive definiteness of the Fisher information matrix is to ensure that $\Delta$ has full column-rank.
To that end, let us calculate $\Delta$, beginning with $\Exp{\mathbf{f}}$:
The expected value of $\mathbf{f}$ is the vector:
\begin{align*}
    \begin{bmatrix}
    E_{11} p_{11} e^{\mathbf{z}_1^T \boldsymbol{\beta}}\lambda_1 \\
    E_{21} p_{21} e^{\mathbf{z}_2^T \boldsymbol{\beta}}\lambda_1 \\
    \vdots \\
    E_{I1} p_{I1} e^{\mathbf{z}_I^T \boldsymbol{\beta}}\lambda_1 \\
    E_{12} p_{12} e^{\mathbf{z}_1^T \boldsymbol{\beta}}\lambda_2 \\
    E_{22} p_{22} e^{\mathbf{z}_2^T \boldsymbol{\beta}}\lambda_2 \\
    \vdots \\
    E_{I2} p_{I2} e^{\mathbf{z}_I^T \boldsymbol{\beta}}\lambda_2 \\
    \vdots \\
    E_{1J} p_{1J} e^{\mathbf{z}_1^T \boldsymbol{\beta}}\lambda_J \\
    E_{2J} p_{2J} e^{\mathbf{z}_2^T \boldsymbol{\beta}}\lambda_J \\
    \vdots \\
    E_{IJ} p_{IJ} e^{\mathbf{z}_I^T \boldsymbol{\beta}}\lambda_J \\
    \sum_j  E_{1j} (1 - p_{1j}) e^{\mathbf{z}_1^T \boldsymbol{\beta}}\lambda_j \\
    \vdots \\
    \sum_j  E_{Ij} (1 - p_{Ij}) e^{\mathbf{z}_I^T \boldsymbol{\beta}}\lambda_j
    \end{bmatrix}
\end{align*}
For efficiency of notation, 
let $c_{ij} = p_{ij} \lambda_j$ $p^{\prime}_{ij} = p_{ij}(1 - p_{ij})$, and $q_{ij} = (1 - p_{ij})$
Let $\mathbf{H}$ by the $(J + 1)I \times 2K$ matrix of partial derivatives of $\Exp{\mathbf{f}}$ with respect to the vector $(\boldsymbol{\beta}, \boldsymbol{\gamma})$:
{\tiny
\begin{align*}
  \mathbf{H} = 
    \begin{bmatrix}
      \mathbf{z}_1^T E_{11} c_{11}e^{\mathbf{z}_1^T \boldsymbol{\beta}} & \mathbf{z}_1^T E_{11} p^{\prime}_{11}\lambda_1e^{\mathbf{z}_1^T \boldsymbol{\beta}} \\
      \mathbf{z}_2^T E_{21} c_{21}e^{\mathbf{z}_2^T \boldsymbol{\beta}} & \mathbf{z}_2^T E_{21} p^{\prime}_{21}\lambda_1e^{\mathbf{z}_2^T \boldsymbol{\beta}} \\
    \vdots & \vdots \\
    \mathbf{z}_I^T E_{I1} c_{I1}e^{\mathbf{z}_I^T \boldsymbol{\beta}} & \mathbf{z}_I^T E_{I1} p^{\prime}_{I1}\lambda_1e^{\mathbf{z}_I^T \boldsymbol{\beta}} \\
    \mathbf{z}_1^T E_{12} c_{12}e^{\mathbf{z}_1^T \boldsymbol{\beta}} & \mathbf{z}_1^T E_{12} p^{\prime}_{12}\lambda_2e^{\mathbf{z}_1^T \boldsymbol{\beta}} \\
    \mathbf{z}_2^T E_{22} c_{22}e^{\mathbf{z}_2^T \boldsymbol{\beta}} & \mathbf{z}_2^T E_{21} p^{\prime}_{22}\lambda_2e^{\mathbf{z}_2^T \boldsymbol{\beta}} \\
    \vdots & \vdots \\
    \mathbf{z}_I^T E_{I2} c_{I2}e^{\mathbf{z}_I^T \boldsymbol{\beta}} & \mathbf{z}_I^T E_{I2} p^{\prime}_{I2}\lambda_2e^{\mathbf{z}_I^T \boldsymbol{\beta}} \\
    \vdots & \vdots \\
    \mathbf{z}_1^T E_{1J} c_{1J}e^{\mathbf{z}_1^T \boldsymbol{\beta}} & \mathbf{z}_1^T E_{1J} p^{\prime}_{1J}\lambda_Je^{\mathbf{z}_1^T \boldsymbol{\beta}} \\
    \mathbf{z}_2^T E_{2J} c_{2J}e^{\mathbf{z}_2^T \boldsymbol{\beta}} & \mathbf{z}_2^T E_{2J} p^{\prime}_{2J}\lambda_J e^{\mathbf{z}_2^T \boldsymbol{\beta}} \\
    \vdots & \vdots \\
    \mathbf{z}_I^T E_{IJ} c_{IJ}e^{\mathbf{z}_I^T \boldsymbol{\beta}} & \mathbf{z}_I^T E_{IJ} p^{\prime}_{IJ}\lambda_Je^{\mathbf{z}_I^T \boldsymbol{\beta}} \\
    \mathbf{z}_1^T e^{\mathbf{z}_1^T \boldsymbol{\beta}}\sum_{j}E_{1j} q_{1j} \lambda_j & -\mathbf{z}_1^T e^{\mathbf{z}_1^T \boldsymbol{\beta}}\sum_{j} E_{1j} p^\prime_{1j} \lambda_j \\
    \vdots & \vdots \\
    \mathbf{z}_I^T e^{\mathbf{z}_I^T \boldsymbol{\beta}}\sum_{j}E_{Ij} q_{Ij} \lambda_j & -\mathbf{z}_I^T e^{\mathbf{z}_I^T \boldsymbol{\beta}}\sum_{j} E_{Ij} p^\prime_{Ij} \lambda_j
    \end{bmatrix}
\end{align*}
}%
Let $\mathbf{T}$ be the $(J + 1)I \times 2J$ matrix of partial derivatives with respect to $(\lambda_1, \lambda_2, \dots, \lambda_{J}, \eta_1,\eta_2, \dots, \eta_{J})$,
{\tiny
\begin{align*}
  \mathbf{T} =   \begin{bmatrix}
      E_{11} p_{11}e^{\mathbf{z}_1^T \boldsymbol{\beta}}  & 0 & \dots  & 0 &  E_{11} p^\prime_{11}\lambda_1e^{\mathbf{z}_1^T \boldsymbol{\beta}} & 0 & \dots & 0\\
      E_{21} p_{21}e^{\mathbf{z}_2^T \boldsymbol{\beta}}  & 0 & \dots  & 0 &  E_{21} p^\prime_{21}\lambda_1e^{\mathbf{z}_2^T \boldsymbol{\beta}} & 0 & \dots & 0\\
     \vdots & \vdots & \vdots & \vdots & \vdots & \vdots  & \vdots & \vdots\\
     E_{I1} p_{I1}e^{\mathbf{z}_I^T \boldsymbol{\beta}}  & 0 & \dots  & 0 &  E_{I1} p^\prime_{I1}\lambda_1e^{\mathbf{z}_I^T \boldsymbol{\beta}} & 0 & \dots & 0\\
     0 &  E_{12} p_{12}e^{\mathbf{z}_1^T \boldsymbol{\beta}}  & \dots & 0 & 0 &  E_{12} p^\prime_{12} \lambda_2e^{\mathbf{z}_1^T \boldsymbol{\beta}}  & \dots & 0\\
     0 &  E_{22} p_{22}e^{\mathbf{z}_2^T \boldsymbol{\beta}}   & \dots & 0 & 0 &  E_{22} p^\prime_{22} \lambda_2e^{\mathbf{z}_2^T \boldsymbol{\beta}} & \dots & 0\\
    \vdots & \vdots & \vdots & \vdots & \vdots & \vdots  & \vdots & \vdots\\
    0 &  E_{I2} p_{I2}e^{\mathbf{z}_I^T \boldsymbol{\beta}}  & \dots & 0 & 0 &  E_{I2} p^\prime_{I2} \lambda_2e^{\mathbf{z}_I^T \boldsymbol{\beta}} & \dots & 0\\
    \vdots & \vdots & \ddots & \vdots & \vdots & \vdots  & \ddots & \vdots\\
    0 & 0 & \dots &  E_{1J} p_{1J}e^{\mathbf{z}_1^T \boldsymbol{\beta}} & 0 & 0 & \dots &  E_{1J} p^\prime_{1J}\lambda_Je^{\mathbf{z}_1^T \boldsymbol{\beta}}\\
    0 & 0 & \dots &  E_{2J} p_{2J}e^{\mathbf{z}_2^T \boldsymbol{\beta}} & 0 & 0 & \dots &  E_{2J} p^\prime_{2J}\lambda_Je^{\mathbf{z}_2^T \boldsymbol{\beta}}\\
    \vdots & \vdots & \vdots & \vdots & \vdots & \vdots  & \vdots & \vdots\\
    0 & 0 & \dots &  E_{IJ} p_{IJ}e^{\mathbf{z}_I^T \boldsymbol{\beta}} & 0 & 0 & \dots &  E_{IJ} p^\prime_{IJ}\lambda_Je^{\mathbf{z}_I^T \boldsymbol{\beta}}\\
    E_{11} q_{11}e^{\mathbf{z}_1^T \boldsymbol{\beta}} &  E_{12} q_{12}e^{\mathbf{z}_1^T \boldsymbol{\beta}} & \dots &  E_{1J} q_{1J}e^{\mathbf{z}_1^T \boldsymbol{\beta}} & -E_{11} p^\prime_{11}\lambda_1e^{\mathbf{z}_1^T \boldsymbol{\beta}} & - E_{12} p^\prime_{12}\lambda_2e^{\mathbf{z}_1^T \boldsymbol{\beta}} & \dots
    & - E_{1J} p^\prime_{1J} \lambda_Je^{\mathbf{z}_1^T \boldsymbol{\beta}}\\
    \vdots & \vdots & \ddots & \vdots & \vdots & \vdots  & \ddots & \vdots\\
    E_{I1} q_{I1} e^{\mathbf{z}_I^T \boldsymbol{\beta}}&  E_{I2} q_{I2}e^{\mathbf{z}_I^T \boldsymbol{\beta}} & \dots &  E_{IJ} q_{IJ}e^{\mathbf{z}_I^T \boldsymbol{\beta}} & -E_{I1} p^\prime_{I1}\lambda_1 e^{\mathbf{z}_I^T \boldsymbol{\beta}}& - E_{I2} p^\prime_{I2}\lambda_2 e^{\mathbf{z}_I^T \boldsymbol{\beta}}& \dots
    & - E_{IJ} p^\prime_{IJ} \lambda_J e^{\mathbf{z}_I^T \boldsymbol{\beta}}\\
    \end{bmatrix}
\end{align*}
}%
Then
\begin{align*}
  \Delta & = \begin{bmatrix} \mathbf{T} & \mathbf{H} \end{bmatrix}
\end{align*}
  
Let the matrix $\mathbf{R}_{i,j}(m)$ be the elementary row-addition matrix.
When a matrix $\mathbf{A} \in \R^{M\times N}$ is left-multiplied by $\mathbf{R}_{ij}(m)$, $\tilde{\mathbf{A}} = \mathbf{E}_{ij}(m) \mathbf{A}$, all rows of $\tilde{\mathbf{A}}$ equal that of $\mathbf{A}$ excepting $\tilde{\mathbf{A}}$'s $i$-th row, which is $\tilde{\mathbf{A}}_{[i,:]} = \mathbf{A}_{[i,:]} + m\mathbf{A}_{[j,:]}$.
Let $\tilde{\mathbf{H}}$ and $\tilde{\mathbf{T}}$ be the result of left-multiplying $\mathbf{H}$ and $\mathbf{T}$ by the same product of elementary row-addition matrices, namely:
{
\begin{align*}
  \prod_{i=1}^I\prod_{j=1}^J \mathbf{R}_{JI + j,(j - 1)I + i}(1) 
\end{align*}
}%
Then let $\tilde{\Delta}$ be the matrix $\Delta$ after applying the product of elementary row-addition matrices:
\begin{align*}
  \tilde{\Delta} & = \prod_{i=1}^I\prod_{j=1}^J \mathbf{R}_{JI + j,(j - 1)I + i}(1) \begin{bmatrix} \mathbf{T} & \mathbf{H} \end{bmatrix} \\
  & = \begin{bmatrix} \prod_{i=1}^I\prod_{j=1}^J \mathbf{R}_{JI + j,(j - 1)I + i}(1) \mathbf{T} & \prod_{i=1}^I\prod_{j=1}^J \mathbf{R}_{JI + j,(j - 1)I + i}(1) \mathbf{H} \end{bmatrix}  \\
                 & = \begin{bmatrix} \tilde{\mathbf{T}} & \tilde{\mathbf{H}} \end{bmatrix}
\end{align*}
where
{\small
\begin{align*}
  \tilde{\mathbf{H}} = 
    \begin{bmatrix}
      \mathbf{z}_1^T E_{11} c_{11}e^{\mathbf{z}_1^T \boldsymbol{\beta}} & \mathbf{z}_1^T E_{11} p^{\prime}_{11}\lambda_1e^{\mathbf{z}_1^T \boldsymbol{\beta}} \\
      \mathbf{z}_2^T E_{21} c_{21}e^{\mathbf{z}_2^T \boldsymbol{\beta}} & \mathbf{z}_2^T E_{21} p^{\prime}_{21}\lambda_1e^{\mathbf{z}_2^T \boldsymbol{\beta}} \\
    \vdots & \vdots \\
    \mathbf{z}_I^T E_{I1} c_{I1}e^{\mathbf{z}_I^T \boldsymbol{\beta}} & \mathbf{z}_I^T E_{I1} p^{\prime}_{I1}\lambda_1e^{\mathbf{z}_I^T \boldsymbol{\beta}} \\
    \mathbf{z}_1^T E_{12} c_{12}e^{\mathbf{z}_1^T \boldsymbol{\beta}} & \mathbf{z}_1^T E_{12} p^{\prime}_{12}\lambda_2e^{\mathbf{z}_1^T \boldsymbol{\beta}} \\
    \mathbf{z}_2^T E_{22} c_{22}e^{\mathbf{z}_2^T \boldsymbol{\beta}} & \mathbf{z}_2^T E_{21} p^{\prime}_{22}\lambda_2e^{\mathbf{z}_2^T \boldsymbol{\beta}} \\
    \vdots & \vdots \\
    \mathbf{z}_I^T E_{I2} c_{I2}e^{\mathbf{z}_I^T \boldsymbol{\beta}} & \mathbf{z}_I^T E_{I2} p^{\prime}_{I2}\lambda_2e^{\mathbf{z}_I^T \boldsymbol{\beta}} \\
    \vdots & \vdots \\
    \mathbf{z}_1^T E_{1J} c_{1J}e^{\mathbf{z}_1^T \boldsymbol{\beta}} & \mathbf{z}_1^T E_{1J} p^{\prime}_{1J}\lambda_Je^{\mathbf{z}_1^T \boldsymbol{\beta}} \\
    \mathbf{z}_2^T E_{2J} c_{2J}e^{\mathbf{z}_2^T \boldsymbol{\beta}} & \mathbf{z}_2^T E_{2J} p^{\prime}_{2J}\lambda_J e^{\mathbf{z}_2^T \boldsymbol{\beta}} \\
    \vdots & \vdots \\
    \mathbf{z}_I^T E_{IJ} c_{IJ}e^{\mathbf{z}_I^T \boldsymbol{\beta}} & \mathbf{z}_I^T E_{IJ} p^{\prime}_{IJ}\lambda_Je^{\mathbf{z}_I^T \boldsymbol{\beta}} \\
    \mathbf{z}_1^T e^{\mathbf{z}_1^T \boldsymbol{\beta}}\sum_{j}E_{1j} \lambda_j & 0_{1 \times K} \\
    \vdots & \vdots \\
    \mathbf{z}_I^T e^{\mathbf{z}_I^T \boldsymbol{\beta}}\sum_{j}E_{Ij} \lambda_j & 0_{1 \times K}
    \end{bmatrix},
\end{align*}
}%
and 
{\tiny
\begin{align*}
  \tilde{\mathbf{T}} =   \begin{bmatrix}
      E_{11} p_{11}e^{\mathbf{z}_1^T \boldsymbol{\beta}}  & 0 & \dots  & 0 &  E_{11} p^\prime_{11}\lambda_1e^{\mathbf{z}_1^T \boldsymbol{\beta}} & 0 & \dots & 0\\
      E_{21} p_{21}e^{\mathbf{z}_2^T \boldsymbol{\beta}}  & 0 & \dots  & 0 &  E_{21} p^\prime_{21}\lambda_1e^{\mathbf{z}_2^T \boldsymbol{\beta}} & 0 & \dots & 0\\
     \vdots & \vdots & \vdots & \vdots & \vdots & \vdots  & \vdots & \vdots\\
     E_{I1} p_{I1}e^{\mathbf{z}_I^T \boldsymbol{\beta}}  & 0 & \dots  & 0 &  E_{I1} p^\prime_{I1}\lambda_1e^{\mathbf{z}_I^T \boldsymbol{\beta}} & 0 & \dots & 0\\
     0 &  E_{12} p_{12}e^{\mathbf{z}_1^T \boldsymbol{\beta}}  & \dots & 0 & 0 &  E_{12} p^\prime_{12} \lambda_2e^{\mathbf{z}_1^T \boldsymbol{\beta}}  & \dots & 0\\
     0 &  E_{22} p_{22}e^{\mathbf{z}_2^T \boldsymbol{\beta}}   & \dots & 0 & 0 &  E_{22} p^\prime_{22} \lambda_2e^{\mathbf{z}_2^T \boldsymbol{\beta}} & \dots & 0\\
    \vdots & \vdots & \vdots & \vdots & \vdots & \vdots  & \vdots & \vdots\\
    0 &  E_{I2} p_{I2}e^{\mathbf{z}_I^T \boldsymbol{\beta}}  & \dots & 0 & 0 &  E_{I2} p^\prime_{I2} \lambda_2e^{\mathbf{z}_I^T \boldsymbol{\beta}} & \dots & 0\\
    \vdots & \vdots & \ddots & \vdots & \vdots & \vdots  & \ddots & \vdots\\
    0 & 0 & \dots &  E_{1J} p_{1J}e^{\mathbf{z}_1^T \boldsymbol{\beta}} & 0 & 0 & \dots &  E_{1J} p^\prime_{1J}\lambda_Je^{\mathbf{z}_1^T \boldsymbol{\beta}}\\
    0 & 0 & \dots &  E_{2J} p_{2J}e^{\mathbf{z}_2^T \boldsymbol{\beta}} & 0 & 0 & \dots &  E_{2J} p^\prime_{2J}\lambda_Je^{\mathbf{z}_2^T \boldsymbol{\beta}}\\
    \vdots & \vdots & \vdots & \vdots & \vdots & \vdots  & \vdots & \vdots\\
    0 & 0 & \dots &  E_{IJ} p_{IJ}e^{\mathbf{z}_I^T \boldsymbol{\beta}} & 0 & 0 & \dots &  E_{IJ} p^\prime_{IJ}\lambda_Je^{\mathbf{z}_I^T \boldsymbol{\beta}}\\
    E_{11} e^{\mathbf{z}_1^T \boldsymbol{\beta}} &  E_{12} e^{\mathbf{z}_1^T \boldsymbol{\beta}} & \dots &  E_{1J} e^{\mathbf{z}_1^T \boldsymbol{\beta}} & 0 & 0 & \dots
    & 0\\
    \vdots & \vdots & \ddots & \vdots & \vdots & \vdots  & \ddots & \vdots\\
    E_{I1} e^{\mathbf{z}_I^T \boldsymbol{\beta}}&  E_{I2} e^{\mathbf{z}_I^T \boldsymbol{\beta}} & \dots &  E_{IJ} e^{\mathbf{z}_I^T \boldsymbol{\beta}} & 0& 0 & \dots
    & 0\\
    \end{bmatrix}.
\end{align*}
}%

We can represent $\tilde{\mathbf{T}}$ and $\tilde{\mathbf{H}}$ as block matrices. 
We let $\mathbf{E}$ be the $I \times J$ matrix with $(i,j)^\text{th}$ elements $E_{ij}$,
and we similarly define the matrix $\mathbf{C}$ to be in $\R^{I \times J}$ with 
its $i,j$ element equal to $c_{ij}$. Furthermore, let $\mathbf{p}^\prime_{ij} = p^\prime_{ij}$. 
Let the matrix $\boldsymbol{\omega}$ be the diagonal matrix in $\R^{I \times I}$ with $i, j$ elements $e^{\mathbf{z}_i^T \boldsymbol{\beta}} \mathbbm{1}(i = j)$. Let $\mathbf{E}_{[:,j]} \odot \mathbf{C}_{[:,j]}$ be the element-wise multiplication between the two matrices $\mathbf{E}_{[:,j]}$ and $ \mathbf{C}_{[:,j]}$.
Let $\mathbf{Z} \in \R^{I \times K}$ with rows $\mathbf{Z}_{[i,:]} = \mathbf{z}_i^T$.
Let $\mathbf{1}$ be the $I$-dimensional vector with each element equal to 1 and let 
\begin{align*}
  \mathbf{D}_j & = \text{diag}(\mathbf{E}_{[:,j]} \odot \mathbf{C}_{[:,j]}) \\
  \mathbf{D}^\prime_j & = \text{diag}(\mathbf{E}_{[:,j]} \odot \mathbf{p}^\prime_{[:,j]}) \\
    \mathbf{F}_j & = \text{diag}(\mathbf{E}_{[:,j]}).
\end{align*}
Let $\boldsymbol{\Omega}$ be the block matrix:
\begin{align}
  \boldsymbol{\Omega} = 
  \begin{bmatrix}
    \boldsymbol{\omega} & 0_{I\times I}& \dots &0_{I\times I}&0_{I\times I}&0_{I\times I}\\
    0_{I\times I} & \boldsymbol{\omega} & \dots &0_{I\times I}&0_{I\times I}&0_{I\times I}\\
    \vdots & \vdots & \ddots & \vdots& \vdots& \vdots\\
    0_{I\times I} & 0_{I\times I} & \dots & \boldsymbol{\omega} & 0_{I\times I}& 0_{I\times I}\\
    0_{I\times I} & 0_{I\times I} & \dots  & 0_{I\times I}& \boldsymbol{\omega}& 0_{I\times I} \\
    0_{I\times I} & 0_{I\times I} & \dots  & 0_{I\times I}& 0_{I\times I} & \boldsymbol{\omega}
  \end{bmatrix}
\end{align}
Then we can write $\tilde{\mathbf{H}}$ as
{\tiny
\begin{align} \label{eq:gradient-modeled-missingness-cat-effects-a}
  \boldsymbol{\Omega}
    \begin{bmatrix}
    \mathbf{D}_1 \mathbf{Z} & \mathbf{D}_1^\prime \mathbf{Z}\lambda_1  \\
    \mathbf{D}_2 \mathbf{Z} & \mathbf{D}_2^\prime \mathbf{Z}\lambda_2  \\
    \vdots & \vdots\\
    \mathbf{D}_{J-1} \mathbf{Z} & \mathbf{D}_{J-1}^\prime \mathbf{Z}\lambda_{J-1}  \\
    \mathbf{D}_J \mathbf{Z} & \mathbf{D}_J^\prime \mathbf{Z} \lambda_{J} \\
    \lp \sum_{j} \lambda_j \mathbf{F}_j \rp \mathbf{Z} & 0_{I \times K}  \\
    \end{bmatrix}
\end{align}
}%

and $\tilde{\mathbf{T}}$ as
{\tiny
\begin{align} \label{eq:gradient-modeled-missingness-cat-effects-b}
  \boldsymbol{\Omega}
    \begin{bmatrix}
    \mathbf{D}_1 \mathbf{1} & 0_{I \times 1} & 
    \dots & 0_{I \times 1} & 0_{I \times 1} & \lambda_{1}\mathbf{D}_1^\prime\mathbf{1}  & 0_{I \times 1}& \dots & 0_{I \times 1} & 0_{I \times 1} \\
    0_{I \times 1} & \mathbf{D}_2 \mathbf{1} & 
    \dots & 0_{I \times 1} & 0_{I \times 1} & 0_{I \times 1} & \lambda_{2}\mathbf{D}^\prime_2\mathbf{1}  & \dots & 0_{I \times 1} & 0_{I \times 1}    \\
    \vdots & \vdots & \ddots & \vdots & \vdots & \vdots & \vdots & \ddots  & \vdots & \vdots \\
    0_{I \times 1} &  0_{I \times 1} & 
    \dots & \mathbf{D}_{J-1}\mathbf{1}  & 0_{I \times 1} & 0_{I \times 1} & 0_{I \times 1}& \dots &  \lambda_{J-1}\mathbf{D}^\prime_{J-1}\mathbf{1} & 0_{I \times 1}\\
    0_{I \times 1} &  0_{I \times 1} & 
    \dots & 0_{I \times 1} & \mathbf{D}_{J}\mathbf{1}  & 0_{I \times 1} & 0_{I \times 1}& \dots &  0_{I \times 1}&  \lambda_{J}\mathbf{D}^\prime_{J} \mathbf{1}\\
    \mathbf{E}_{[:,1]} & \mathbf{E}_{[:,2]}  & \dots & \mathbf{E}_{[:,J-1]}  & \mathbf{E}_{[:,J]}  & 
    0_{I \times 1} & 0_{I \times 1} & \dots & 0_{I \times 1} & 0_{I \times 1}
    \end{bmatrix}
\end{align}
}%
Rearranging the columns of $\tilde{\Delta}$ to form $\tilde{\Delta}^\prime$ does not change the rank of the matrix, and clarifies the conditions needed for the matrix to be full-rank:
{\tiny
\begin{align} \label{eq:gradient-modeled-missingness-cat-effects-comb}
  \tilde{\Delta}^\prime = \boldsymbol{\Omega}
    \begin{bmatrix}
    \mathbf{D}_1 \mathbf{Z} & \mathbf{D}_1\mathbf{1}  & 0_{I \times 1} &  \dots & 0_{I \times 1} & 0_{I \times 1} & \lambda_1\mathbf{D}_1^\prime \mathbf{Z} & 0_{I \times 1}& \lambda_1\mathbf{D}_1^\prime \mathbf{1} & 0_{I \times 1}& \dots & 0_{I \times 1} \\
    \mathbf{D}_2 \mathbf{Z} & 0_{I \times 1} & \mathbf{D}_2 \mathbf{1}& \dots & 0_{I \times 1} & 0_{I \times 1} & \lambda_2\mathbf{D}_2^\prime \mathbf{Z} & 0_{I \times 1} & 0_{I \times 1} & \lambda_2\mathbf{D}^\prime_2\mathbf{1}  & \dots & 0_{I \times 1}\\
    \vdots & \vdots & \vdots & \ddots & \vdots & \vdots & \vdots & \vdots & \vdots & \vdots & \ddots  & \vdots\\
    \mathbf{D}_{J-1} \mathbf{Z}  & 0_{I \times 1} & 0_{I \times 1} &  \dots &  \mathbf{D}_{J-1}\mathbf{1}  & 0_{I \times 1} & \lambda_{J-1}\mathbf{D}^\prime_{J-1} \mathbf{Z} & 0_{I \times 1} & 0_{I \times 1} & 0_{I \times 1} & \dots & \lambda_{J-1}\mathbf{D}_{J-1}^\prime\mathbf{1}\\
   \mathbf{D}_J \mathbf{Z} & 0_{I \times 1} & 0_{I \times 1} & \dots & 0_{I \times 1} &\mathbf{D}_J \mathbf{1} &
   \lambda_{J}\mathbf{D}^\prime_J \mathbf{Z}  & \lambda_{J}\mathbf{D}_{J}^\prime\mathbf{1} & 0_{I \times 1} & 0_{I \times 1} & \dots & 0_{I \times 1}\\
  \lp \sum_{j} \lambda_j\mathbf{F}_j \rp \mathbf{Z} & \mathbf{E}_{[:,1]}  & \mathbf{E}_{[:,2]}  & \dots & \mathbf{E}_{[:,J-1]}  & \mathbf{E}_{[:,J]}  & 0_{I \times K}  &  0_{I \times 1} & 0_{I \times 1}  & 0_{I \times 1}& \dots & 0_{I \times 1}  
    \end{bmatrix}
\end{align}
}%
is $\in \R^{\lp I(J + 1)\rp \times \lp 2K + 2J\rp}$. The first matrix $\boldsymbol{\Omega}$
is a $I(J + 1) \times I(J + 1)$ block diagonal matrix of diagonal matrices and by condition \ref{cond:full-model-E} is rank $I(J + 1)$. 
Then the product in equation \eqref{eq:gradient-modeled-missingness-cat-effects-comb} is rank $2K + 2J$ if the second matrix above
is rank $2K + 2J$ by Sylvester's rank inequality. 
The second matrix above is rank $2K + 2J$ if the following three sub-blocks are full column rank:
{\tiny
\begin{align} \label{eq:gradient-modeled-missingness-cat-effects-block-1}
\mathbf{L}_1  =   \begin{bmatrix}
    \lambda_1\mathbf{D}_1^\prime\mathbf{1}  & 0_{I \times 1}& \dots & 0_{I \times 1} \\
    0_{I \times 1} & \lambda_2\mathbf{D}^\prime_2\mathbf{1}  & \dots & 0_{I \times 1}    \\
    \vdots & \vdots & \ddots  & \vdots \\
    0_{I \times 1} & 0_{I \times 1} & \dots &  \lambda_{J-1} \mathbf{D}^\prime_{J-1} \mathbf{1}
    \end{bmatrix}
\end{align}
}%
and 
{\tiny
\begin{align} \label{eq:gradient-modeled-missingness-cat-effects-block-2}
\mathbf{L}_2  =    \begin{bmatrix}
    \lp \sum_{j} \lambda_j\mathbf{F}_j \rp \mathbf{Z} & \mathbf{E}_{[:,1]}  & \mathbf{E}_{[:,2]} & \dots & \mathbf{E}_{[:,J-1]}  & \mathbf{E}_{[:,J]}
    \end{bmatrix}
\end{align}
}%
and 
{\tiny
\begin{align} \label{eq:gradient-modeled-missingness-cat-effects-block-3}
\mathbf{L}_3  =    \begin{bmatrix}
    \lambda_{J}\mathbf{D}^\prime_J \mathbf{Z} & \lambda_{J}\mathbf{D}^\prime_J \mathbf{1} 
    \end{bmatrix}
\end{align}
}%
Sufficient conditions for $\mathbf{L}_1$ to be full column rank is 
\begin{enumerate}
    \item $I \geq (J-1)S$
    \item $\text{diag}(\mathbf{E}_{[:,j]})\mathbf{1}$ is full column rank for all $j$
    \item $\lambda_j > 0 \forall j \leq J$
    \item $\eta_{j} \in (-\infty, \infty) \forall j \leq J$
    \item $\beta_k, \gamma_k \in (-\infty, \infty) \forall k \leq J$
\end{enumerate}
Sufficient conditions for $\mathbf{L}_2$ to be full column rank is
\begin{enumerate}
    \item $I \geq K + J$
    \item $\forall i \in 1, \dots, I$ $\sum_{j=1}^J E_{ij} > 0$
    \item $\text{rank}\lp\begin{bmatrix} \mathbf{Z} & \mathbf{E}\end{bmatrix}\rp = J + K$
    \item 
    ${\rm rank}\lp \begin{bmatrix}
    \mathbf{F}_1\mathbf{Z} & \dots & \mathbf{F}_J\mathbf{Z} & \mathbf{E}_{[:,1]}  & \mathbf{E}_{[:,2]} & \dots & \mathbf{E}_{[:,J-1]}  & \mathbf{E}_{[:,J]}
    \end{bmatrix}\rp > J + K$ 
\end{enumerate}
Sufficient conditions for $\mathbf{L}_3$ to be full column rank is
\begin{enumerate}
    \item $I \geq K + S$
    \item $\mathbf{F}_J \begin{bmatrix} \mathbf{Z} & \mathbf{1}\end{bmatrix}$ 
    is full column rank
\end{enumerate}
We recognize matrix $\mathbf{L}_2$ as the same matrix as in \cref{lemma:model-block-rank}
and that \crefrange{cond:full-model-rank-E}{conds:full-model-rank-big-mat} are a superset of
the \crefrange{conds:rank-lemma-dim}{conds:rank-lemma-rank-big-mat} in \cref{conds:rank-lemma}. Then by \cref{lemma:model-block-rank}, matrix $\mathbf{L}_2$ is rank $J + K$.
\Crefrange{cond:full-model-rank-E}{conds:full-model-rank-big-mat} ensure that $\mathbf{L}_1$ and $\mathbf{L}_3$ are full rank as well, so $\text{rank}(\tilde{\Delta})$ is full rank.

Given that $\text{rank}(\tilde{\Delta}^\prime) = \text{rank}(\tilde{\Delta}) = \text{rank}(\Delta) = 2J + 2K$, the column dimension of $\Delta$, $\text{Cov}(\mathbf{f})$ is positive definite, and that the observational density, $\pi_\theta(\mathbf{f})$ is Poisson so $\frac{\partial}{\partial \theta_j} \int f_i \pi_\theta(\mathbf{x}) d\nu = \int f_i \frac{\partial}{\partial \theta_j} \pi_\theta(\mathbf{x}) d\nu$, we can apply lemma \ref{thm:pos-def-info}, which bounds the positive definiteness below by 0 for the Fisher information matrix $\mathcal{I}$.
In other words, the Fisher Information matrix is positive definite.
By Theorem 1 in \cite{rothenbergIdentificationParametricModels1971}, model \ref{model:full} is locally identifiable for any $(\mathbf{p}, \boldsymbol{\lambda}, \boldsymbol{\beta}, \boldsymbol{\gamma}) \in ((0, 1)^J \times (0,\infty)^J \times \R^K \times \R^K)$, where $((a,b)^n)$ is the $n$-fold Cartesian product of the set $(a,b)$.
\end{proof}

\section{Further simulation study results}

\subsection{Root mean squared error plots}\label{app:mse-plots}

\begin{figure}[H]
    \centering
    \includegraphics[scale=0.5]{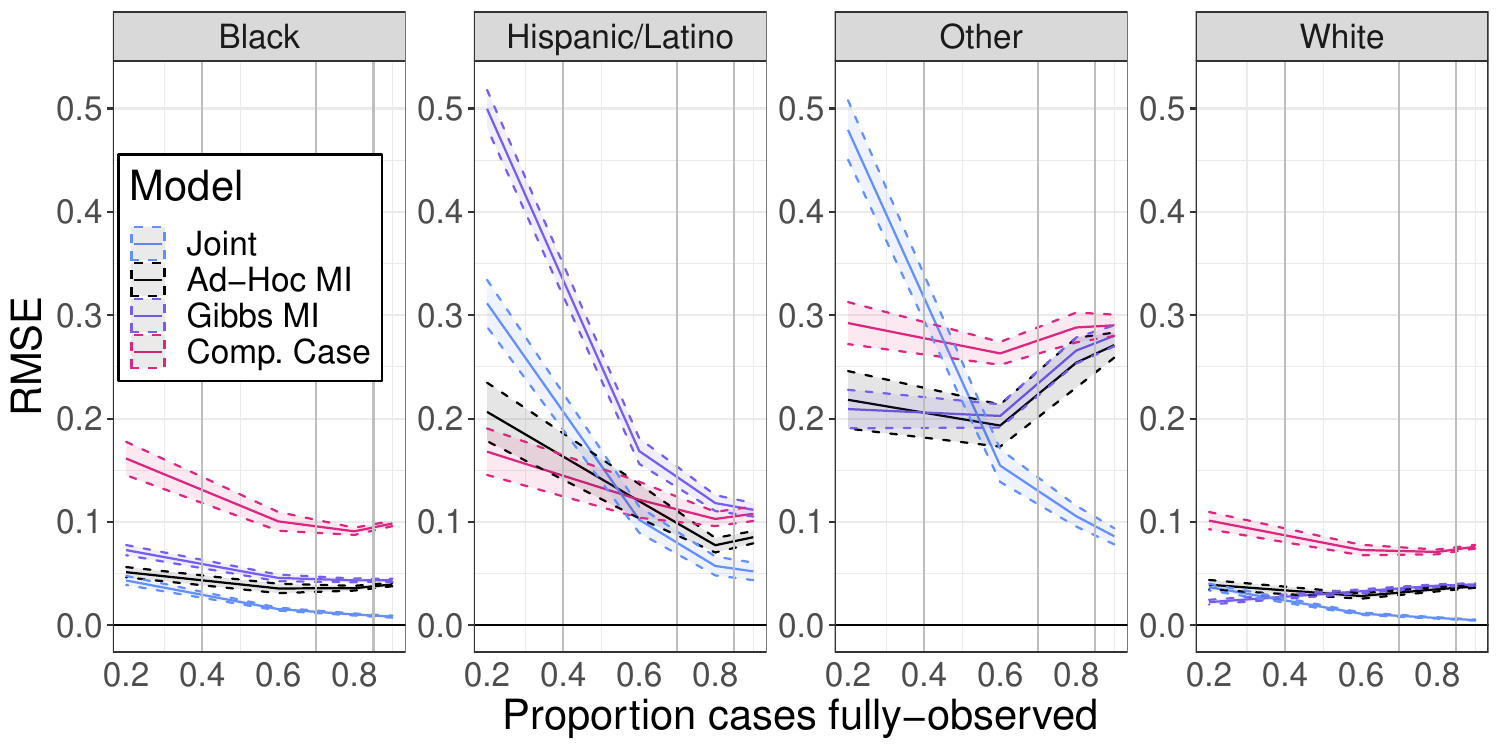}
    \caption{Root mean squared error across simulated datasets for the standardized incidence ratio, or $\text{SIR}_j$ for Blacks, Hispanic/Latinos, Others, and Whites plotted against the proportion of cases observed with race data. The blue color corresponds to the joint model in equation \eqref{model:full-geo}, while the red color corresponds to a the model defined in equation \eqref{model:comp-case}, or a  complete-case analysis. Smaller magnitude is better.}
    \label{fig:mse-sir}
\end{figure}

\begin{figure}[H]
    \centering
    \includegraphics[scale=0.5]{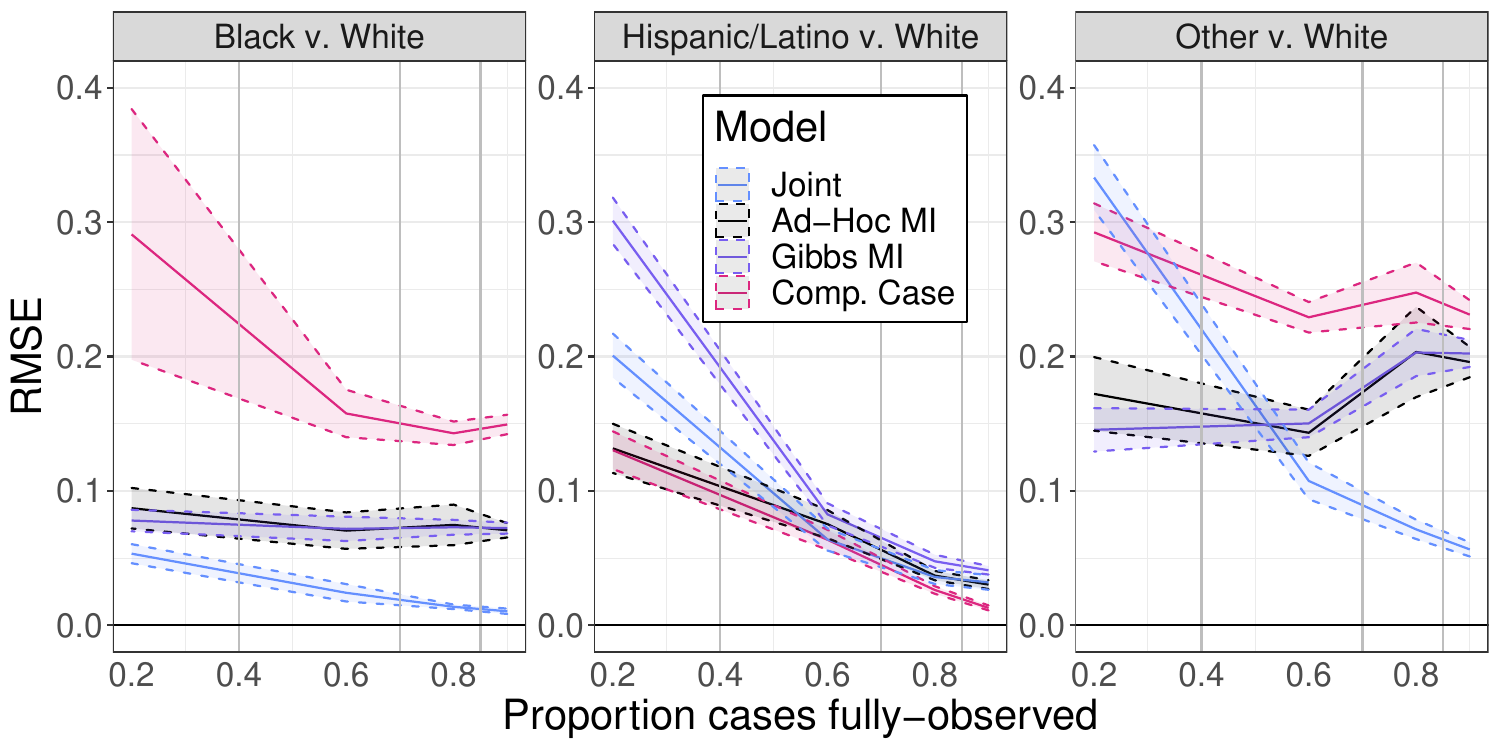}
    \caption{Root mean squared error across simulated datasets for the relative risk ratio, or $\mathbbm{I}_j / \mathbbm{I}_J$ for Blacks, Hispanic/Latinos, Others relative to Whites plotted against the proportion of cases observed with race data. The blue color corresponds to the joint model in equation \eqref{model:full-geo}, while the red color corresponds to a the model defined in equation \eqref{model:comp-case}, or a  complete-case analysis. Smaller magnitude is better.}
    \label{fig:mse-rr}
\end{figure}

\begin{figure}[H]
    \centering
    \includegraphics[scale=0.5]{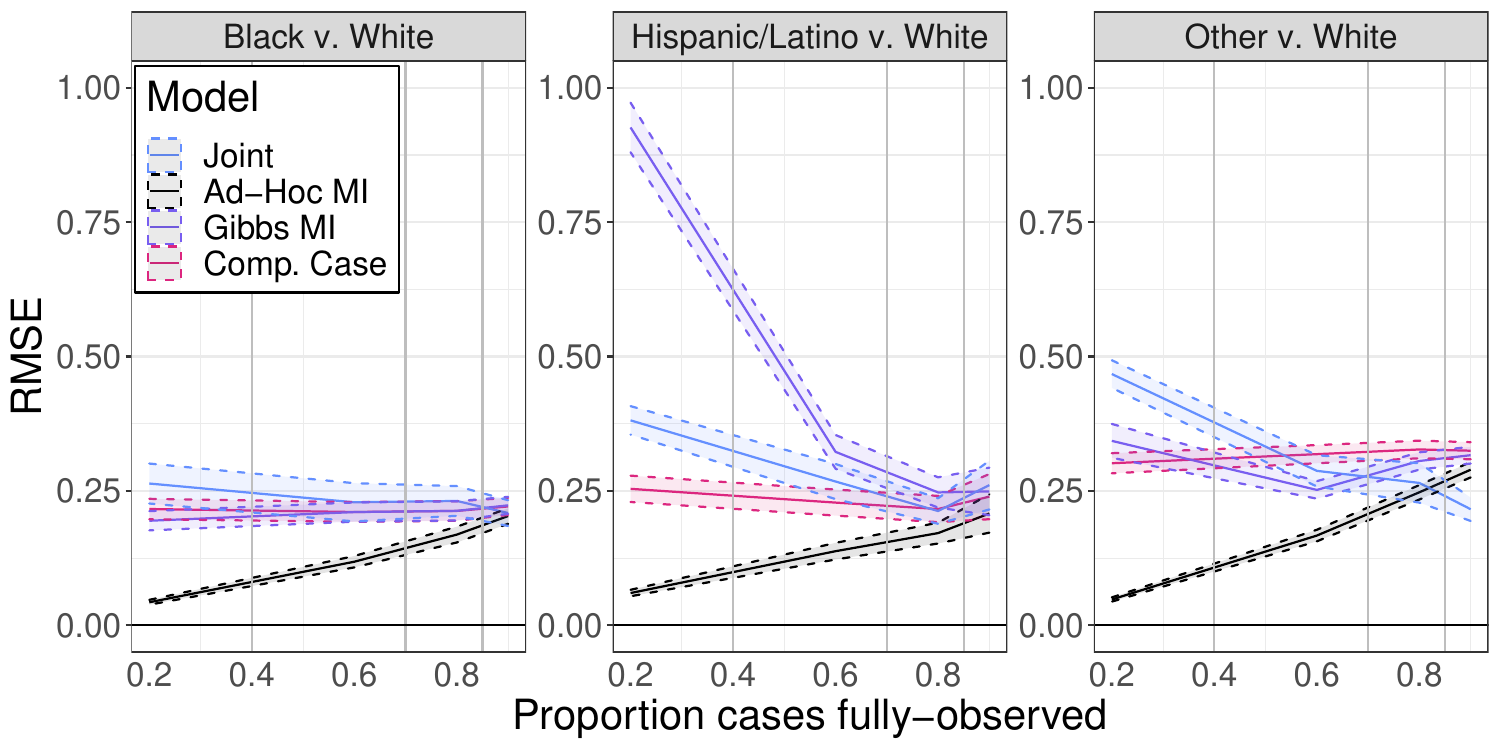}
    \caption{Mean squared error across simulated datasets for the population relative risk ratio, or $\exp((\boldsymbol{\alpha}_\lambda)_j - (\boldsymbol{\alpha}_\lambda)_J)$ for Blacks, Hispanic/Latinos, Others relative to Whites plotted against the proportion of cases observed with race data. The blue color corresponds to the joint model in equation \eqref{model:full-geo}, while the red color corresponds to a the model defined in equation \eqref{model:comp-case}, or a  complete-case analysis. Smaller magnitude is better.}
    \label{fig:mse-exp-alpha}
\end{figure}

\subsection{80\% posterior interval coverage}\label{app:post-intervals}

\begin{table}[ht]
\caption{Table shows 80\% posterior credible interval coverages and lengths for estimands of interest from the simulation study. Coverage proportion is calculated across 200 simulated datasets for each model/simulation scenario. Column headers for percentages (e.g. 20\%) indicate the missing-data simulation scenario which corresponds to the statistic calculated in the table column; the simulation scenario corresponds to the proportion of cases observed with completely observed race covariates.}
\label{table:80pct-intervals}
\tabcolsep=0.10cm
\centering
\resizebox{\textwidth}{!}{%
\begin{tabular}{@{}llcccccccc@{}}
  \hline
  &  & \multicolumn{4}{c}{80\% interval coverage} & \multicolumn{4}{c}{80\% mean interval length} \\ 
 Parameter & Model  & 20\% & 60\% & 80\% & 90\% & 20\% & 60\% & 80\% & 90\% \\ 
   \hline
  $\exp\big((\boldsymbol{\alpha}_\lambda)_\text{Blacks}$ & Complete Case & 0.70 & 0.70 & 0.68 & 0.60 & 0.51 & 0.47 & 0.47 & 0.44 \\ 
$ \quad \quad - (\boldsymbol{\alpha}_\lambda)_\text{Whites}\big)$   & Joint & 0.78 & 0.81 & 0.81 & 0.78 & 0.62 & 0.55 & 0.55 & 0.53 \\ 
      & MI-Ad-hoc& 1.00 & 0.88 & 0.77 & 0.66 & 0.41 & 0.40 & 0.43 & 0.42 \\ 
  & MI-Gibbs & 0.75 & 0.70 & 0.67 & 0.61 & 0.52 & 0.46 & 0.46 & 0.44 \\ [6pt]
  $\exp\big((\boldsymbol{\alpha}_\lambda)_\text{Hispanics/Latinos}$ & Complete Case & 0.77 & 0.80 & 0.83 & 0.77 & 0.63 & 0.56 & 0.56 & 0.53 \\ 
$\quad \quad - (\boldsymbol{\alpha}_\lambda)_\text{Whites}\big)$   & Joint & 0.56 & 0.77 & 0.84 & 0.78 & 0.68 & 0.61 & 0.57 & 0.56 \\ 
      & MI-Ad-hoc& 0.99 & 0.91 & 0.89 & 0.82 & 0.42 & 0.44 & 0.49 & 0.49 \\ 
   & MI-Gibbs & 0.07 & 0.68 & 0.82 & 0.78 & 1.13 & 0.65 & 0.58 & 0.55 \\ [6pt]
  $\exp\big((\boldsymbol{\alpha}_\lambda)_\text{Others}$ & Complete Case & 0.42 & 0.33 & 0.27 & 0.28 & 0.44 & 0.38 & 0.37 & 0.36 \\ 
 $\quad \quad - (\boldsymbol{\alpha}_\lambda)_\text{Whites}\big)$  & Joint & 0.37 & 0.73 & 0.77 & 0.82 & 0.63 & 0.67 & 0.60 & 0.58 \\ 
      & MI-Ad-hoc& 1.00 & 0.62 & 0.41 & 0.30 & 0.40 & 0.36 & 0.36 & 0.35 \\ 
   & MI-Gibbs & 0.70 & 0.48 & 0.32 & 0.29 & 0.71 & 0.42 & 0.39 & 0.37 \\ [6pt]
  $\mathbbm{SI}_\text{Blacks}$ & Complete Case & 0.06 & 0.04 & 0.01 & 0.00 & 0.03 & 0.02 & 0.01 & 0.01 \\ 
   & Joint & 0.67 & 0.77 & 0.80 & 0.78 & 0.09 & 0.04 & 0.03 & 0.02 \\ 
      & MI-Ad-hoc& 0.09 & 0.20 & 0.06 & 0.00 & 0.01 & 0.01 & 0.01 & 0.01 \\ 
   & MI-Gibbs & 0.03 & 0.01 & 0.00 & 0.00 & 0.02 & 0.01 & 0.01 & 0.01 \\ [6pt]
  $\mathbbm{SI}_\text{Hispanics/Latinos}$ & Complete Case & 0.34 & 0.24 & 0.12 & 0.01 & 0.13 & 0.07 & 0.06 & 0.06 \\ 
   & Joint & 0.46 & 0.74 & 0.81 & 0.70 & 0.47 & 0.23 & 0.14 & 0.10 \\ 
     & MI-Ad-hoc& 0.12 & 0.23 & 0.27 & 0.06 & 0.07 & 0.06 & 0.06 & 0.06 \\ 
   & MI-Gibbs & 0.00 & 0.05 & 0.01 & 0.01 & 0.16 & 0.08 & 0.06 & 0.06 \\ [6pt]
  $\mathbbm{SI}_\text{Others}$ & Complete Case & 0.03 & 0.00 & 0.00 & 0.00 & 0.11 & 0.06 & 0.05 & 0.05 \\ 
   & Joint & 0.27 & 0.70 & 0.76 & 0.77 & 0.49 & 0.36 & 0.25 & 0.20 \\ 
      & MI-Ad-hoc& 0.17 & 0.14 & 0.01 & 0.00 & 0.07 & 0.06 & 0.05 & 0.05 \\ 
  & MI-Gibbs & 0.20 & 0.01 & 0.00 & 0.00 & 0.15 & 0.06 & 0.05 & 0.05 \\ [6pt]
  $\mathbbm{SI}_\text{Whites}$ & Complete Case & 0.04 & 0.01 & 0.00 & 0.00 & 0.02 & 0.01 & 0.01 & 0.01 \\ 
   & Joint & 0.50 & 0.80 & 0.82 & 0.84 & 0.06 & 0.03 & 0.02 & 0.01 \\ 
      & MI-Ad-hoc& 0.14 & 0.17 & 0.00 & 0.00 & 0.01 & 0.01 & 0.01 & 0.01 \\ 
  & MI-Gibbs & 0.21 & 0.01 & 0.00 & 0.00 & 0.02 & 0.01 & 0.01 & 0.01 \\ [6pt]
  $\mathbbm{I}_\text{Blacks}/\mathbbm{I}_\text{Whites}$ & Complete Case & 0.06 & 0.01 & 0.00 & 0.00 & 0.04 & 0.02 & 0.02 & 0.02 \\ 
   & Joint & 0.73 & 0.79 & 0.81 & 0.81 & 0.12 & 0.05 & 0.04 & 0.03 \\ 
     & MI-Ad-hoc& 0.10 & 0.20 & 0.02 & 0.00 & 0.02 & 0.02 & 0.02 & 0.02 \\ 
   & MI-Gibbs & 0.12 & 0.01 & 0.00 & 0.00 & 0.03 & 0.02 & 0.02 & 0.02 \\ [6pt]
  $\mathbbm{I}_\text{Hispanics/Latinos}/$ & Complete Case & 0.17 & 0.18 & 0.57 & 0.83 & 0.07 & 0.04 & 0.04 & 0.03 \\ 
$\mathbbm{I}_\text{Whites}$   & Joint & 0.43 & 0.73 & 0.81 & 0.73 & 0.28 & 0.14 & 0.09 & 0.06 \\ 
     & MI-Ad-hoc& 0.12 & 0.21 & 0.32 & 0.41 & 0.04 & 0.04 & 0.04 & 0.03 \\ 
  & MI-Gibbs & 0.00 & 0.14 & 0.24 & 0.17 & 0.10 & 0.05 & 0.04 & 0.04 \\ [6pt]
  $\mathbbm{I}_\text{Others}/\mathbbm{I}_\text{Whites}$ & Complete Case & 0.00 & 0.00 & 0.00 & 0.00 & 0.06 & 0.03 & 0.03 & 0.03 \\ 
   & Joint & 0.26 & 0.72 & 0.74 & 0.76 & 0.32 & 0.24 & 0.17 & 0.13 \\ 
  & MI-Ad-hoc& 0.14 & 0.11 & 0.01 & 0.00 & 0.05 & 0.04 & 0.03 & 0.03 \\ 
  & MI-Gibbs & 0.21 & 0.00 & 0.00 & 0.00 & 0.10 & 0.04 & 0.03 & 0.03 \\ [6pt]
   \hline
\end{tabular}}
\end{table}

\begin{table}[ht]
\caption{Table shows 50\% posterior credible interval coverages and lengths for estimands of interest from the simulation study. Coverage proportion is calculated across 200 simulated datasets for each model/simulation scenario. Column headers for percentages (e.g. 20\%) indicate the missing-data simulation scenario which corresponds to the statistic calculated in the table column; the simulation scenario corresponds to the proportion of cases observed with completely observed race covariates.}
\label{table:full-50pct-intervals}
\tabcolsep=0.10cm
\centering
\resizebox{\textwidth}{!}{%
\begin{tabular}{@{}llcccccccc@{}}
  \hline
  &  & \multicolumn{4}{c}{50\% interval coverage} & \multicolumn{4}{c}{50\% mean interval length} \\ 
 Parameter & Model  & 20\% & 60\% & 80\% & 90\% & 20\% & 60\% & 80\% & 90\% \\ 
    \hline
$\exp\big((\boldsymbol{\alpha}_\lambda)_\text{Blacks}$ & Complete Case & 0.39 & 0.35 & 0.41 & 0.32 & 0.27 & 0.24 & 0.24 & 0.23 \\ 
 $ \quad \quad - (\boldsymbol{\alpha}_\lambda)_\text{Whites}\big)$  & Joint & 0.52 & 0.53 & 0.49 & 0.47 & 0.32 & 0.28 & 0.29 & 0.27 \\ 
   & MI-Ad-hoc& 0.96 & 0.56 & 0.47 & 0.30 & 0.21 & 0.21 & 0.22 & 0.22 \\ 
   & MI-Gibbs & 0.48 & 0.38 & 0.38 & 0.29 & 0.27 & 0.24 & 0.24 & 0.23 \\ [6pt]
  $\exp\big((\boldsymbol{\alpha}_\lambda)_\text{Hispanics/Latinos}$ & Complete Case & 0.48 & 0.47 & 0.47 & 0.44 & 0.33 & 0.29 & 0.29 & 0.28 \\ 
 $\quad \quad - (\boldsymbol{\alpha}_\lambda)_\text{Whites}\big)$  & Joint & 0.27 & 0.48 & 0.47 & 0.43 & 0.35 & 0.31 & 0.30 & 0.29 \\ 
   & MI-Ad-hoc& 0.96 & 0.62 & 0.53 & 0.47 & 0.22 & 0.23 & 0.25 & 0.25 \\ 
   & MI-Gibbs & 0.01 & 0.39 & 0.46 & 0.45 & 0.58 & 0.33 & 0.30 & 0.28 \\ [6pt]
  $\exp\big((\boldsymbol{\alpha}_\lambda)_\text{Others}$ & Complete Case & 0.20 & 0.10 & 0.12 & 0.09 & 0.23 & 0.20 & 0.19 & 0.19 \\ 
 $ \quad \quad -(\boldsymbol{\alpha}_\lambda)_\text{Whites}\big)$  & Joint & 0.17 & 0.41 & 0.47 & 0.51 & 0.32 & 0.35 & 0.31 & 0.30 \\ 
  & MI-Ad-hoc& 0.91 & 0.29 & 0.15 & 0.10 & 0.21 & 0.18 & 0.19 & 0.18 \\ 
   & MI-Gibbs & 0.38 & 0.23 & 0.12 & 0.10 & 0.36 & 0.22 & 0.20 & 0.19 \\ [6pt]
  $\mathbbm{SI}_\text{Blacks}$ & Complete Case & 0.04 & 0.02 & 0.00 & 0.00 & 0.01 & 0.01 & 0.01 & 0.01 \\ 
   & Joint & 0.39 & 0.49 & 0.48 & 0.49 & 0.04 & 0.02 & 0.01 & 0.01 \\ 
   & MI-Ad-hoc& 0.04 & 0.09 & 0.03 & 0.00 & 0.01 & 0.01 & 0.01 & 0.01 \\ 
   & MI-Gibbs & 0.03 & 0.01 & 0.00 & 0.00 & 0.01 & 0.01 & 0.01 & 0.01 \\ [6pt]
  $\mathbbm{SI}_\text{Hispanics/Latinos}$ & Complete Case & 0.17 & 0.12 & 0.05 & 0.00 & 0.07 & 0.04 & 0.03 & 0.03 \\ 
   & Joint & 0.27 & 0.47 & 0.51 & 0.41 & 0.24 & 0.12 & 0.07 & 0.05 \\ 
   & MI-Ad-hoc& 0.07 & 0.15 & 0.18 & 0.01 & 0.04 & 0.03 & 0.03 & 0.03 \\ 
   & MI-Gibbs & 0.00 & 0.01 & 0.01 & 0.00 & 0.08 & 0.04 & 0.03 & 0.03 \\ [6pt]
 $\mathbbm{SI}_\text{Others}$ & Complete Case & 0.01 & 0.00 & 0.00 & 0.00 & 0.06 & 0.03 & 0.03 & 0.02 \\ 
  & Joint & 0.12 & 0.45 & 0.48 & 0.41 & 0.25 & 0.18 & 0.13 & 0.10 \\ 
   & MI-Ad-hoc& 0.08 & 0.07 & 0.01 & 0.00 & 0.04 & 0.03 & 0.03 & 0.03 \\ 
   & MI-Gibbs & 0.09 & 0.00 & 0.00 & 0.00 & 0.08 & 0.03 & 0.03 & 0.02 \\ [6pt]
  $\mathbbm{SI}_\text{Whites}$ & Complete Case & 0.03 & 0.01 & 0.00 & 0.00 & 0.01 & 0.01 & 0.00 & 0.00 \\ 
   & Joint & 0.28 & 0.51 & 0.54 & 0.51 & 0.03 & 0.02 & 0.01 & 0.01 \\ 
   & MI-Ad-hoc& 0.07 & 0.08 & 0.00 & 0.00 & 0.01 & 0.00 & 0.00 & 0.00 \\ 
   & MI-Gibbs & 0.08 & 0.00 & 0.00 & 0.00 & 0.01 & 0.01 & 0.00 & 0.00 \\ [6pt]
  $\mathbbm{I}_\text{Blacks}/\mathbbm{I}_\text{Whites}$ & Complete Case & 0.03 & 0.01 & 0.00 & 0.00 & 0.02 & 0.01 & 0.01 & 0.01 \\ 
   & Joint & 0.48 & 0.54 & 0.52 & 0.48 & 0.06 & 0.03 & 0.02 & 0.01 \\ 
   & MI-Ad-hoc& 0.04 & 0.10 & 0.01 & 0.00 & 0.01 & 0.01 & 0.01 & 0.01 \\ 
   & MI-Gibbs & 0.07 & 0.01 & 0.00 & 0.00 & 0.02 & 0.01 & 0.01 & 0.01 \\ [6pt]
  $\mathbbm{I}_\text{Hispanics/Latinos}/$ & Complete Case & 0.09 & 0.08 & 0.33 & 0.53 & 0.03 & 0.02 & 0.02 & 0.02 \\ 
$\mathbbm{I}_\text{Whites}$  & Joint & 0.24 & 0.46 & 0.51 & 0.45 & 0.14 & 0.07 & 0.05 & 0.03 \\ 
    & MI-Ad-hoc& 0.09 & 0.12 & 0.17 & 0.24 & 0.02 & 0.02 & 0.02 & 0.02 \\ 
   & MI-Gibbs & 0.00 & 0.09 & 0.07 & 0.05 & 0.05 & 0.02 & 0.02 & 0.02 \\ [6pt]
  $\mathbbm{I}_\text{Others}/\mathbbm{I}_\text{Whites}$ & Complete Case & 0.00 & 0.00 & 0.00 & 0.00 & 0.03 & 0.02 & 0.02 & 0.01 \\ 
   & Joint & 0.12 & 0.43 & 0.47 & 0.41 & 0.16 & 0.12 & 0.09 & 0.07 \\ 
     & MI-Ad-hoc& 0.09 & 0.06 & 0.00 & 0.00 & 0.02 & 0.02 & 0.02 & 0.02 \\ 
   & MI-Gibbs & 0.09 & 0.00 & 0.00 & 0.00 & 0.05 & 0.02 & 0.02 & 0.02 \\ 
   \hline
\end{tabular}}
\end{table}

\section{Prior sensitivity graphs}\label{app:prior-sens-graphs}
Graphs to support conclusions in \ref{sec:sim-study-sens}
\begin{figure}[H]
    \centering
    \includegraphics[width=0.49\textwidth]{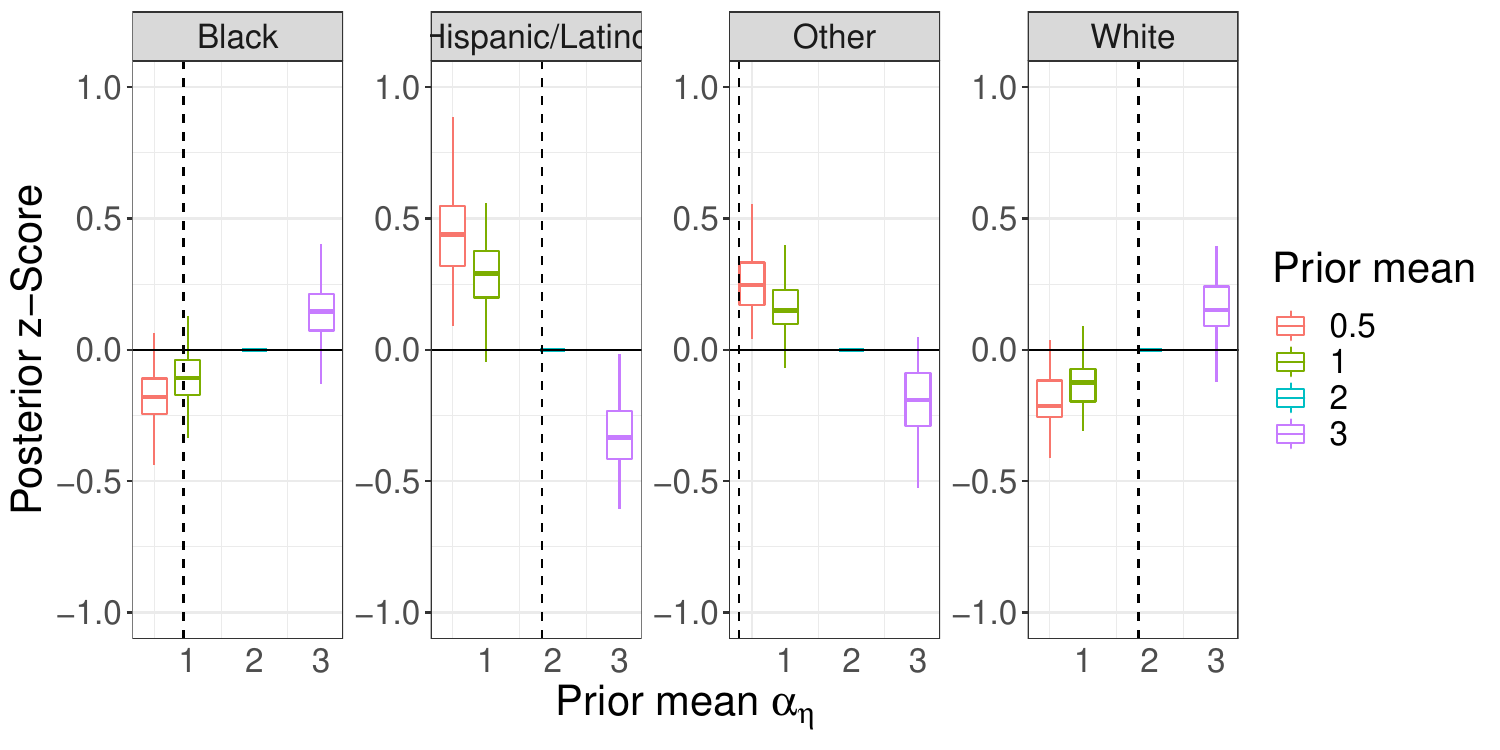}\hfill
    \includegraphics[width=0.49\textwidth]{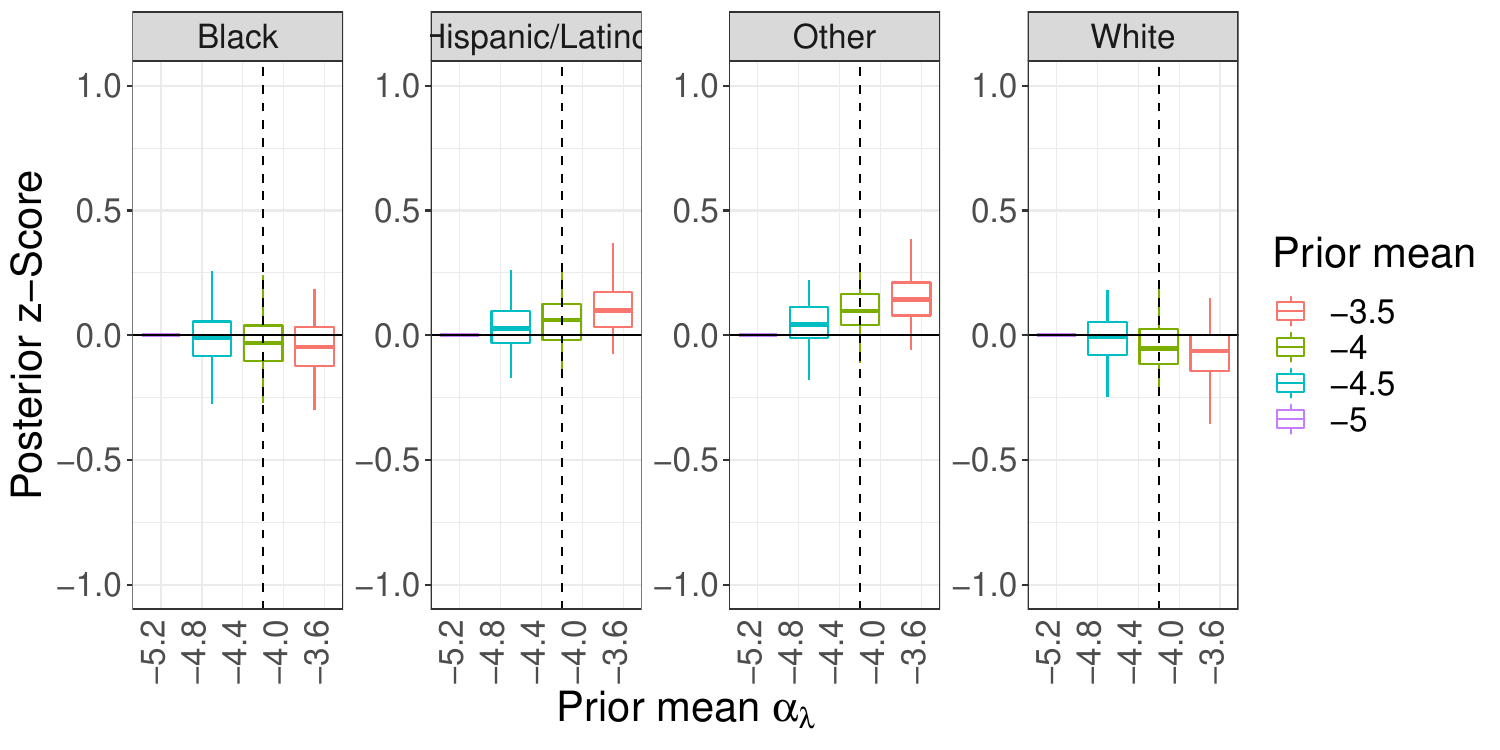}
    \\[\smallskipamount]
    \includegraphics[width=0.49\textwidth]{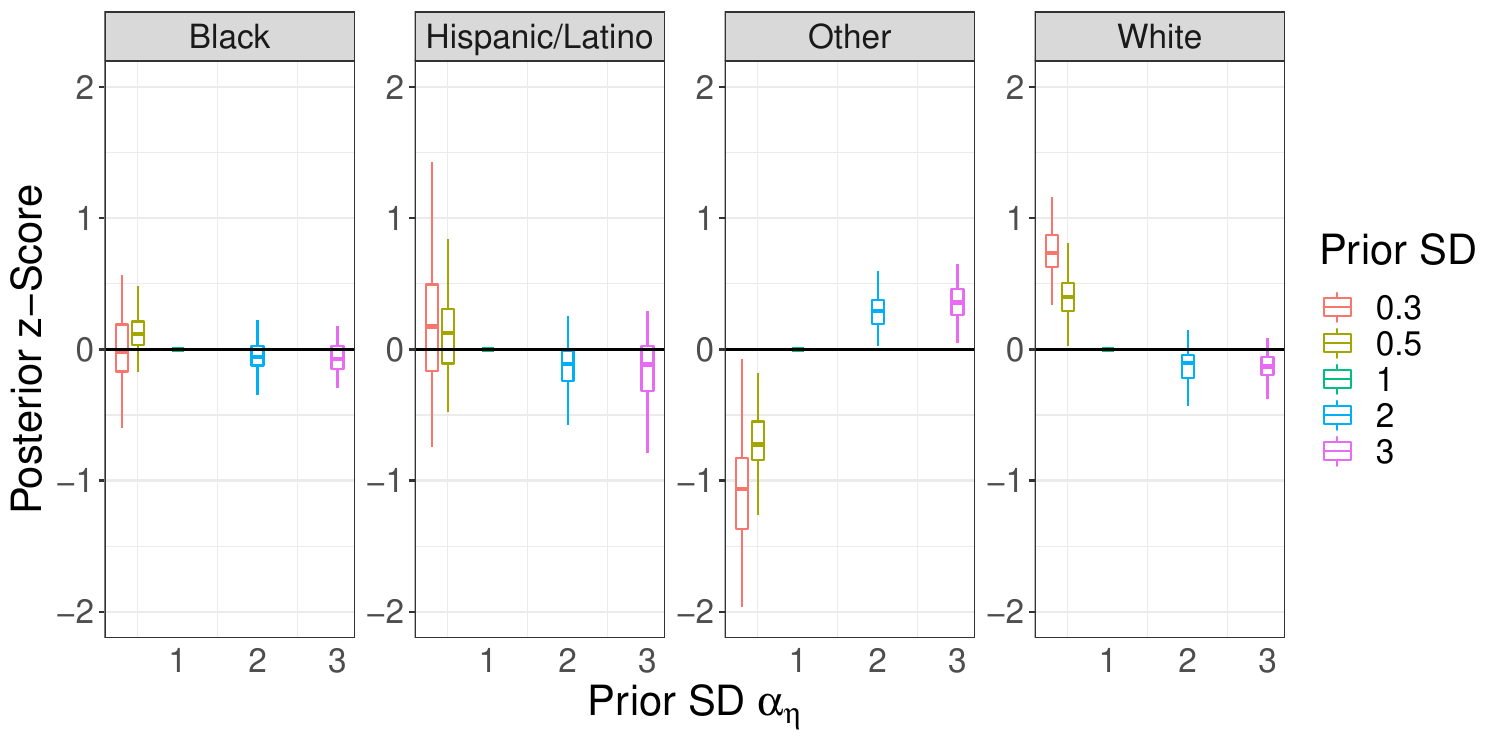}\hfill
    \includegraphics[width=0.49\textwidth]{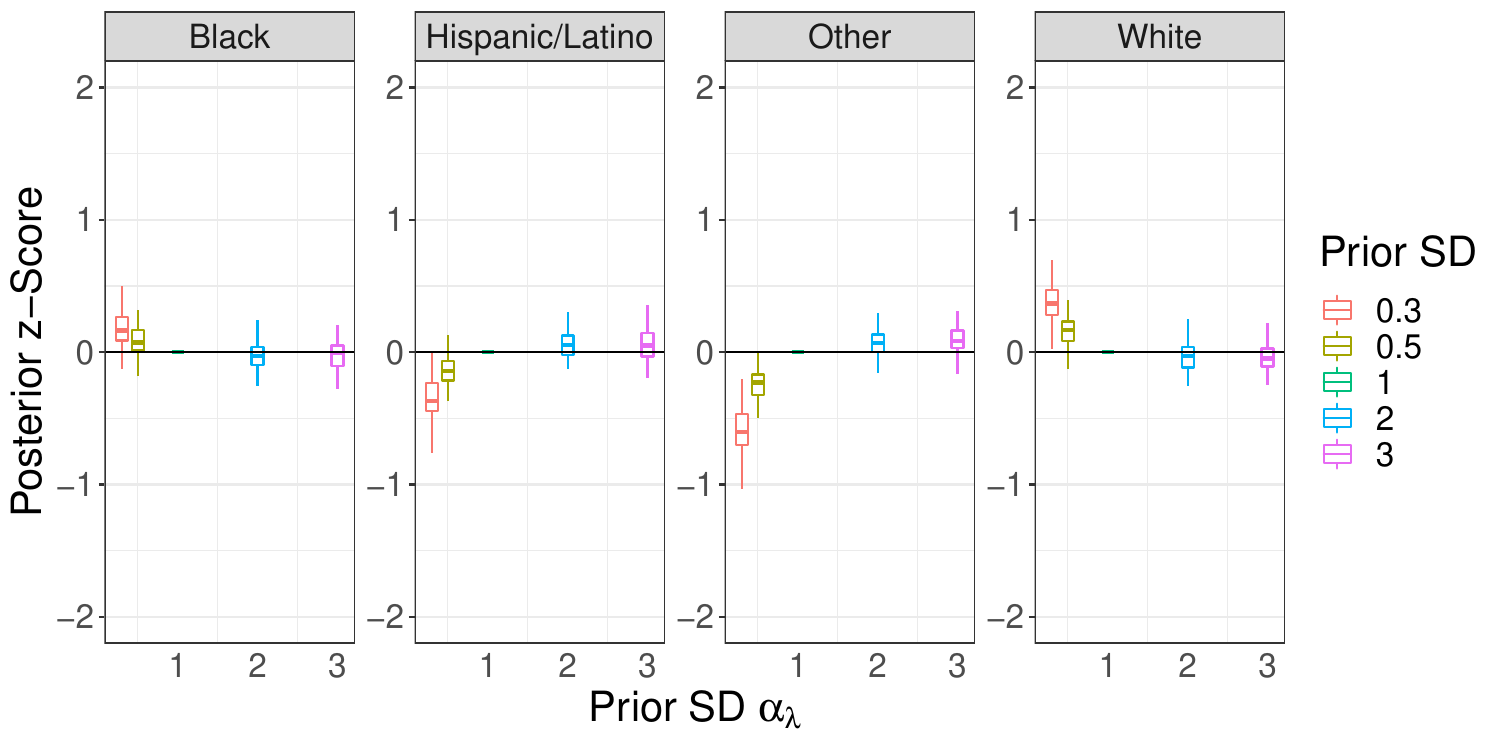}\hfill
    \caption{Graphs above show box plots of posterior-standard-deviation-scaled differences in posterior mean incidences with respect to a baseline prior for various priors over population hyperparameters, or $(\mathbb{E}_{\pi_{\mathrm{a}}(\boldsymbol{\theta} | \mathrm{Data})}[g(\boldsymbol{\theta})] - \mathbb{E}_{\pi_{\mathrm{b}}(\boldsymbol{\theta} | \mathrm{Data})}[g(\boldsymbol{\theta})])/{\sqrt{\mathrm{Var}_{\pi_{\mathrm{b}}(\boldsymbol{\theta} | \mathrm{Data})}(g(\boldsymbol{\theta}))}}$. The graphs quantify how sensitive posterior mean incidence for each race/ethnicity group is to priors over population parameters $\boldsymbol{\alpha}_{\boldsymbol{\lambda}}$ and $\boldsymbol{\alpha}_{\boldsymbol{\eta}}$.}
    \label{fig:prior-sens-z-score-alpha}
\end{figure}
\begin{figure}[H]
    \centering
    \includegraphics[width=0.49\textwidth]{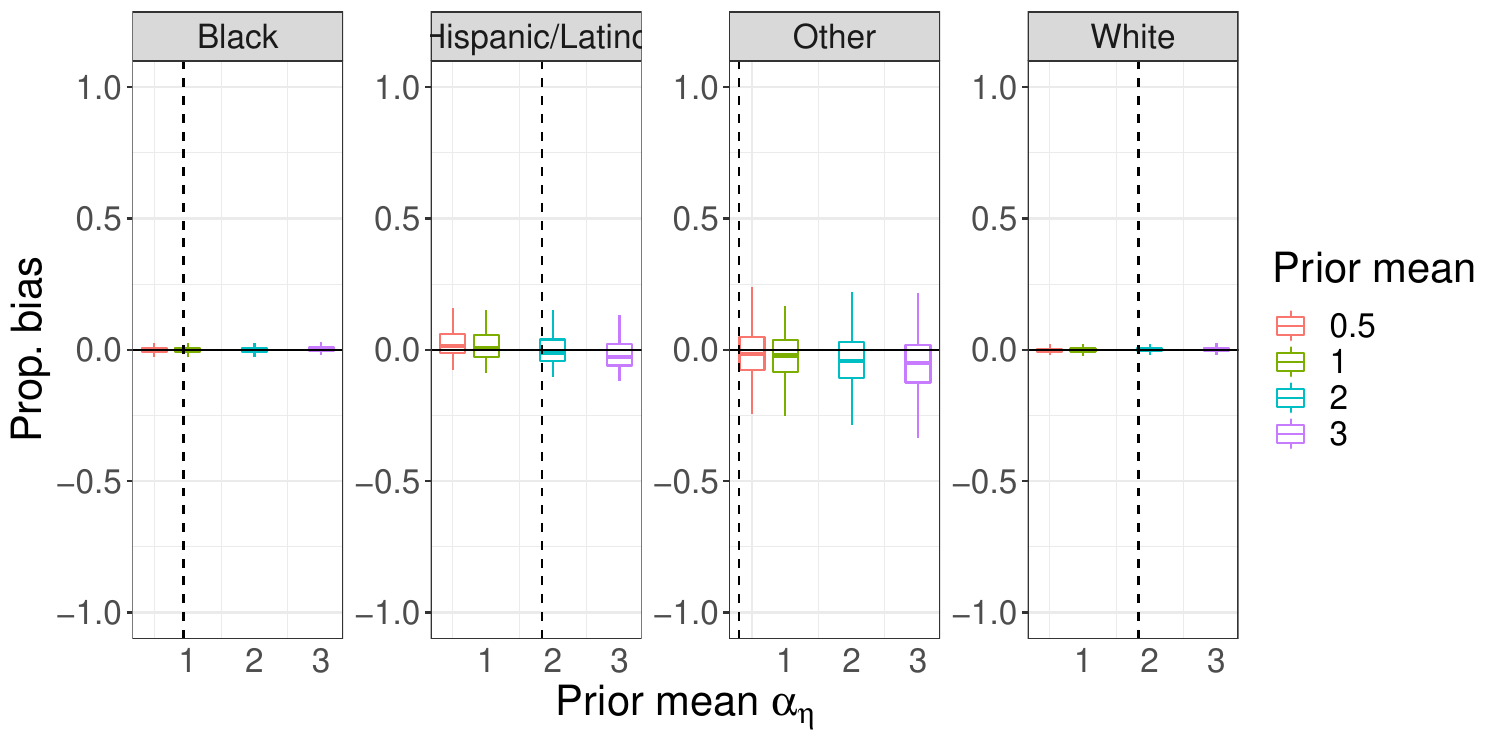}\hfill
    \includegraphics[width=0.49\textwidth]{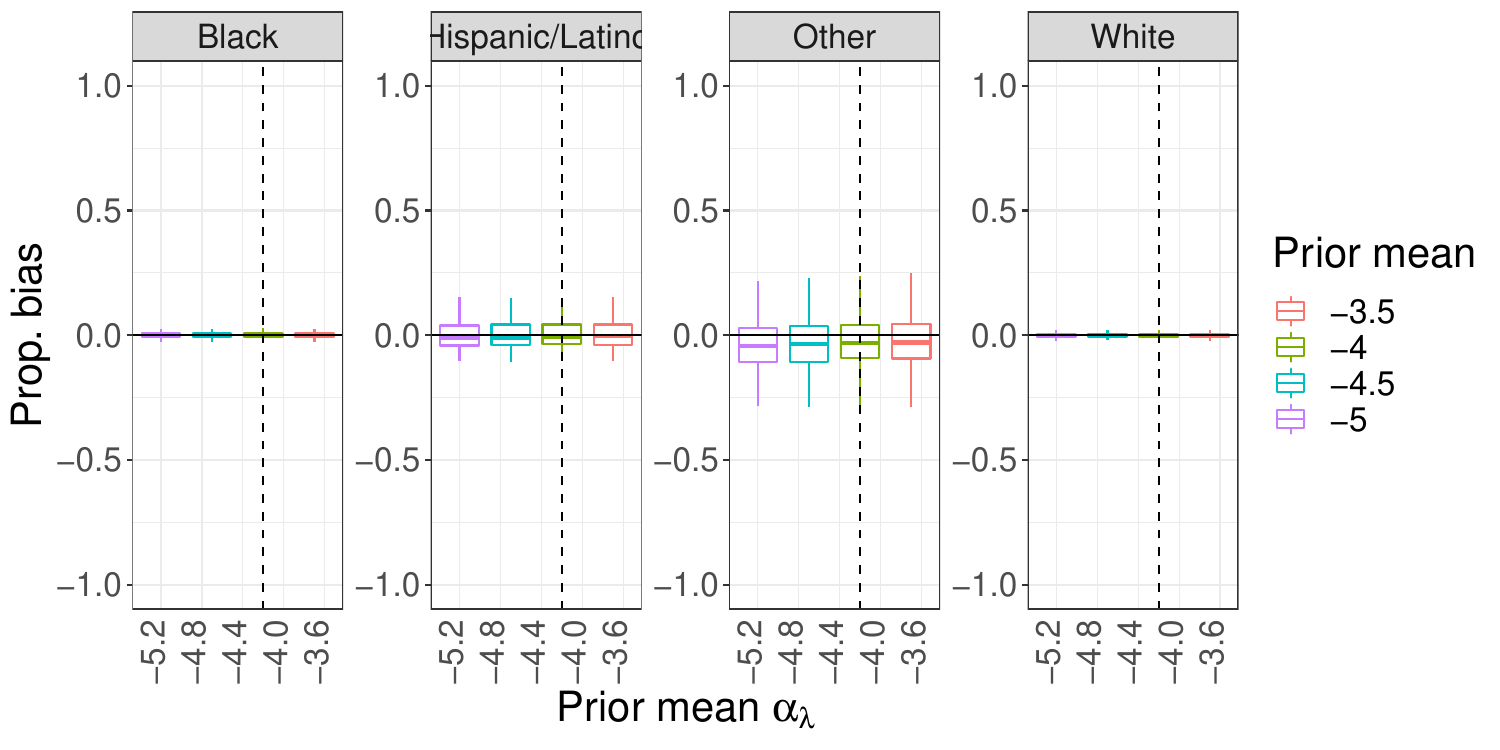}
    \\[\smallskipamount]
    \includegraphics[width=0.49\textwidth]{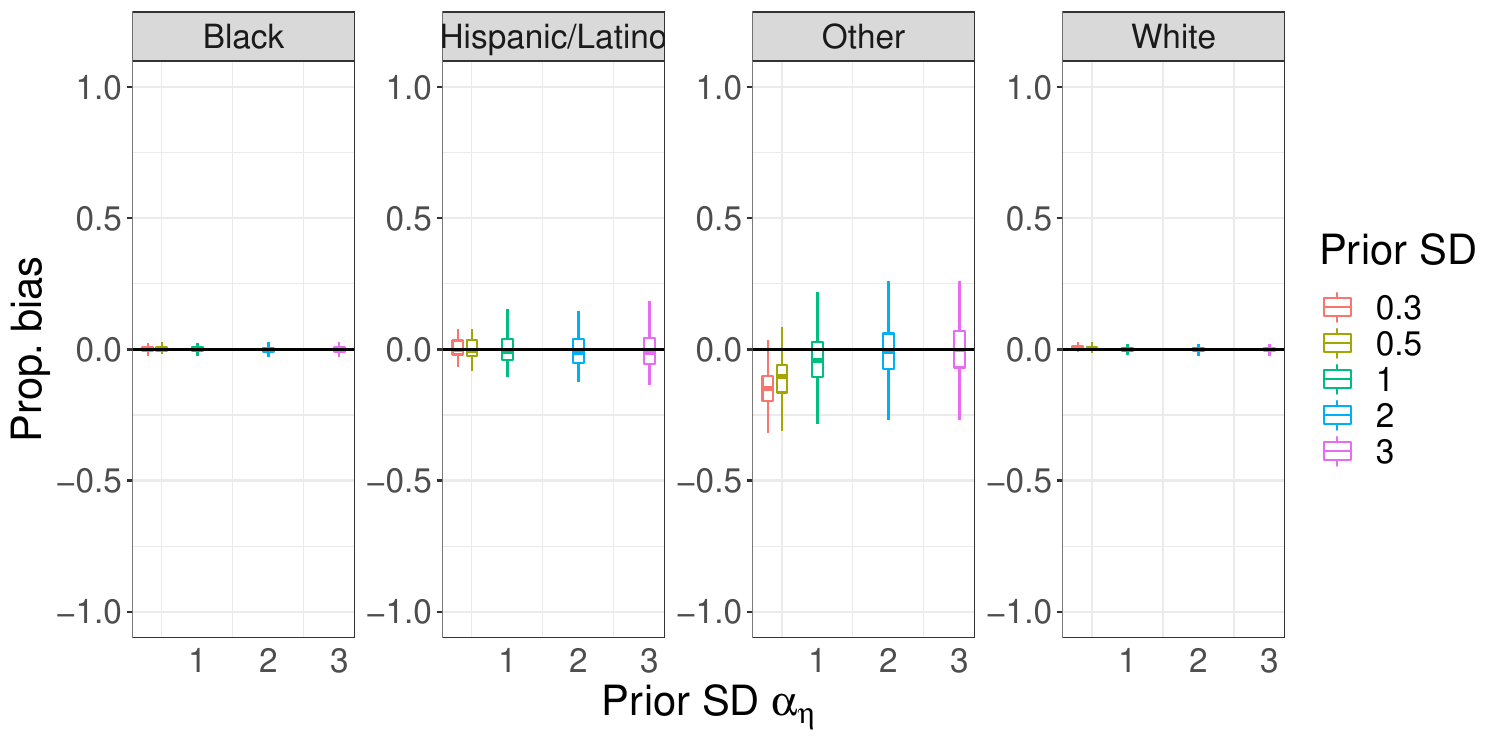}\hfill
    \includegraphics[width=0.49\textwidth]{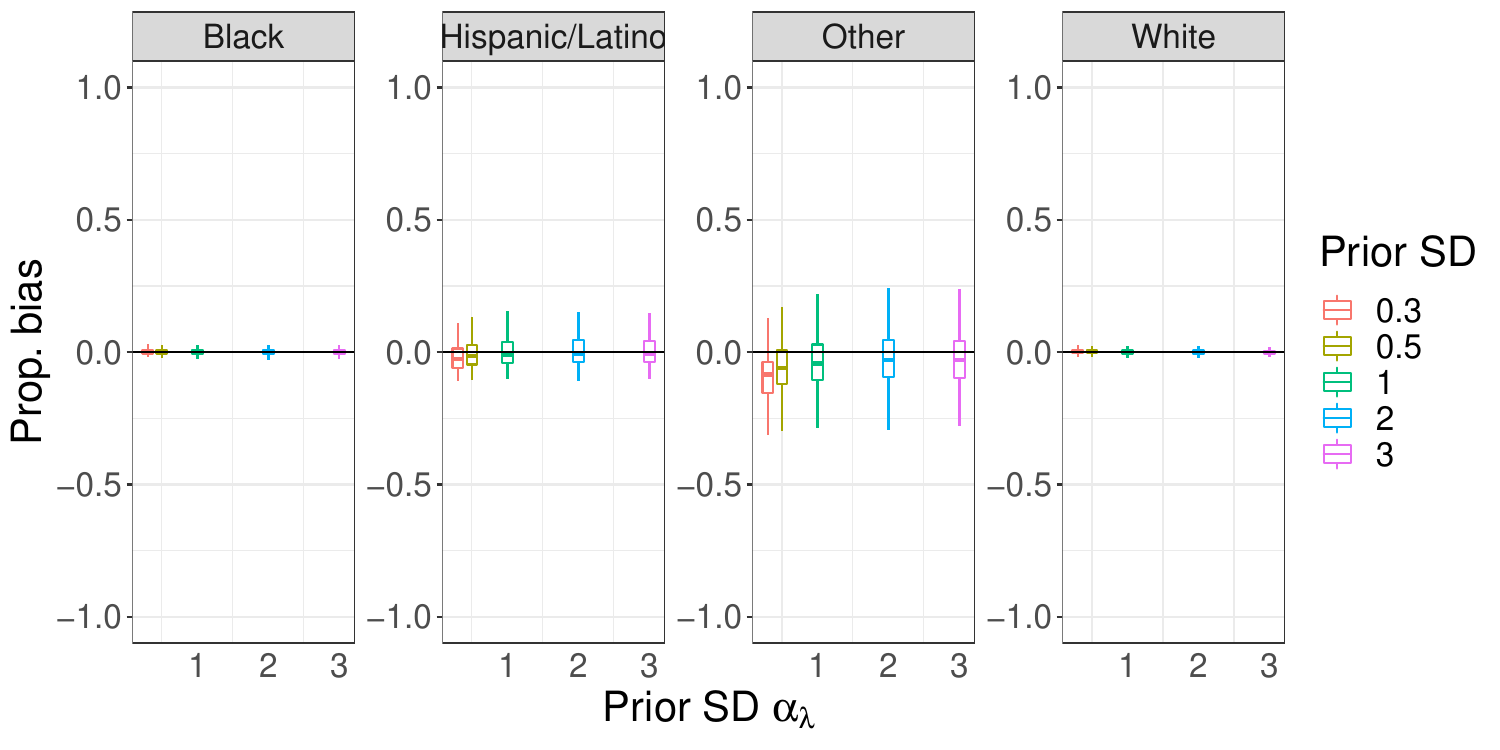}\hfill
    \caption{Graphs above show box plots of scaled biases in the posterior mean for true incidences $g(\theta^\dagger)$, or $(\mathbb{E}_{\pi_{\mathrm{a}}(\boldsymbol{\theta} | \mathrm{Data})}[g(\boldsymbol{\theta})]-g(\boldsymbol{\theta}^\dagger))/{g(\boldsymbol{\theta}^\dagger)}$.
    The graphs quantify how priors over population parameters $\boldsymbol{\alpha}_{\boldsymbol{\lambda}}$ and $\boldsymbol{\alpha}_{\boldsymbol{\eta}}$ influence the bias of the posterior mean estimator.}
    \label{fig:prior-sens-bias-alpha}
\end{figure}
\begin{figure}[H]
    \centering
    \includegraphics[width=0.49\textwidth]{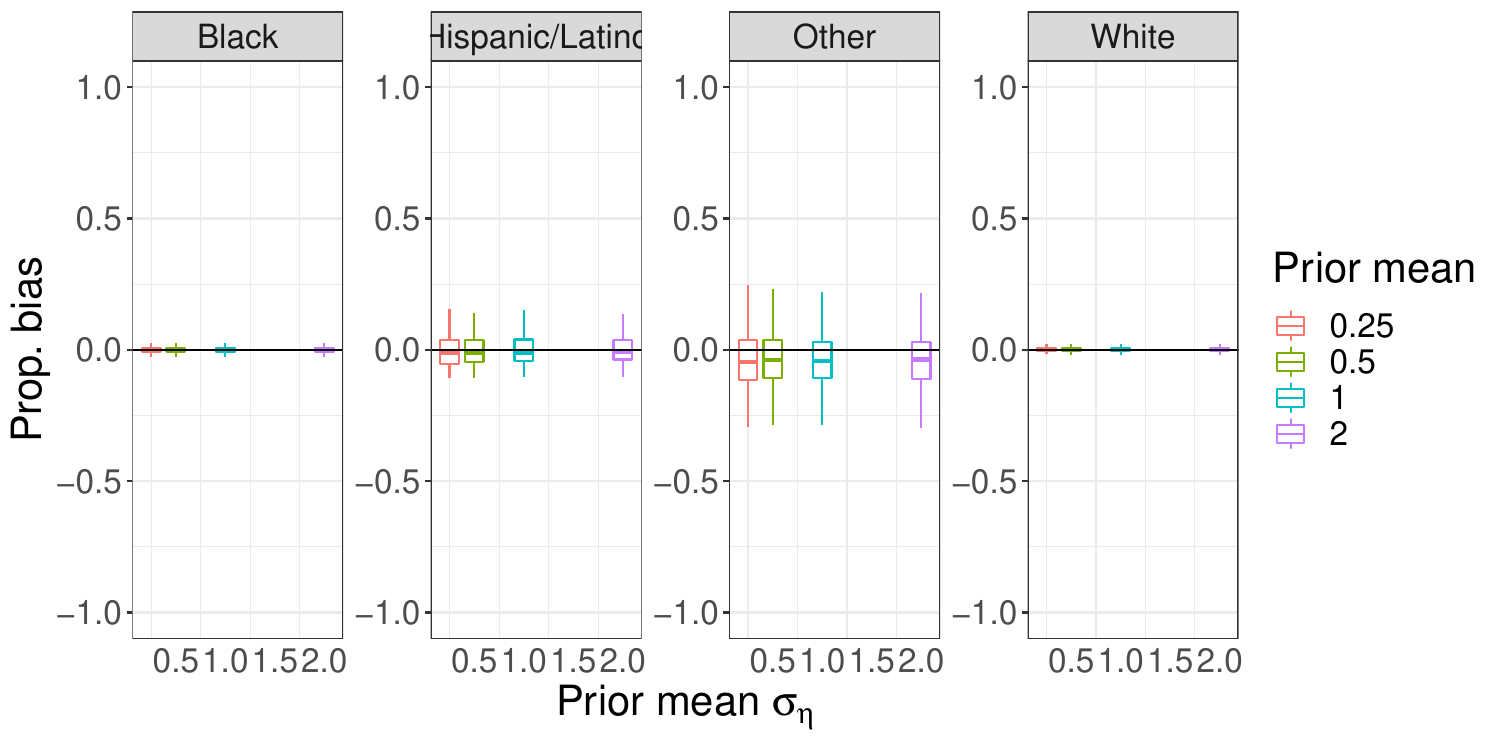}\hfill
    \includegraphics[width=0.49\textwidth]{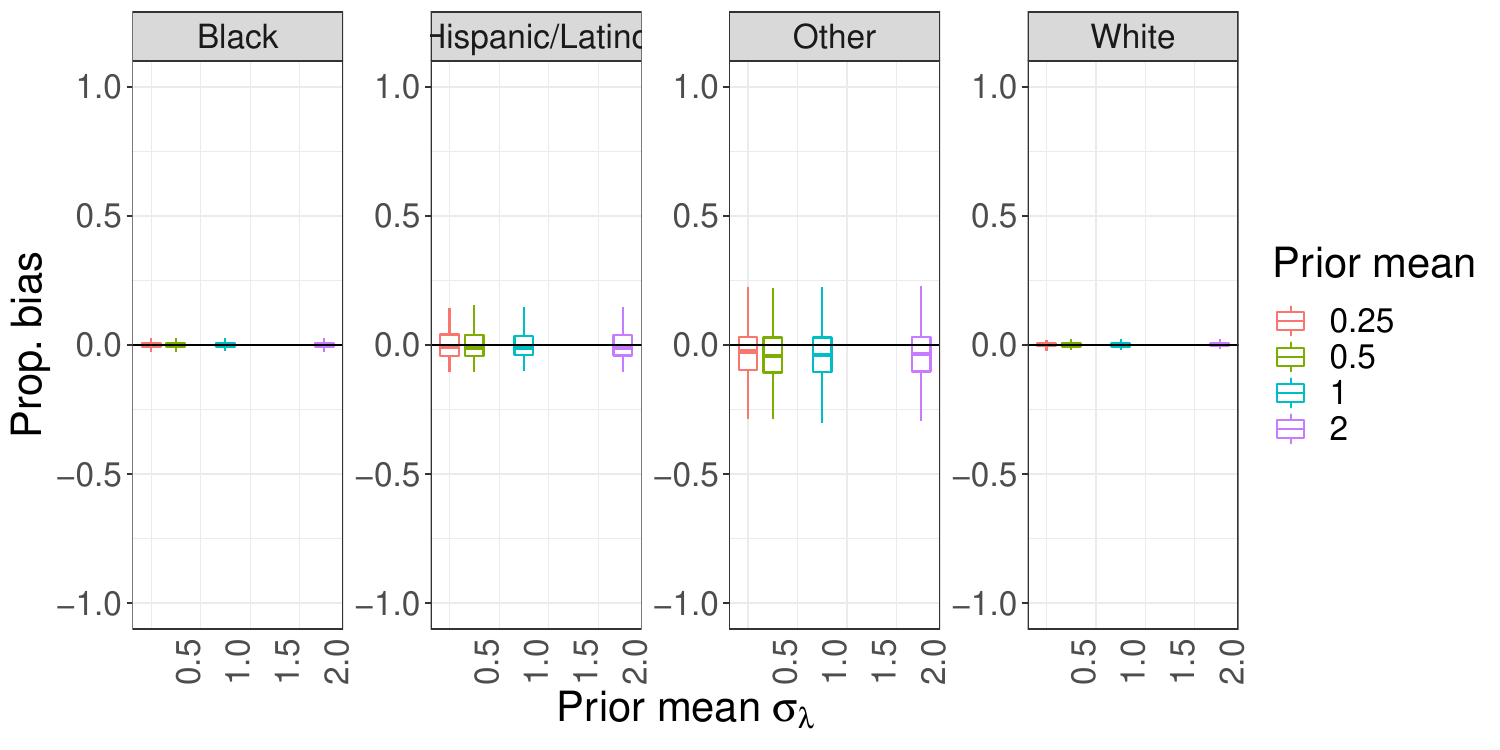}
    \\[\smallskipamount]
    \includegraphics[width=0.49\textwidth]{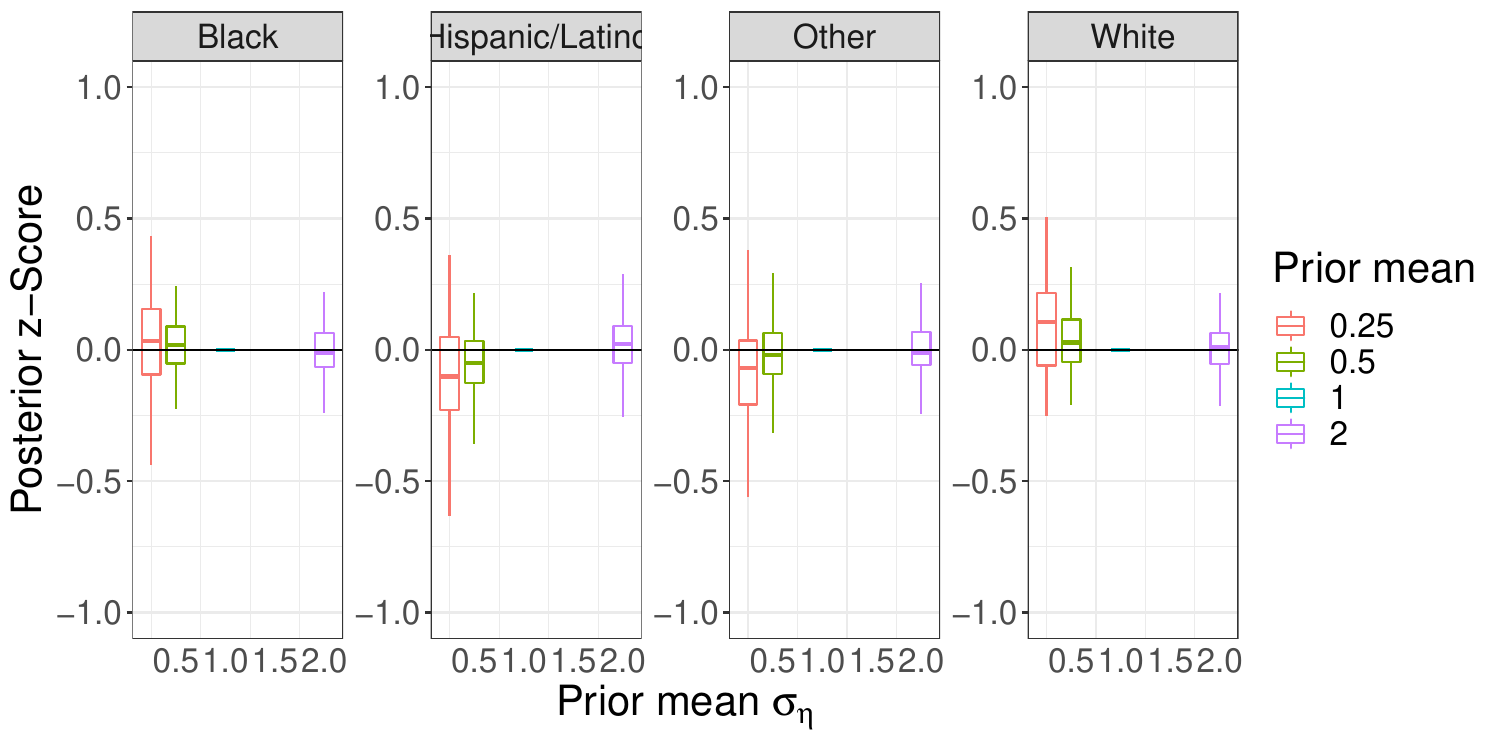}\hfill
    \includegraphics[width=0.49\textwidth]{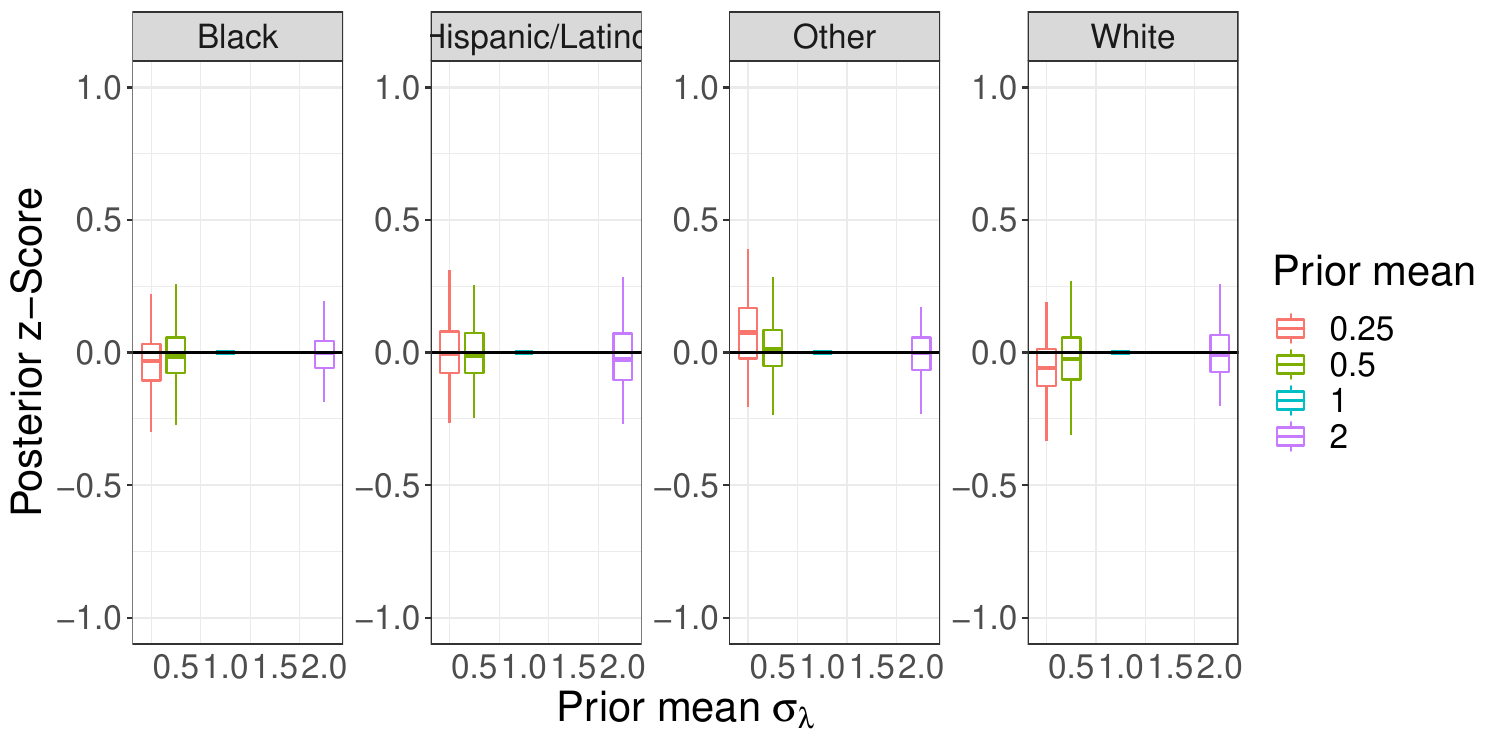}\hfill
    \caption{The graphs above show the posterior bias (\cref{eq:post-mean-bias}) and posterior z-scores (\cref{eq:post-mean-z}) for $\boldsymbol{\sigma}_{\boldsymbol{\lambda}}$ and $\boldsymbol{\sigma}_{\boldsymbol{\eta}}$ }
    \label{fig:prior-sens-sigma}
\end{figure}

\section{Further Wayne County applied data analysis results}
\subsection{Age-Race/Ethnicity posterior predictive checks}\label{app:ppc-age}

\begin{figure}[ht] 
\centering
    \includegraphics[width=.75\linewidth]{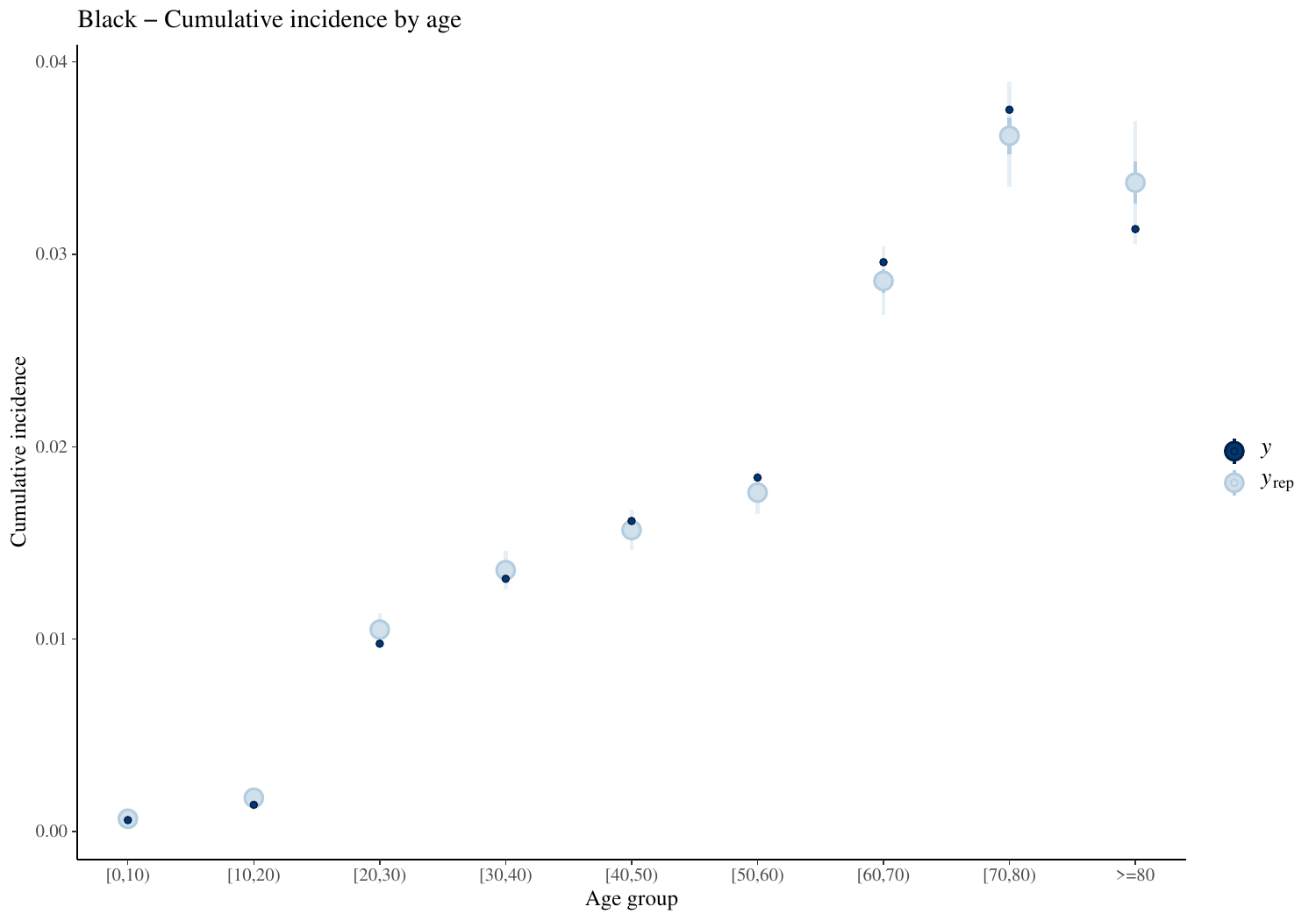} 
    \includegraphics[width=.75\linewidth]{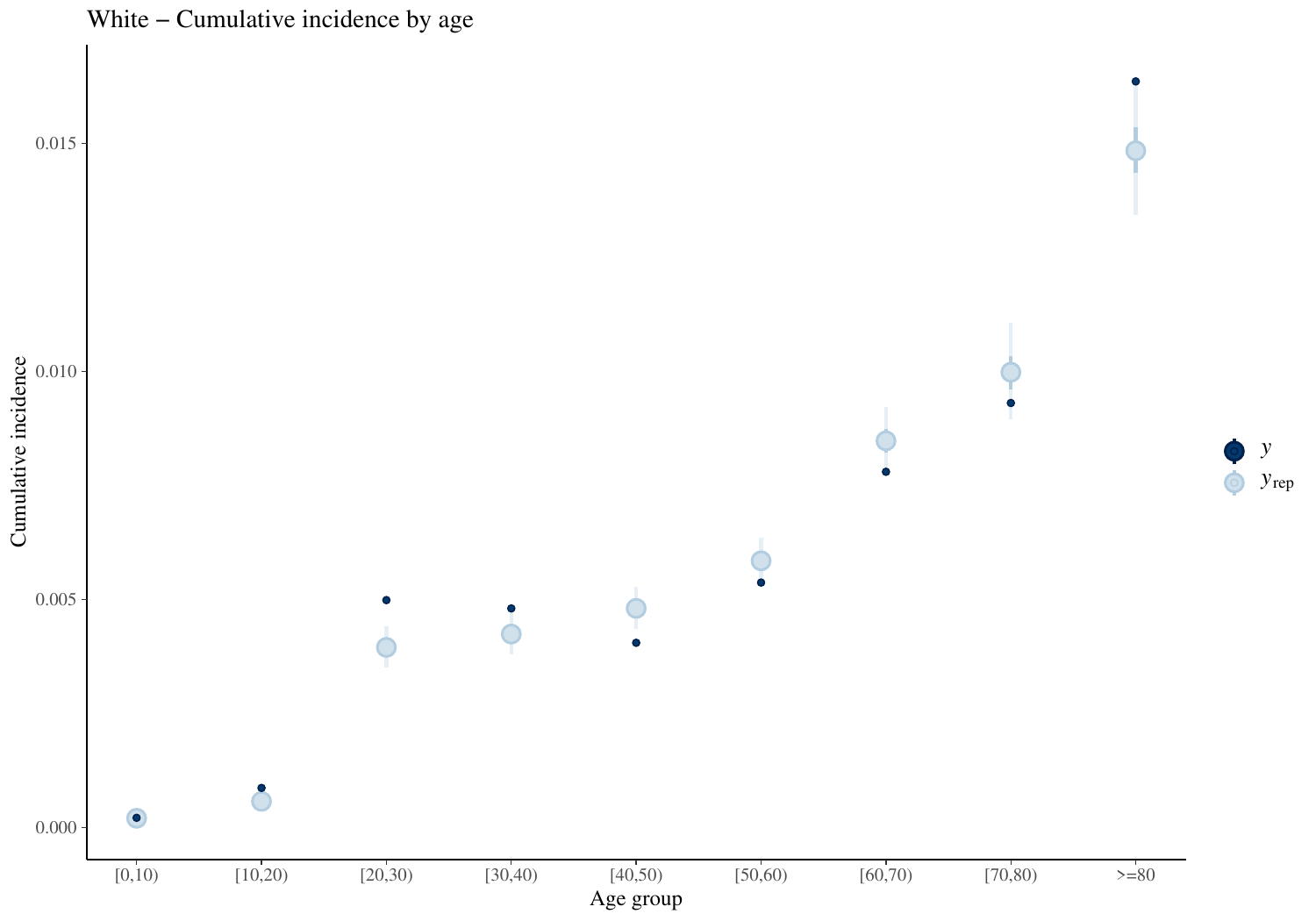} 
    \caption{Posterior predictive checks for cumulative incidence by age group by race for Blacks and Whites.}
\end{figure}

\subsection{Rootogram}\label{app:root}

\begin{figure}[ht] 
\centering
    \includegraphics[width=.75\linewidth]{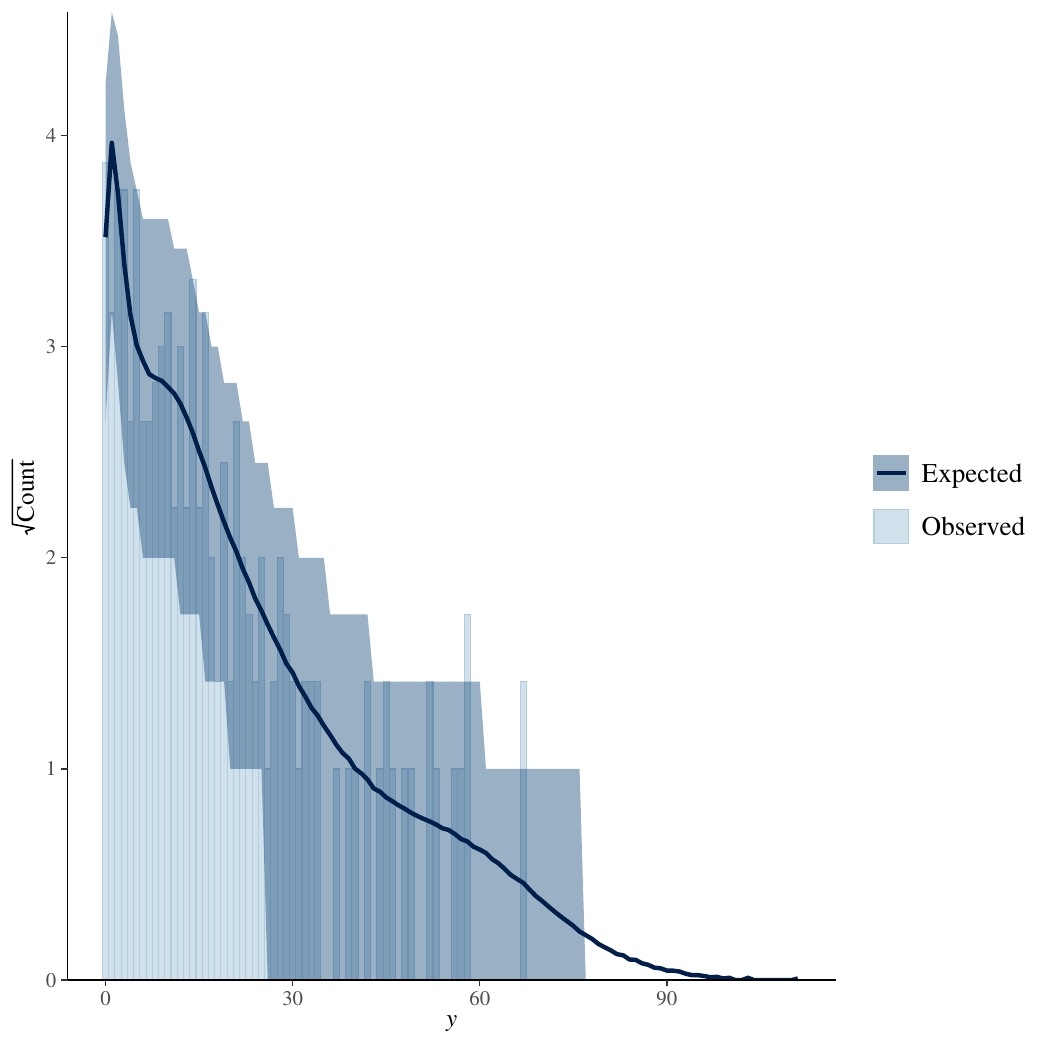} 
    \caption{Posterior predictive rootogram for missing case counts.}
\end{figure}

\subsection{Tables for posterior summaries for estimands of interest} \label{app:post-table-applied}

\begin{table}[ht]
\caption{This table presents posterior summary statistics for the Wayne-County estimands of interest. Post. mean stands for Posterior Mean, and MCSE stands for Monte Carlo Standard Error, which is the standard error in the posterior estimator, which can be estimated assuming that the MCMC central limit theorem holds. See \cite{betancourt2015hamiltonian} and \cite{vehtarirhat} for more details}
\label{tab:estimands-applied}
\centering
\tabcolsep=0.10cm
\resizebox{\textwidth}{!}{%
\begin{tabular}{@{}lccccccc@{}}
  \hline
 &  & Post. &  & 10\% Post. &  & 90\% Post. &  \\ 
Estimand & Model & Mean & MCSE& quant. & MCSE & quant. & MCSE \\ 
   \hline
$\exp\big((\boldsymbol{\alpha}_\lambda)_\text{Blacks}$
& Joint & 3.93 & 1.28e-02 & 3.16 & 1.40e-02 & 4.76 & 2.06e-02 \\ 
 $\quad \quad - (\boldsymbol{\alpha}_\lambda)_\text{Whites}\big)$  & Complete Case model & 3.56 & 1.02e-02 & 2.89 & 1.09e-02 & 4.29 & 1.79e-02 \\ 
   & Ad-Hoc MI & 2.96 & 1.80e-03 & 2.43 & 1.60e-03 & 3.52 & 2.15e-03 \\ 
   & Gibbs MI & 3.49 & 1.75e-03 & 2.78 & 1.84e-03 & 4.24 & 2.59e-03 \\ [6pt]
  $\exp\big((\boldsymbol{\alpha}_\lambda)_\text{Hispanics/Latinos}$
   & Joint & 2.10 & 6.34e-03 & 1.65 & 6.39e-03 & 2.59 & 8.22e-03 \\ 
 $ \quad \quad- (\boldsymbol{\alpha}_\lambda)_\text{Whites}\big)$  & Complete Case model & 1.93 & 4.35e-03 & 1.58 & 4.66e-03 & 2.29 & 8.52e-03 \\ 
   & Ad-Hoc MI & 1.67 & 1.83e-03 & 1.35 & 1.72e-03 & 2.01 & 1.55e-03 \\ 
   & Gibbs MI & 2.09 & 3.22e-03 & 1.65 & 2.62e-03 & 2.56 & 3.87e-03 \\ [6pt]
$\exp\big((\boldsymbol{\alpha}_\lambda)_\text{Others}$
  & Joint & 8.88 & 4.78e-02 & 6.09 & 5.51e-02 & 11.81 & 6.13e-02 \\ 
$\quad \quad - (\boldsymbol{\alpha}_\lambda)_\text{Whites}\big)$   & Complete Case model & 5.21 & 1.30e-02 & 4.26 & 1.55e-02 & 6.22 & 1.91e-02 \\ 
   & Ad-Hoc MI & 4.08 & 2.75e-03 & 3.32 & 2.52e-03 & 4.90 & 2.79e-03 \\ 
   & Gibbs MI & 5.27 & 5.03e-03 & 4.18 & 4.14e-03 & 6.42 & 4.88e-03 \\ [6pt]
  $\exp\big((\boldsymbol{\alpha}_\lambda)_\text{Asians/Pacific Islanders}$ 
   & Joint & 1.35 & 4.79e-03 & 1.01 & 5.14e-03 & 1.73 & 7.37e-03 \\ 
$ \quad \quad - (\boldsymbol{\alpha}_\lambda)_\text{Whites}\big)$   & Complete Case model & 1.29 & 3.30e-03 & 1.00 & 3.59e-03 & 1.61 & 6.06e-03 \\ 
   & Ad-Hoc MI & 1.21 & 1.98e-03 & 0.96 & 1.63e-03 & 1.48 & 2.43e-03 \\ 
   & Gibbs MI & 1.66 & 4.44e-03 & 1.22 & 3.10e-03 & 2.15 & 6.17e-03 \\ [6pt]
  $\mathbbm{SI}_\text{Blacks}$ 
   & Joint & 1.59 & 6.62e-04 & 1.55 & 1.20e-03 & 1.64 & 7.33e-04 \\ 
   & Complete Case model & 1.58 & 8.02e-05 & 1.57 & 1.44e-04 & 1.60 & 1.53e-04 \\ 
   & Ad-Hoc MI & 1.56 & 1.47e-04 & 1.55 & 1.63e-04 & 1.57 & 1.37e-04 \\ 
  & Gibbs MI & 1.60 & 1.65e-04 & 1.58 & 1.72e-04 & 1.61 & 1.65e-04 \\ [6pt]
  $\mathbbm{SI}_\text{Hispanics/Latinos}$ 
   & Joint & 1.16 & 1.59e-03 & 1.02 & 1.55e-03 & 1.30 & 2.20e-03 \\ 
   & Complete Case model & 1.17 & 3.29e-04 & 1.12 & 6.04e-04 & 1.23 & 6.36e-04 \\ 
   & Ad-Hoc MI & 1.15 & 7.02e-04 & 1.10 & 6.59e-04 & 1.20 & 7.41e-04 \\ 
  & Gibbs MI & 1.23 & 9.89e-04 & 1.17 & 1.05e-03 & 1.28 & 1.01e-03 \\ [6pt]
  $\mathbbm{SI}_\text{Others}$ 
   & Joint & 4.64 & 1.81e-02 & 3.50 & 2.24e-02 & 5.64 & 1.38e-02 \\ 
   & Complete Case model & 3.06 & 8.03e-04 & 2.93 & 1.33e-03 & 3.19 & 1.77e-03 \\ 
   & Ad-Hoc MI & 2.68 & 1.21e-03 & 2.57 & 1.03e-03 & 2.80 & 1.38e-03 \\ 
   & Gibbs MI & 3.15 & 2.40e-03 & 3.01 & NA & 3.28 & 2.69e-03 \\ [6pt]
  $\mathbbm{SI}_\text{Asians/Pacific Islanders}$ 
   & Joint & 0.61 & 1.09e-03 & 0.53 & 4.95e-04 & 0.71 & 2.52e-03 \\ 
   & Complete Case model & 0.65 & 3.27e-04 & 0.60 & 7.57e-04 & 0.71 & 7.64e-04 \\ 
   & Ad-Hoc MI & 0.66 & 8.58e-04 & 0.61 & 8.63e-04 & 0.72 & 9.36e-04 \\ 
   & Gibbs MI & 0.72 & 1.14e-03 & 0.66 & NA & 0.78 & 1.34e-03 \\ [6pt]
  $\mathbbm{SI}_\text{Whites}$ 
   & Joint & 0.46 & 4.25e-04 & 0.44 & 3.98e-04 & 0.49 & 4.65e-04 \\ 
   & Complete Case model & 0.52 & 5.20e-05 & 0.52 & 7.63e-05 & 0.53 & 1.08e-04 \\ 
   & Ad-Hoc MI & 0.55 & 1.00e-04 & 0.54 & 9.79e-05 & 0.56 & 1.02e-04 \\ 
   & Gibbs MI & 0.50 & 9.26e-05 & 0.49 & 1.11e-04 & 0.51 & 8.00e-05 \\ [6pt]
$\mathbbm{I}_\text{Blacks}/\mathbbm{I}_\text{Whites}$ 
   & Joint & 3.01 & 2.86e-03 & 2.84 & 2.98e-03 & 3.19 & 3.31e-03 \\ 
   & Complete Case model & 2.63 & 3.77e-04 & 2.57 & 5.53e-04 & 2.70 & 7.52e-04 \\ 
   & Ad-Hoc MI & 2.49 & 6.53e-04 & 2.44 & 6.89e-04 & 2.55 & 6.51e-04 \\ 
   & Gibbs MI & 2.79 & 7.12e-04 & 2.73 & 6.31e-04 & 2.85 & 8.37e-04 \\ [6pt]
  $\mathbbm{I}_\text{Hispanics/Latinos}/$ 
   & Joint & 1.69 & 3.26e-03 & 1.47 & 3.26e-03 & 1.92 & 4.36e-03 \\ 
$\mathbbm{I}_\text{Whites}$   & Complete Case model & 1.50 & 4.66e-04 & 1.42 & 9.17e-04 & 1.58 & 1.13e-03 \\ 
   & Ad-Hoc MI & 1.42 & 9.31e-04 & 1.35 & 9.26e-04 & 1.49 & 9.86e-04 \\ 
   & Gibbs MI & 1.66 & 1.43e-03 & 1.57 & 1.44e-03 & 1.74 & 1.50e-03 \\ [6pt]
  $\mathbbm{I}_\text{Others}/\mathbbm{I}_\text{Whites}$ 
   & Joint & 6.55 & 2.96e-02 & 4.74 & 3.24e-02 & 8.12 & 2.06e-02 \\ 
   & Complete Case model & 3.78 & 1.11e-03 & 3.60 & 2.05e-03 & 3.96 & 2.31e-03 \\ 
   & Ad-Hoc MI & 3.20 & 1.67e-03 & 3.04 & 1.40e-03 & 3.35 & 1.90e-03 \\ 
   & Gibbs MI & 4.10 & 3.36e-03 & 3.90 & NA & 4.30 & 4.10e-03 \\ [6pt]
  $\mathbbm{I}_\text{Asians/Pacific Islanders}/$
     & Joint & 1.08 & 2.59e-03 & 0.92 & 1.60e-03 & 1.27 & 5.65e-03 \\ 
$\mathbbm{I}_\text{Whites}$   & Complete Case model & 1.01 & 5.26e-04 & 0.92 & 1.04e-03 & 1.10 & 1.02e-03 \\ 
   & Ad-Hoc MI & 0.99 & 1.32e-03 & 0.91 & NA & 1.07 & 1.39e-03 \\ 
  & Gibbs MI & 1.18 & 1.89e-03 & 1.08 & NA & 1.28 & 2.25e-03 \\ 
   \hline
\end{tabular}}
\end{table}

\begin{table}[ht]
\caption{The table shows sampling efficiency for population estimands of interest presented in table \ref{tab:estimands-applied}. ESS stands for effective sample size; Bulk ESS and Tail ESS are measures of the equivalent number of independent samples generated from a MCMC procedure. See \cite{vehtarirhat} for more detail. Bulk and Tail ESS efficiency are the Bulk and Tail ESS figures divided by the total number of MCMC samples, which is $16,000$. As noted in \cite{vehtarirhat} it is possible for MCMC samplers to generate Tail and Bulk ESS values that are larger than the total number of samples.}
\label{tab:sampling-eff}
\centering
\resizebox{\textwidth}{!}{%
\begin{tabular}{@{}lllrrcc@{}}
  \hline
Estimand & Model & $\hat{R}$ & Bulk ESS & Tail ESS & Bulk ESS eff. & Tail ESS eff. \\ 
  \hline
 $\exp\big((\boldsymbol{\alpha}_\lambda)_\text{Blacks}$ 
  & Joint & 1.00 & 2465 & 4753 & 0.15 & 0.30 \\ 
$\quad \quad - (\boldsymbol{\alpha}_\lambda)_\text{Whites}\big)$   & Complete Case model & 1.00 & 3037 & 5645 & 0.19 & 0.35 \\ 
    & Ad-Hoc MI & 1.01 & 57036 & 223009 & 0.07 & 0.28 \\ 
   & Gibbs MI & 1.00 & 114009 & 259368 & 0.14 & 0.32 \\ [6pt]
$\exp\big((\boldsymbol{\alpha}_\lambda)_\text{Hispanics/Latinos}$    
   & Joint & 1.00 & 3506 & 6994 & 0.22 & 0.44 \\ 
$\quad \quad - (\boldsymbol{\alpha}_\lambda)_\text{Whites}\big)$   & Complete Case model & 1.00 & 4440 & 6910 & 0.28 & 0.43 \\ 
    & Ad-Hoc MI & 1.01 & 19965 & 91952 & 0.02 & 0.11 \\ 
   & Gibbs MI & 1.02 & 12435 & 68983 & 0.02 & 0.09 \\ [6pt]
$\exp\big((\boldsymbol{\alpha}_\lambda)_\text{Others}$  
   & Joint & 1.00 & 2098 & 4315 & 0.13 & 0.27 \\ 
$\quad \quad - (\boldsymbol{\alpha}_\lambda)_\text{Whites}\big)$   & Complete Case model & 1.00 & 3742 & 6528 & 0.23 & 0.41 \\ 
     & Ad-Hoc MI & 1.01 & 51519 & 256045 & 0.06 & 0.32 \\ 
   & Gibbs MI & 1.01 & 30320 & 220330 & 0.04 & 0.28 \\ [6pt]
$\exp\big((\boldsymbol{\alpha}_\lambda)_\text{Asians/Pacific Islanders}$ 
   & Joint & 1.00 & 3850 & 7799 & 0.24 & 0.49 \\ 
$\quad \quad - (\boldsymbol{\alpha}_\lambda)_\text{Whites}\big)$   & Complete Case model & 1.00 & 5511 & 8283 & 0.34 & 0.52 \\ 
    & Ad-Hoc MI & 1.02 & 11078 & 45632 & 0.01 & 0.06 \\ 
   & Gibbs MI & 1.03 & 7298 & 30332 & 0.01 & 0.04 \\ [6pt]
  $\mathbbm{SI}_\text{Blacks}$ 
    & Joint & 1.00 & 2531 & 4877 & 0.16 & 0.30 \\ 
   & Complete Case model & 1.00 & 16204 & 14672 & 1.01 & 0.92 \\ 
     & Ad-Hoc MI & 1.05 & 4322 & 15046 & 0.01 & 0.02 \\ 
   & Gibbs MI & 1.06 & 3490 & 13738 & 0.00 & 0.02 \\ [6pt]
  $\mathbbm{SI}_\text{Hispanics/Latinos}$ 
     & Joint & 1.00 & 4523 & 9442 & 0.28 & 0.59 \\ 
   & Complete Case model & 1.00 & 16596 & 12657 & 1.04 & 0.79 \\ 
      & Ad-Hoc MI & 1.07 & 3392 & 14038 & 0.00 & 0.02 \\ 
   & Gibbs MI & 1.12 & 1990 &  & 0.00 & NA \\ [6pt]
 $\mathbbm{SI}_\text{Others}$ 
   & Joint & 1.00 & 2034 & 4928 & 0.13 & 0.31 \\ 
   & Complete Case model & 1.00 & 16338 & 11881 & 1.02 & 0.74 \\ 
      & Ad-Hoc MI & 1.04 & 5584 & 18672 & 0.01 & 0.02 \\ 
   & Gibbs MI & 1.12 & 1938 &  & 0.00 & NA \\ [6pt]
 $\mathbbm{SI}_\text{Asians/Pacific Islanders}$  
    & Joint & 1.00 & 6412 & 6513 & 0.40 & 0.41 \\ 
   & Complete Case model & 1.00 & 17496 & 11891 & 1.09 & 0.74 \\ 
      & Ad-Hoc MI & 1.10 & 2439 &  & 0.00 & NA \\ 
   & Gibbs MI & 1.15 & 1673 &  & 0.00 & NA \\ [6pt]
$\mathbbm{SI}_\text{Whites}$   
   & Joint & 1.00 & 1722 & 5572 & 0.11 & 0.35 \\ 
   & Complete Case model & 1.00 & 16277 & 13527 & 1.02 & 0.85 \\ 
      & Ad-Hoc MI & 1.05 & 4083 & 17557 & 0.01 & 0.02 \\ 
   & Gibbs MI & 1.05 & 4488 &  & 0.01 & NA \\ [6pt]
$\mathbbm{I}_\text{Blacks}/\mathbbm{I}_\text{Whites}$   
   & Joint & 1.00 & 2197 & 5606 & 0.14 & 0.35 \\ 
   & Complete Case model & 1.00 & 16184 & 13675 & 1.01 & 0.85 \\ 
      & Ad-Hoc MI & 1.05 & 4225 & 16200 & 0.01 & 0.02 \\ 
   & Gibbs MI & 1.05 & 4673 &  & 0.01 & NA \\ [6pt]
$\mathbbm{I}_\text{Hispanics/Latinos}/$ 
  & Joint & 1.00 & 2928 & 6147 & 0.18 & 0.38 \\ 
$\mathbbm{I}_\text{Whites}$   & Complete Case model & 1.00 & 16631 & 12370 & 1.04 & 0.77 \\ 
      & Ad-Hoc MI & 1.06 & 3557 & 14586 & 0.00 & 0.02 \\ 
   & Gibbs MI & 1.11 & 2144 &  & 0.00 & NA \\ [6pt]
$\mathbbm{I}_\text{Others}/\mathbbm{I}_\text{Whites}$ 
  & Joint & 1.00 & 1933 & 4788 & 0.12 & 0.30 \\ 
  & Complete Case model & 1.00 & 16306 & 11978 & 1.02 & 0.75 \\ 
     & Ad-Hoc MI & 1.04 & 5156 &  & 0.01 & NA \\ 
   & Gibbs MI & 1.11 & 2114 &  & 0.00 & NA \\ [6pt]
$\mathbbm{I}_\text{Asians/Pacific Islanders}/$ 
   & Joint & 1.00 & 3827 & 6402 & 0.24 & 0.40 \\ 
$\mathbbm{I}_\text{Whites}$   & Complete Case model & 1.00 & 17423 & 11926 & 1.09 & 0.75 \\ 
      & Ad-Hoc MI & 1.10 & 2447 &  & 0.00 & NA \\ 
   & Gibbs MI & 1.14 & 1731 &  & 0.00 & NA \\ 
   \hline
\end{tabular}}
\end{table}

\begin{table}[ht]
\caption{The table shows the posterior means, 80\% credible interval endpoints and the Monte Carlo standard errors of these estimates. $\text{CC}$ stands for the complete-case model while $\text{J}$ stands for the joint model.}
\label{tab:miss-estimands}
\centering
\resizebox{\textwidth}{!}{%
\begin{tabular}{@{}lcccccc@{}}
  \hline
 & Post. &  & 10\% Post. &  & 90\% Post. & \\ 
Estimand  & Mean & MCSE & quant. & MCSE  & quant. & MCSE \\ 
  \hline
$\mathbb{I}^{\text{CC}}_\text{Blacks} / \mathbb{I}^\text{J}_\text{Blacks}$ 
& 0.81 & 3.50e-04 & 0.79 & 3.55e-04 & 0.84 & 5.80e-04 \\ [2pt]
 $\mathbb{I}^{\text{CC}}_\text{Hispanics/Latinos} / \mathbb{I}^\text{J}_\text{Hispanics/Latinos}$  
  & 0.83 & 1.14e-03 & 0.73 & 1.14e-03 & 0.94 & 1.70e-03 \\[2pt]
  $\mathbb{I}^{\text{CC}}_\text{Others} / \mathbb{I}^\text{J}_\text{Others}$
   & 0.55 & 2.41e-03 & 0.44 & 1.36e-03 & 0.71 & 4.26e-03 \\ [2pt]
  $\mathbb{I}^{\text{CC}}_\text{Asians/Pacific Islanders} / \mathbb{I}^\text{J}_\text{Asians/Pacific Islanders}$ 
   & 0.88 & 1.42e-03 & 0.73 & 2.50e-03 & 1.03 & 1.71e-03 \\ [2pt]
  $\mathbb{I}^{\text{CC}}_\text{Whites} / \mathbb{I}^\text{J}_\text{Whites}$ 
   & 0.93 & 8.51e-04 & 0.88 & 7.85e-04 & 0.98 & 7.79e-04 \\[2pt]
  $\Prob{\text{Race observed}}_\text{Blacks}$ 
   & 0.85 & 7.55e-04 & 0.81 & 7.25e-04 & 0.91 & 1.35e-03 \\ [2pt]
  $\Prob{\text{Race observed}}_\text{Hispanics/Latinos}$ 
   & 0.87 & 8.63e-04 & 0.78 & 1.43e-03 & 0.95 & 6.59e-04 \\ [2pt]
  $\Prob{\text{Race observed}}_\text{Others}$ 
   & 0.58 & 2.93e-03 & 0.45 & 1.20e-03 & 0.77 & 5.31e-03 \\ [2pt]
  $\Prob{\text{Race observed}}_\text{Asians/Pacific Islanders}$ 
   & 0.90 & 8.25e-04 & 0.81 & 2.36e-03 & 0.97 & 3.80e-04 \\ [2pt]
  $\Prob{\text{Race observed}}_\text{Whites}$ 
   & 0.94 & 7.09e-04 & 0.89 & 8.70e-04 & 0.98 & 4.33e-04 \\ 
   \hline
\end{tabular}}
\end{table}

\clearpage

\section{Stan code for negative binomial likelihood} \label{app:dynamic}

The following Stan code computes the likelihood related to the following generative model using an efficient dynamic programming algorithm :
\begin{align*}
Y_{ij} & \sim \text{Binomial}(E_{ij}, \theta_{ij}) \\
X_{ij} | Y_{ij} & \sim \text{Binomial}(Y_{ij}, p_{ij})
\end{align*}
after marginalizing over all combinations of $Y_{ij}$ such that $\sum_{j} Y_{ij} = T$ where $T$ is the total identified cases of disease in stratum $i$, a known quantity.  

The code was derived from \cite{poisbinom}.

\begin{lstlisting}[language=Stan]
functions {
  real binomial_2_lpmf(int y_obs, int y_miss,
                              real p, real theta, int E) {
    return binomial_lpmf(y_obs | y_miss, p)
           + binomial_lpmf(y_miss | E, theta);
  }
  real miss_lpmf(int[] y, int n_miss, 
                 vector p, vector theta,
                 int[] E) {
    int N = rows(theta);
    real alpha[N + 1, n_miss + 1];
  
  // alpha[n + 1, tot + 1] = log p of tot missing cases
  // distributed among first n categories 
    alpha[1, 1:(n_miss + 1)] = rep_array(0, n_miss+1);
    for (n in 1:N) {
      // tot = 0
      alpha[n + 1, 1] = alpha[n, 1] 
          + binomial_2_lpmf(y[n]|y[n],p[n],theta[n], E[n]);
      
      // 0 < tot < n
        
        for (tot in 1:n_miss) {
          if (n > 1) {
            vector[tot + 1] vec;
              for (i in 1:(tot + 1)) {
                vec[i] = alpha[n,i] 
                + binomial_2_lpmf(y[n] | 
                    y[n] + tot - (i - 1), 
                    p[n],theta[n], E[n]);
              }
              alpha[n + 1, tot + 1] = log_sum_exp(vec);
          } else {
              alpha[n + 1,tot + 1] 
               = binomial_2_lpmf(y[n]| y[n] 
               + tot,p[n],theta[n], E[n]);
          }
        }
    }
    return alpha[N + 1, n_miss + 1];
  }
}
\end{lstlisting}

\end{appendix}

\begin{acks}[Acknowledgments]
 We would like to thank Mitzi Morris, Andrew Gelman, and Bob Carpenter for their feedback on an earlier draft of the paper. 
\end{acks}

\begin{funding}
JZ \& RT were supported by award \#6 U01 IP00113801-01 from the U.S. Centers for Disease Control and Prevention, and award \#812255 from the Simons Foundation.
This research was supported in part through computational resources and services provided by Advanced Research Computing (ARC), a division of Information and Technology Services (ITS) at the University of Michigan, Ann Arbor.
\end{funding}

\clearpage

\bibliographystyle{imsart-nameyear} 
\bibliography{missing-categorical-data.bib}       

\end{document}